\documentclass[preprint]{aastex_noabbrev}
\usepackage{natbib}
\usepackage{mathrsfs}
\usepackage{rotating}
\usepackage{graphicx}
\usepackage{epstopdf}
\usepackage{longtable,lscape}

\newcommand{\ud}{\mathrm{d}}

\newcommand{\fn}[1]{\footnote{\scriptsize{#1}}}
\newcommand{\Eqn}[1]{Eq{#1}.}  
\newcommand{\Fig}[1]{Fig{#1}.}  
\newcommand{\Cassit}{\textit{Cassini}}  
\newcommand{\Galit}{\textit{Galileo}}  
\newcommand{\Voyit}[1]{\textit{Voyager{#1}}}  
\newcommand{\Pionit}[1]{\textit{Pioneer{#1}}}  
\newcommand{\NHit}{\textit{New~Horizons}}  
\newcommand{\MGSit}{\textit{Mars Global Surveyor}}  
\newcommand{\Junoit}{\textit{Juno}}  

\newcommand\areps{Ann.~Rev.~Earth~Planet.~Sci.}%
\newcommand\cmda{Cel.~Mech.~Dyn.~Astron.}%

\shorttitle{}
\shortauthors{Tiscareno}

\begin{document}

\title{\vspace{-0.15in}Planetary Rings}
\author{Matthew~S.~Tiscareno}
\affil{Center for Radiophysics and Space Research, Cornell University, Ithaca, NY 14853\\\texttt{matthewt@astro.cornell.edu}}

\begin{abstract}

\end{abstract}

Planetary rings are the only nearby astrophysical disks, and the only disks that have been investigated by spacecraft (especially the \Cassit{} spacecraft orbiting Saturn).  Although there are significant differences between rings and other disks, chiefly the large planet/ring mass ratio that greatly enhances the flatness of rings (aspect ratios as small as $10^{-7}$), understanding of disks in general can be enhanced by understanding the dynamical processes observed at close-range and in real-time in planetary rings. 

We review the known ring systems of the four giant planets, as well as the prospects for ring systems yet to be discovered.  We then review planetary rings by type.  The A, B, and C rings of Saturn, plus the Cassini Division, comprise our solar system's only dense broad disk and host many phenomena of general application to disks including spiral waves, gap formation, self-gravity wakes, viscous overstability and normal modes, impact clouds, and orbital evolution of embedded moons.  Dense narrow rings are found both at Uranus (where they comprise the main rings entirely) and at Saturn (where they are embedded in the broad disk), and are the primary natural laboratory for understanding shepherding and self-stability.  Narrow dusty rings, likely generated by embedded source bodies, are surprisingly found to sport azimuthally-confined arcs at Neptune, Saturn, and Jupiter.  Finally, every known ring system includes a substantial component of diffuse dusty rings. 

Planetary rings have shown themselves to be useful as detectors of planetary processes around them, including the planetary magnetic field and interplanetary impactors as well as the gravity of nearby perturbing moons.  Experimental rings science has made great progress in recent decades, especially numerical simulations of self-gravity wakes and other processes but also laboratory investigations of coefficient of restitution and spectroscopic ground truth.  The age of self-sustained ring systems is a matter of debate; formation scenarios are most plausible in the context of the early solar system, while signs of youthfulness indicate at least that rings have never been static phenomena. 


\section{Introduction \label{Intro}}

Planetary rings come in a diverse array of shapes and sizes.  They may be broad or narrow, dense or tenuous, dusty or not, and they may contain various kinds of structures including arcs, wavy edges, embedded moonlets, and radial variations.  Rings share the defining characteristic of a swarm of objects orbiting a central planet with vertical motions that are small compared to their motions within a common plane.  The latter arises because planets in our solar system (with the exceptions of Mercury and Venus, which have no known natural material in orbit) are fast-enough rotaters that their shapes are dominated by an equatorial bulge that adds a strong quadrupole moment ($J_2$) to their gravity fields (see Section~\ref{OrbElems}).  

This is a major contrast between rings and other astrophysical disks, which are not defined by asymmetry in an external gravity field but by the average angular momentum of the disk itself (in both cases, once a preferred plane is established, collisions among particles damp out the motions perpendicular to it).  However, rings do have a number of similarities with other astrophysical disks, which add to the motivation for studying them.  Unlike other known disk systems that are either many light-years away or (like the early stages of our solar system) far back in time, planetary rings can be studied up close and in real time.  Thus, it is worthwhile to consider the parallels that can be drawn between planetary rings and the study of other disks (see Section~\ref{OtherDisks}).  

In this chapter, after some further introductory notes on important concepts (Sections~\ref{OrbElems} 
through~\ref{OpticalDepth}), we will give an overview of the known ring systems, as well as systems where rings are unconfirmed but plausible, in Section~\ref{RingsByPlanet}.  More detailed descriptions can be found in Section~\ref{RingsByType}, which contains a discussion of various ring structures organized by type, with a focus on finding commonalities among rings in different locations that share certain qualities.  In Section~\ref{Experiments} we will discuss experimental methods of learning about rings, and in Section~\ref{AgeOrigin} we will discuss the age and origin of ring systems.  Finally, in Section~\ref{OtherDisks}, we will discuss ways that planetary rings can illuminate the study of other astrophysical disks. 

\subsection{Orbital elements \label{OrbElems}}

Rings are fundamentally populations of orbiting material.  Therefore, to discuss structure within rings, we will occasionally refer to one or more of the six \textit{orbital elements} that describe the orbit of any object around another object.  These six parameters are simply a transformation of the six Cartesian parameters for position and velocity ($x$, $y$, $z$, $\dot{x}$, $\dot{y}$, and $\dot{z}$) under the assumption that the object moves in the gravity field of a point mass (hereafter the ``planet'').  The derivation can be found in any textbook on orbital mechanics \citep[e.g., Section~2.8 in][]{MD99}.  

\begin{figure*}[!t]
\begin{center}
\includegraphics[height=5cm]{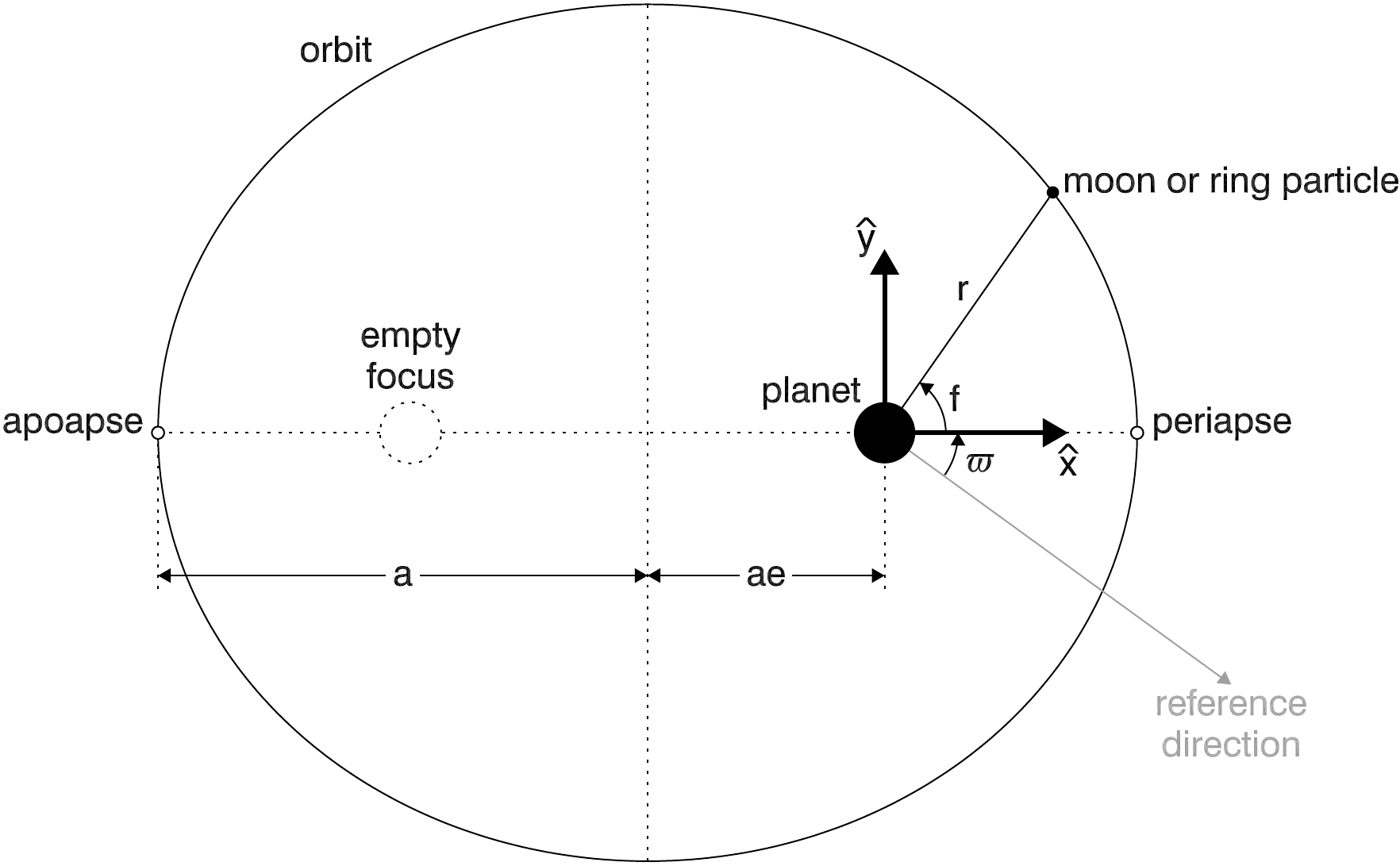}
\includegraphics[height=5cm]{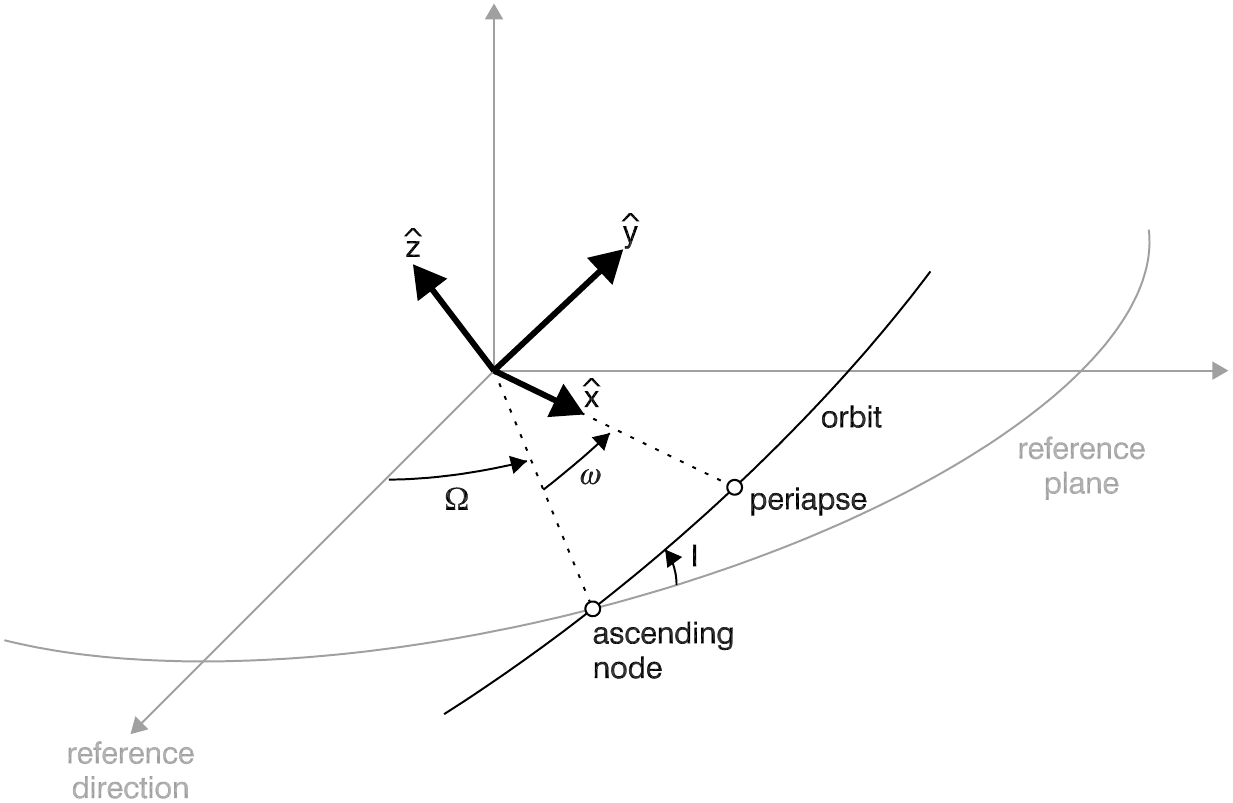}
\caption{The geometry of (left) an elliptical orbit within the orbit plane and (right) the orbit plane within 3-D space.  
\label{OrbElemsFig}}
\end{center}
\end{figure*}

A diagram of the orbit in space is found in \Fig{}~\ref{OrbElemsFig}.  The size of the orbit, and its gravitational potential energy, is described by the semimajor axis $a$, which is the mean distance between the orbiting particle and the planet.  The shape of the orbit is described by the eccentricity $e$; for a circular orbit $e=0$, but real orbits that remain bound to the planet take elliptical shapes with $0 < e < 1$, with most ring particles having $e \ll 1$.  Unbound orbits, either parabolic or hyperbolic, have $e \geq 1$.  The orbit plane may be inclined with respect to the reference plane (for ring applications, this is often the planet's equatorial plane), by an angle known as the inclination $I$.  A non-zero inclination requires an account of the orbit plane's orientation, and thus its line of intersection with the reference plane (the ``line of nodes'') is described by the longitude of the ascending node $\Omega$, measured with respect to a reference axis.  Similarly, a non-zero eccentricity requires an account of the orientation of the ellipse within the orbit plane, and thus the line connecting the planet to the location of the particle's closest approach (its ``periapse'') is described by the argument of periapse $\omega$.  Finally, once the orbit has been defined by the five parameters already mentioned, the particle's position along the orbit can be given by its actual position (the true anomaly $f$) or its time-averaged position (the mean anomaly $M$) relative to periapse.  Also commonly used are the longitude of periapse $\varpi = \Omega + \omega$ and the mean longitude $\lambda = \Omega + \omega + M$, which are not physical angles since they are the sums of angles not necessarily in the same plane, but they have the virtue of being reckoned from a stationary reference axis rather than a moving line and are useful as long as $I$ is not too large.  

The osculating orbital elements, which are most simply calculated and most often used, assume that the planet's gravity field is that of a point mass.  But for ring applications, the known planets are oblate (or bulged at the equators) due to their fast rotation.  This is adequately described by adding to the account of the gravity field a positive quadrupole moment $J_2$ \citep[for details see, e.g., Section 4.5 of][]{MD99}, though it may be necessary to further include higher moments for applications requiring great precision.  The presence of a non-zero $J_2$, in addition to defining the Laplace plane\fn{The Laplace plane is the plane about which orbits precess.  When the vertical motions of objects are damped by mutual collisions, material will settle into a ring centered on the Laplace plane.} for orbits near the planet, causes orbits to precess, in the prograde direction for apses ($\dot{\varpi} > 0$) and in the retrograde direction for nodes ($\dot{\Omega} < 0$).  

A non-zero $J_2$ also compromises the physical meaningfulness of the osculating elements, especially for low-eccentricity orbits, introducing fast (i.e., orbit-frequency) variations in all six elements.  The physical meaningfulness of orbital elements can be restored using a revised system of \textit{epicyclic orbital elements} \citep{BL87,LB91,BRL94}, which are based on the geometrical shape of streamlines.  These put the orbit-frequency variations back into an analogue of $\lambda$, leaving the other five elements to again describe a static (or at least slowly-varying) orbit.  A useful algorithm for converting Cartesian coordinates into epicyclic orbital elements was devised by \citet{RS06}. 

\subsection{Roche limits, Roche lobes, and Roche critical densities \label{Roche}}

The ``Roche limit'' is the distance from a planet within which its tides can pull apart a compact object.  Simply speaking, a ring would be expected to reside inside its planet's Roche limit, while any disk of material beyond that distance would be expected to accrete into one or more moons.  However, the Roche limit does not actually have a single value, but depends particularly on the density and internal material strength of the moon that may or may not get pulled apart \citep{CE95}.  A simple value for the Roche limit can be calculated from a balance between the tidal force (i.e., the difference between the planet's gravitational pull on one side of the moon and its pull on the other side) that would tend to pull a moon apart, and the moon's own gravity that would tend to hold it together.  This works out to \citep[e.g., \Eqn{}~4.131 in][]{MD99}
\begin{equation}
\label{RocheLimitEqn}
a_{\mathrm{Roche}} = R_{\mathrm{p}} \left( \frac{4 \pi \rho_\mathrm{p}}{\gamma \rho} \right)^{1/3} , 
\end{equation}
where $R$ is radius and $\rho$ is internal density, and the subscript ``p'' denotes the central planet.   

The dimensionless geometrical parameter $\gamma = 4 \pi / 3 \approx 4.2$ for a sphere, but is smaller for an object that takes a non-spherical shape with its long axis pointing toward Saturn, as one would expect for an actively accreting body and as at least several of Saturn's ring-moons appear to do \citep{PorcoSci07,Charnoz07}.  Simply distributing the moon's material into the shape of its Roche lobe, with uniform density, yields $\gamma \approx 1.6$ \citep{PorcoSci07}.  However, fully accounting for the feedback between the moon's distorted shape and its (now non-point-mass) gravity field \citep{Chandra69,MD99} leads to a rather smaller value, $\gamma \approx 0.85$; on the other hand, some central mass concentration and the failure of a rubble pile to exactly take its equilibrium shape will likely prevent $\gamma$ from becoming quite this low.

Note that the moon's internal density $\rho$ appears in \Eqn{}~\ref{RocheLimitEqn}.  Thus the Roche limit is variable; a denser object can venture closer to the planet without danger than can an object that is less dense.  The moon's diameter, on the other hand, does not appear in \Eqn{}~\ref{RocheLimitEqn}.  Why then do we commonly imagine a large object getting pulled into smaller pieces when it ventures inside the Roche limit?  This is because the Roche limit has been defined here as the distance within which an object can no longer be held together \textit{by its own gravity}.  Size becomes relevant for objects small enough to be held together by their internal material strength (which is not considered in \Eqn{}~\ref{RocheLimitEqn}) in spite of the tidal forces that enter into the Roche calculation.

\begin{figure*}[!t]
\begin{center}
\includegraphics[width=8cm]{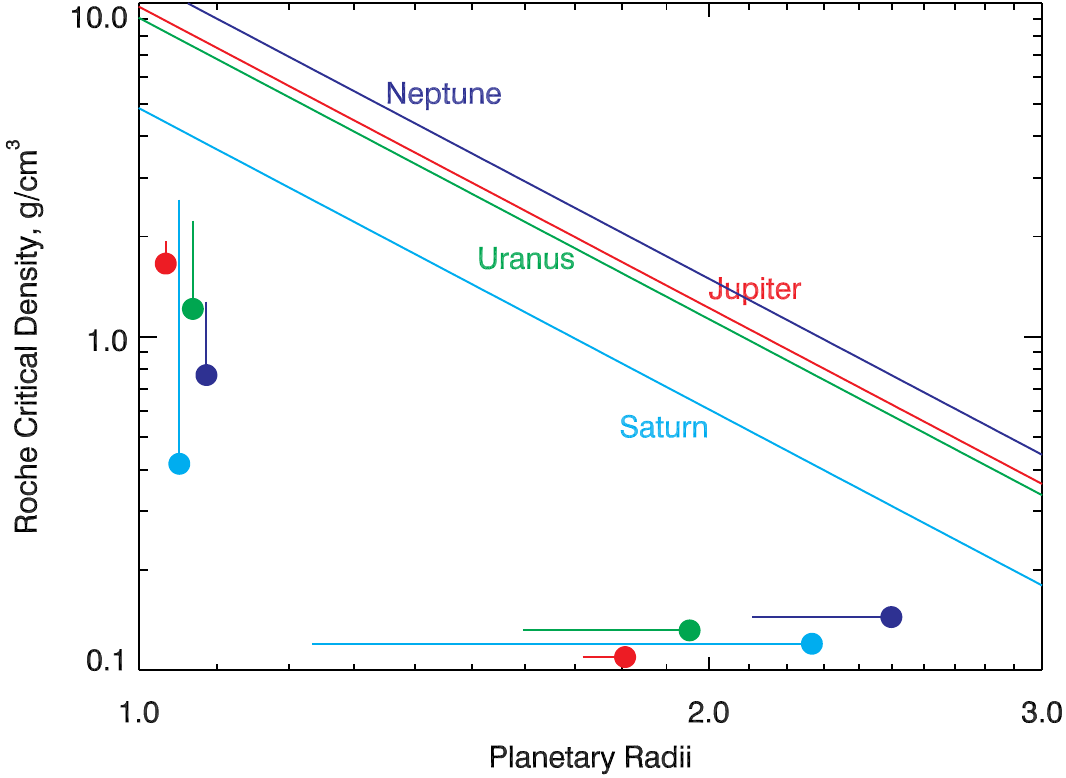}
\caption{Roche critical density $\rho_{\mathrm{Roche}}$ (\Eqn{}~\ref{RocheDensEqn}, with $\gamma = 1.6$) plotted against planetary radii for Jupiter (red), Saturn (cyan), Uranus (green), and Neptune (blue).  An object must have density higher than $\rho_{\mathrm{Roche}}$ to be held together by its own gravity; conversely, in the presence of abundant disk material, an embedded object will actively accrete as long as its density remains higher than $\rho_{\mathrm{Roche}}$.  The colored bars along the bottom and along the left-hand side show the extent of each planet's main ring system.  For each, a solid circle indicates the outermost extent, and the corresponding minimum $\rho_\mathrm{Roche}$, of the main rings. 
\label{rochedens}}
\end{center}
\end{figure*}

In fact, it is often more useful in the context of rings to consider the limit from planetary tides not as a critical distance but as a critical density.  At any given distance $a$ from the planet, there is a Roche critical density $\rho_{\mathrm{Roche}}$ at which the moon's size entirely fills its region of gravitational dominance (its ``Roche lobe'' or ``Hill sphere'' of characteristic radius $r_\mathrm{Hill}$).  We can rearrange \Eqn{}~\ref{RocheLimitEqn} to obtain
\begin{equation}
\label{HillRadiusEqn}
r_{\mathrm{Hill}} = a \left( \frac{m}{3M_{\mathrm{p}}} \right)^{1/3}
\end{equation}
and
\begin{equation}
\label{RocheDensEqn}
\rho_{\mathrm{Roche}} = \frac{4 \pi \rho_\mathrm{p}}{\gamma (a/R_{\mathrm{p}})^3}= \frac{3 M_{\mathrm{p}}}{\gamma a^3} , 
\end{equation}
where $m$ and $M_\mathrm{p}$ are the masses of the moon and planet, respectively.  The first of these two expressions is the most useful for interpreting \Fig{}~\ref{rochedens}. 

Within a ring, where material for accretion is plentiful, any pre-existing solid chunk with internal density greater than $\rho_{\mathrm{Roche}}$ should accrete a mantle of porous ring material until its density decreases to match $\rho_{\mathrm{Roche}}$.  This process should govern the size and density of the largest disk-embedded objects \citep{PorcoSci07,Charnoz07}.  On the other hand, the density naturally achieved by transient clumps should be compared to $\rho_{\mathrm{Roche}}$ in order to predict whether disruption (rings) or accretion (discrete moons) will dominate in a particular location, and the persistent existence of a ring implies that the densities of transient clumps do not exceed $\rho_{\mathrm{Roche}}$ (that is, we expect $\rho \lesssim \rho_{\mathrm{Roche}}$).  As seen in \Fig{}~\ref{rochedens}, Saturn's rings extend outward to significantly lower values of $\rho_{\mathrm{Roche}}$, approaching 0.4~g~cm$^{-3}$, than are seen in any of the other three known ring systems, probably reflecting their much lower rock fraction (and higher fraction of water ice) as already known from spectroscopy and photometry \citep{CuzziChapter09}.  That $\rho_{\mathrm{Roche}}$ for Saturn's rings reaches values much lower even than the density of solid water ice indicates a high degree of porosity, which is not surprising for a system in balance between disruption and accretion.  

If the outer edge of a ring system is taken to be the transition between disruption-dominated and accretion-dominated regions, which is probably true at least for Saturn and Uranus given the large number of moons immediately outward of their main rings, and if the porosity of accreting objects is relatively constant among the different systems, then differences in $\rho_{\mathrm{Roche}}$ at the transition location probably reflect differences in bulk composition.  Since Uranus has a transitional $\rho_{\mathrm{Roche}}$ three times that of Saturn (\Fig{}~\ref{rochedens}), we may well infer that its rings are made of material with a higher grain density, i.e. a significantly higher rock/ice ratio.  Neptune's transitional $\rho_{\mathrm{Roche}}$ is intermediate between Saturn's and Uranus', possibly indicating an intermediate rock/ice ratio.  Our inference, from the Roche critical density at the ring/moon transition, that the Uranus system is rockier overall than the Saturn system is consistent with the fact that the average density of Saturn's mid-size moons \citep{MatsonChapter09} is 1.2~g~cm$^{-3}$, while that of Uranus' major moons \citep{Jake92} is 1.6~g~cm$^{-3}$.  We cannot test our inference that Neptune's rock/ice ratio is intermediate in this way, as Neptune has no indigenous major moons due to the cataclysm of Triton's capture \citep{GoldreichNeptune89}. 

The extent of Jupiter's Main ring, in contrast to the other three ring systems, is clearly limited by the availability of material (which originates at source moons Metis and Adrastea and evolves inward, and which is not abundant) rather than by a disruption/accretion balance.  However, its high value of $\rho_{\mathrm{Roche}} \sim 1.7$~g~cm$^{-3}$ places the only known limit on the densities (and thus masses) of the source moons.  However, it may not be valid to assume that Metis and Adrastea are held together by gravity, as accreting masses must be, given the large gap in particle size between the $\sim 10$-km moons and other Main ring particles, which observationally cannot be larger than 1~km \citep{Show07}.  This large gap in particle size might be explained if Metis and Adrastea are solid bodies originating further out, now held together by material strength, while no bodies of similar size are now able to form through \textit{in~situ} accretion. 

\subsection{Optical depth \label{OpticalDepth}}

The amount of material in a system with general disk morphology can be measured in several ways.  The most straight-forward is the surface density, the mass per unit surface area of the disk, though it must be borne in mind that a disk with greater vertical thickness will have proportionately lower volume density than a vertically-thin disk, even if both have the same surface density.  However, surface density can be difficult to measure directly.  A much more common observable is the optical depth $\tau$, which can be thought of as the attenuation of a beam of light passing through the disk, measured in $e$-folding terms.  That is, 
\begin{equation}
\tau = - \ln \left( \frac{I}{I_*} \right) \equiv - \ln T ,
\label{TauDef}
\end{equation}
\noindent where $I$ is the observed intensity, $I_*$ is the unocculted intensity (e.g., from a background star), and the ratio between them is defined as the ring's transparency $T$.  

The optical depth is sensitive to both the number density and the size of ring particles, which can be obtained when both the optical depth and the surface density are known \citep[e.g.,][]{Colwell09}.  For a given value of the surface mass density, the optical depth scales inversely with particle size, that is, with the ratio of volume to surface area.  

\begin{figure*}[!t]
\begin{center}
\includegraphics[width=16cm]{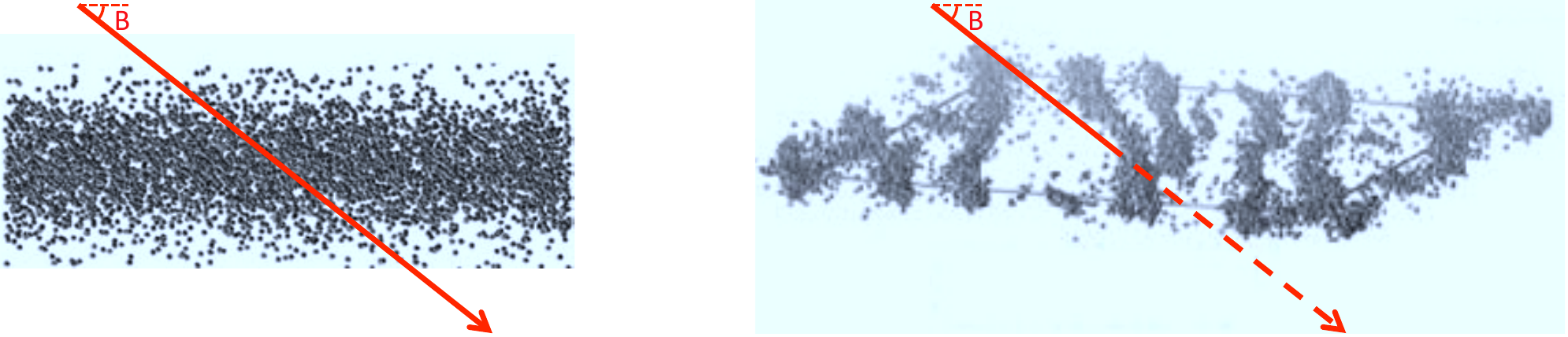}
\caption{Schematic showing a slanted path through (a) a homogeneous slab and (b) flattened SGWs.  The measured optical depth $\tau$ is proportional to $\sin B$ in the first case, but is relatively insensitive to elevation angle in the second.  
\label{TauSlant}}
\end{center}
\end{figure*}

For a vertically-thick homogeneous disk, the optical depth is proportional to the path length through the disk, which in turn is proportional to $\mu \equiv \sin B$, where the elevation angle $B$ is the angle between the line-of-sight and the ring-plane (\Fig{}~\ref{TauSlant}a).  In order to compare observations taken over a range of elevation angles, the normal optical depth, $\tau_\perp \equiv \mu \tau$, is often used.  However, this parameter must be used with caution for disks that lack homogeneity and/or are close to a single layer thick.  \citet{Colwell07} found that $\tau_\perp$ varies strongly with $\mu$ in the B~ring, and that the uncorrected $\tau$ is a more robust parameter in that case, indicating that the B~ring is composed of vertically-thin nearly-opaque clumps with nearly-transparent gaps between them (\Fig{}~\ref{TauSlant}b), and that the optical depth is controlled by the relative abundance of clumps and gaps (see Section~\ref{SGWs}). 

In numerical simulations (see Section~\ref{NumSims}), the photometric optical depth $\tau$ (\Eqn{}~\ref{TauDef}) is cumbersome to calculate, but a useful proxy known as the dynamical optical depth $\tau_\mathrm{dyn}$ can be found by summing the total cross-section area of all simulated particles and dividing by the area of the simulation patch.  This quantity turns out to be equal to the photometric optical depth as long as particles are randomly distributed, as the Gaussian probability of particle overlap plays the same role when using $\tau_\mathrm{dyn}$ to calculate the total transparency that the exponential plays when using \Eqn{}~\ref{TauDef}.  However, for high values of $\tau$, when the distance between particles becomes comparable to the particle size, particles become constrained as to the locations in space they can occupy and the two quantities diverge.  Specifically, 
\begin{equation}
\tau / \tau_\mathrm{dyn} \simeq 1 + kD , 
\end{equation}
\noindent where the volume filling factor $D$ is calculated from particle radius, disk scale height, and $\tau_\mathrm{dyn}$, and $k$ is a scalar of order unity \citep{SK03,Anparsgw10}.  Furthermore, the existence of microstructure such as self-gravity wakes causes the distribution of particles to be strongly non-random, and can cause $\tau_\mathrm{dyn}$ to diverge strongly from the photometrically observed $\tau$. 

\section{Rings by planetary system \label{RingsByPlanet}}

\subsection{The rings of Jupiter \label{Jupiter}}

Jupiter is adorned by the simplest of the known ring systems.  All of its rings are tenuous and composed of dust-sized\fn{Throughout this work, we will use the word ``dust'' to refer to $\mu$m-sized particles regardless of their composition.} particles.  As the only confirmed ring system without any dense component, and by far the least massive \citep{BurnsChapter04}, Jupiter's is the only ring system to have been discovered by spacecraft without having previously been seen from Earth either directly (as Saturn's) or by stellar occultations (as Uranus' and Neptune's).  The Main ring was first clearly described from \Voyit{~1} images \citep{OwenJupiterRings79} after initial hints from charged-particle detectors aboard \Pionit{~11} \citep{FMM75,AN76,BurnsChapter04}

\begin{figure*}[!t]
\begin{center}
\includegraphics[width=16cm]{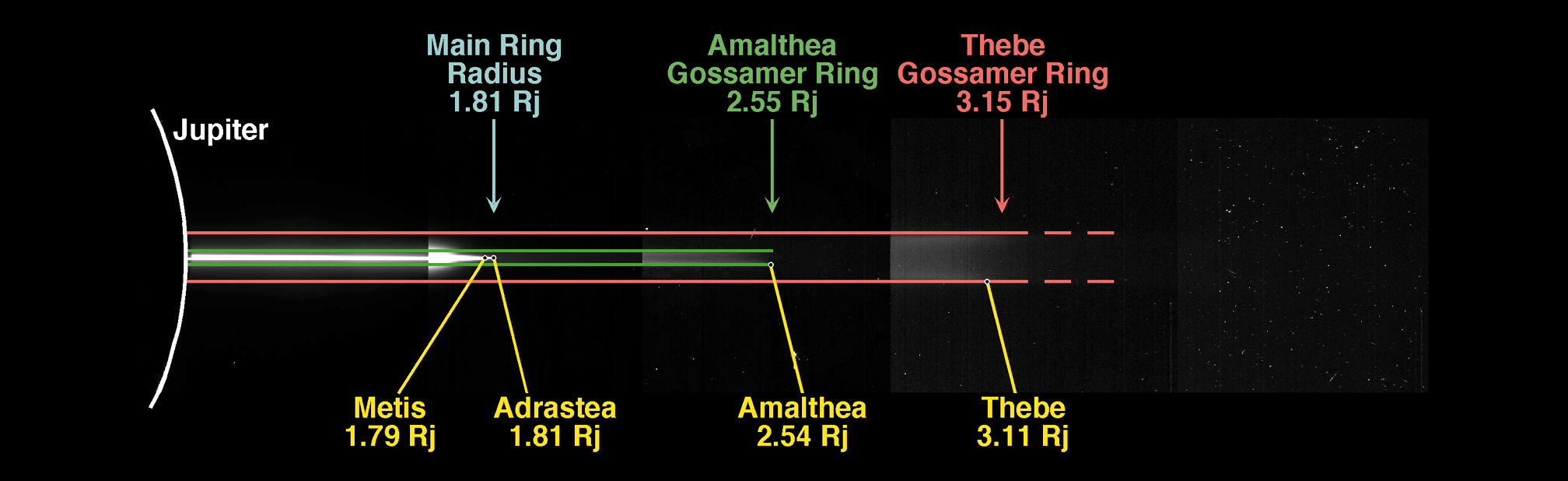}
\caption{\Galit{} image mosaic of Jupiter's rings, seen nearly edge-on at very high phase angle, annotated to show the primary components of the ring system.  Image sensitivity increases from left to right, in order to show the increasingly faint structure.  Figure from \citet{Ockert-Bell99}.  
\label{JupiterRingsFig}}
\end{center}
\end{figure*}

The basic structure of Jupiter's rings is well understood \citep[see][for a recent comprehensive review]{BurnsChapter04}.  The Main ring and the two Gossamer rings are like three nested ``tuna cans'' (\Fig{}~\ref{JupiterRingsFig}), with radius set by the semimajor axis ($a$) of the ring's source moon and vertical height by the moon's vertical excursions relative to Jupiter's equatorial plane ($a \sin I$, for inclination $I$).  Particles enter the ring as ejecta from micro-meteoroid impacts onto the moon \citep{Burns99}, and begin with orbital parameters $a$, $e$, and $I$ (see Section~\ref{OrbElems}) similar to the moon's.  The tuna-can structure arises as the orientations of particles' orbit planes ($\Omega$) become quickly randomized due to small variations in $a$, and thus in the precession rate.  Particles evolve inward under Poynting-Robertson drag \citep{Burns99}, thus filling out the cylindrical shape.  A double-layered vertical structure arises dynamically because any orbiting particle spends more time at its vertical maxima than it does in the midplane.  The increasing vertical thickness of the rings arises because Amalthea's inclination is larger than that of Metis or Adrastea, while Thebe in turn has an even larger inclination. 

\begin{figure*}[!t]
\begin{center}
\includegraphics[width=10cm]{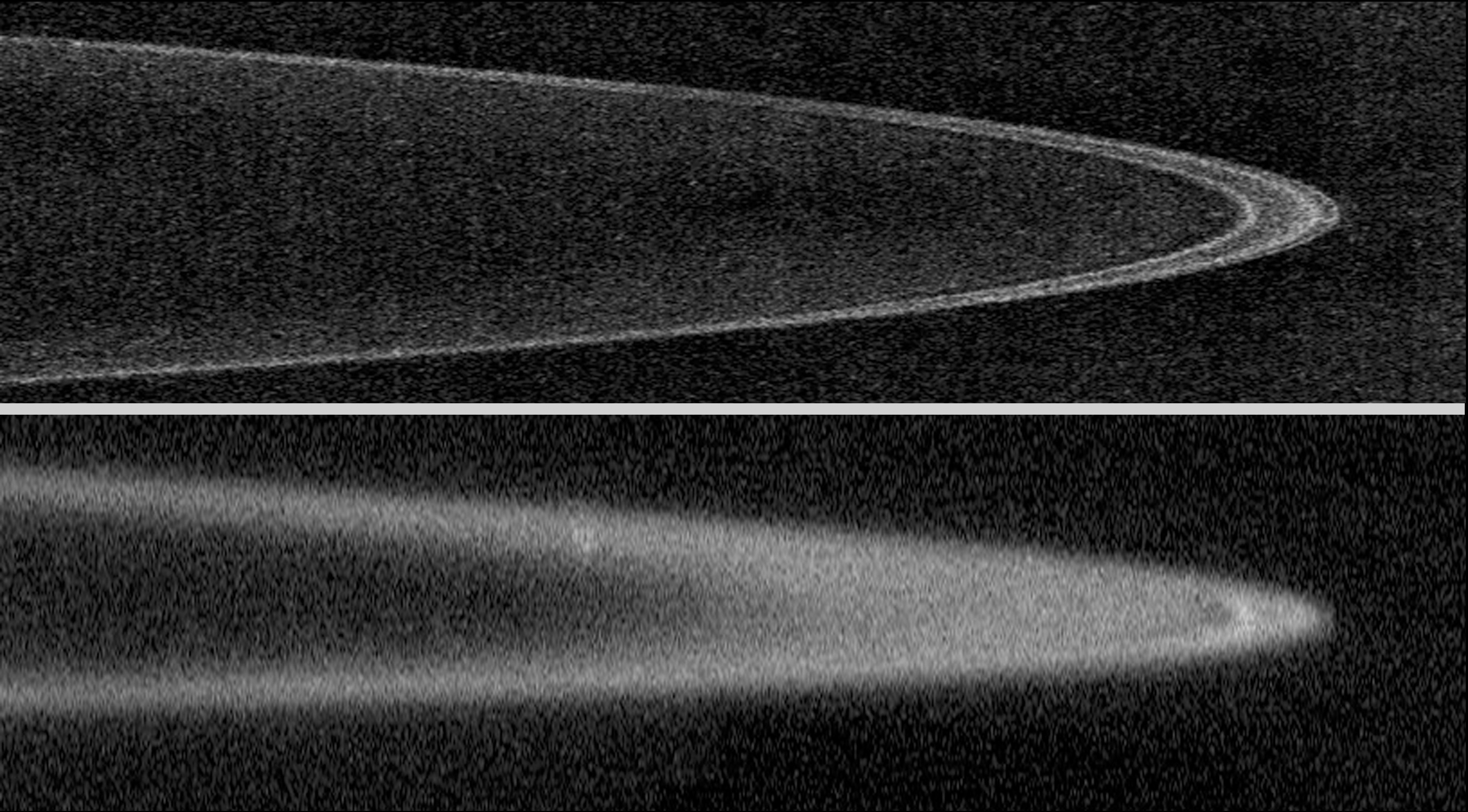}
\caption{\NHit{} images of Jupiter's Main ring at low phase (upper panel) and at high phase (lower panel), respectively showing the structures composed of macroscopic particles and the dusty envelope.  Credit: NASA.  
\label{JupiterMainRings}}
\end{center}
\end{figure*}

The two Gossamer rings consist entirely of material evolving inward in this way, though the Thebe Gossamer ring has an additional segment extending slightly \textit{outward} from Thebe's orbit, which has been attributed to charging and discharging of grains as they pass into and out of the planet's shadow \citep{HK08}.  Further inward, the $\sim$1000-km-wide core of the Main ring contains a significant population of cm-size and larger particles lying between the orbits of Adrastea and Metis.  This core, which is the only component of the ring system not composed of dust and the only one that appears bright at low phase angles,\fn{The phase angle is formed by the Sun-object-observer lightpath.  Dust-sized particles, having size comparable to the wavelength of visible light, tend to diffract light forward and are brightest at high phase angles.  Larger objects tend to reflect light and are brightest at low phase angles.} is composed of several ringlets (including one lying just outside Adrastea's orbit), whose cause is not known.  \NHit{} imaging limits the sizes of objects in this belt to be $<$1~km \citep{Show07} --- other than Adrastea and Metis themselves, which have mean radii of 8~and 22~km, respectively.  However, \citet{Show07} did find several azimuthal clumps in a ringlet just inward of Adrastea (see Section~\ref{Arcs}).  A dusty component of the Main ring extends inward of its core, also evolving under Poynting-Robertson drag (\Fig{}~\ref{JupiterMainRings}).  

The dusty component of the Main ring has a vertical scale height that increases monotonically with decreasing distance to Jupiter \citep{Ockert-Bell99}.  When the inward-moving material reaches a radius of 122,800~km from Jupiter's center, it becomes strongly affected by a 3:2 ``Lorentz resonance'' \citep{Burns85} between its orbital period and the rotation period of Jupiter's magnetic field.  Inward of this location, the vertical extent of the ring increases dramatically, forming the toroidal Halo ring.  Material in the Halo ranges tens of thousands of km above Jupiter's ring plane, though most of its material is concentrated within just a few hundred km \citep{BurnsChapter04}.  

\subsection{The rings of Saturn \label{Saturn}}

\begin{figure*}[!t]
\begin{center}
\includegraphics[width=16cm]{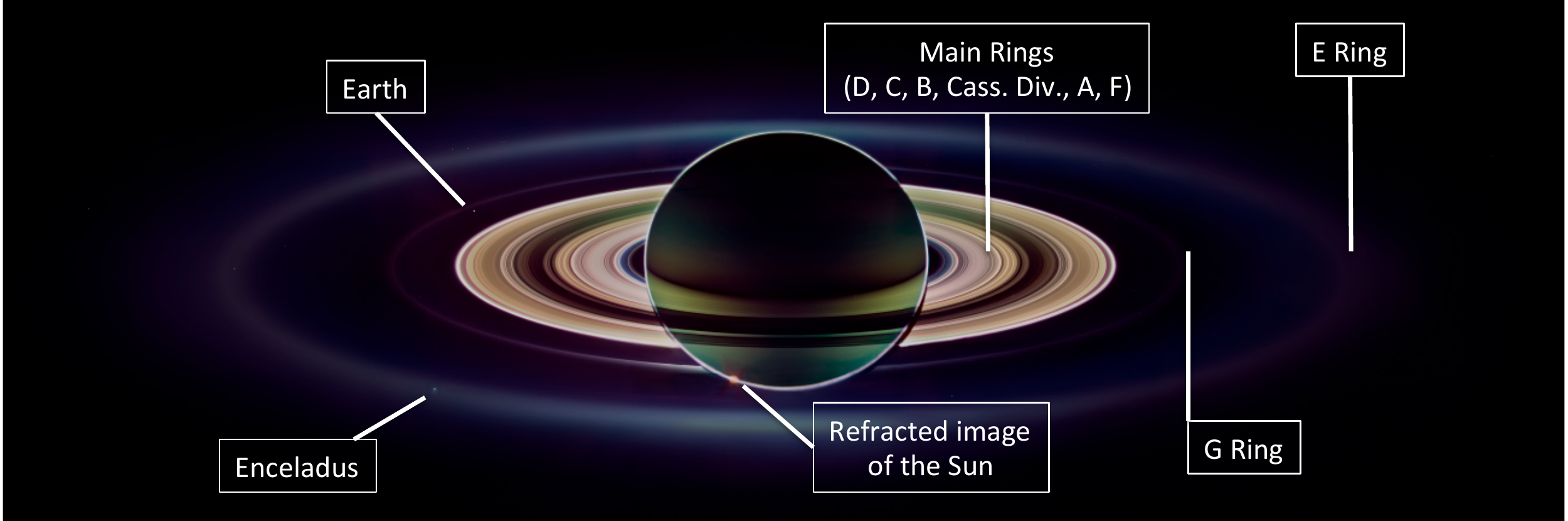}
\caption{This \Cassit{} image mosaic shows Saturn's tenuous D, E~and G~rings with comparable brightness to the main disk, which occurs because the viewing geometry is at high phase angle (in fact, in eclipse) and also views the unlit face of the main disk.  The darkest part of the main disk is actually the densest and most opaque, namely the mid- to outer-B~ring.  The Cassini Division is difficult to distinguish from the A~ring in this view.  The markings on the planet do not line up with those on the rings because the latter are due to sunlight filtering directly through the rings while the former are due to light reflected off the rings, then reflected off the planet, and then filtered through the rings again.  The Sun, which is actually behind Saturn, can be seen refracted through the planet's atmosphere at 7~o'clock.  Enceladus (actually, only its geyser plume is bright in this geometry) can be seen embedded in the E~ring at 8~o'clock.  The Earth can be seen as a pinpoint of light between the F~and G~rings at 10~o'clock.  Credit: NASA and M.~Hedman, annotated by the author.  
\label{SaturnBacklit}}
\end{center}
\end{figure*}

Saturn possesses by far the most massive and the most diverse of the known ring systems (\Fig{}~\ref{SaturnBacklit}).  The only ring system known before recent decades, Saturn's rings were among the first objects observed through a telescope, by Galileo Galilei in 1610, explicated as a disk by Christiaan Huygens, proved to consist of individual particles on independent orbits by James Clerk Maxwell, and have in general been the focus of much productive study by astronomers over the past four centuries \citep{Alexander62,VanHelden84,MWC07}.  The main part of the rings comprises the solar system's only known broad and dense disk (Section~\ref{DenseBroad}), which was found by G.~D.~Cassini to be divided into two parts --- now called the A~and B~rings, separated by what is now called the Cassini Division.  Furthermore, the latter is now known to be not an empty gap but simply a region of the disk with more moderate surface density, similar in character to the C~ring, which lies inward of the B~ring and was discovered in 1850.  

There are a small number of truly empty radial gaps in the dense disk of the main rings, most of them in the C~ring and Cassini Division but two in the outer A~ring, all of them sharp-edged.  These gaps are named for scientists who have made contributions to the study of Saturn's rings.  The two A-ring gaps are held open by moons at their centers, and the C~ring's Colombo Gap is known to be held open by a resonance with Titan, but most of the gaps remain unexplained.  Recent work by \citet{HedmanCassDiv10} suggested that all the Cassini Division gaps are due to a secondary resonance associated with the Mimas~2:1 resonant mechanism that defines the nearby outer edge of the B~ring, though this idea has yet to be successfully worked out in detail \citep{SP10}.  A diverse array of narrow rings and ringlets resides within ring gaps, some of them dense and sharp-edged and others diffuse and/or dusty.  They often are given the same name as the gap within which they reside, though several have been given nicknames.  These structures, which can be compared with narrow rings around other planets, are discussed in Section~\ref{DenseNarrow}. 

The Saturn system also contains the most diverse retinue of tenuous dusty ring structures known in the solar system, discussed in Section~\ref{ThinDusty}.  The main components, given letters in order of their discovery, are the D~ring situated innermost between the main rings and Saturn's atmosphere, the dense F~ring (Section~\ref{ThickDusty}) just off the edge of the A~ring, and the G~and E~rings farther out.  The region between the A~and F~rings, now known as the Roche Division,\fn{As recently formalized by the IAU, a ``division'' is defined as a region between two lettered rings that contains a sheet of material, while a ``gap'' is a clear region within a lettered ring that may or may not contain one or more ringlets (\texttt{http://planetarynames.wr.usgs.gov/append8.html}).} contains a tenuous dusty sheet, and several other rings or ring arcs are named for moons whose orbits they share. 

The largest known ring in the solar system, the Phoebe~ring, was recently discovered by the Spitzer Space Telescope \citep{VSH09}.  This ring is also the only known ring to be tilted from its planet's equatorial plane (it lies in the plane of Saturn's orbit, as solar perturbations are much more important than Saturn's $J_2$ at its distance) and is likely the only known ring whose particles orbit in the retrograde direction (if indeed its material is primarily derived from Phoebe, which orbits retrograde).  As particles in the Phoebe~ring spiral inward under Poynting-Robertson drag, they preferentially impact the leading hemisphere of Iapetus \citep{Soter74,Tamayo11}, which, together with solar-driven thermal processing, appears likely to explain the strong brightness dichotomy on the surface of that moon \citep{SD10,Denk10}. 

The end of \Cassit{}'s initial 4-year mission at Saturn (though its extended mission continues) has occasioned several recent reviews of Saturn's rings, including articles by \citet{Cuzzi10} and \citet{EspoAnnRev10}.  A recent comprehensive review in five parts discussed the rings' structure \citep{ColwellChapter09}, dynamics \citep{SchmidtChapter09}, particle sizes and composition \citep{CuzziChapter09}, diffuse rings \citep{HoranyiChapter09}, and origins \citep{CharnozChapter09}, in addition to a review of pre-\Cassit{} understanding \citep{OrtonChapter09}. 

\subsection{The rings of Uranus \label{Uranus}}

\begin{figure*}[!t]
\begin{center}
\includegraphics[width=8cm]{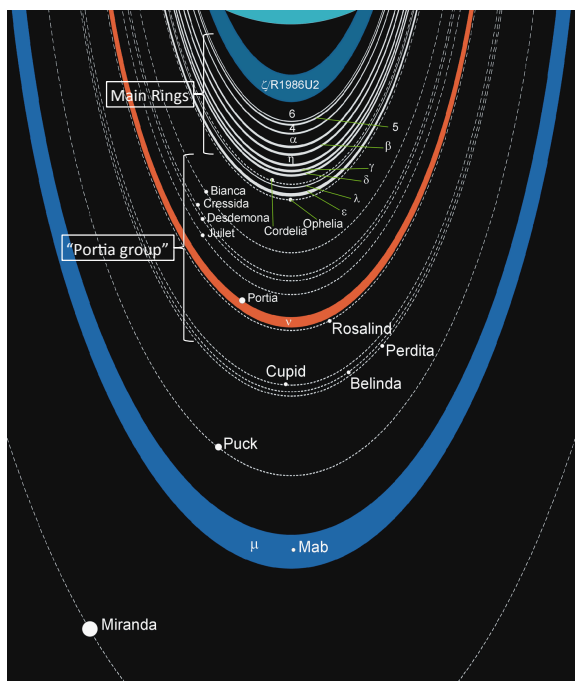}
\caption{Uranus' main rings are situated immediately inward of a retinue of small moons.  If one were to spread the mass of Uranus' ``Portia~group'' of moons evenly over the annulus they occupy, the surface density would be similar to that of Saturn's A~ring.  This moon system may be very similar in origin to the known ring systems, except that the natural density of accreted objects is larger than the Roche critical density (i.e., it is beyond the ``Roche limit,'' see \Fig{}~\ref{rochedens}) so that any moon that gets disrupted by a collision (which ought to have happened many times over the age of the solar system) will simply re-accrete.  Credit: Wikimedia Foundation, annotated by the author. 
\label{UranusRingsMoons}}
\end{center}
\end{figure*}

\begin{figure*}[!t]
\begin{center}
\includegraphics[width=6cm]{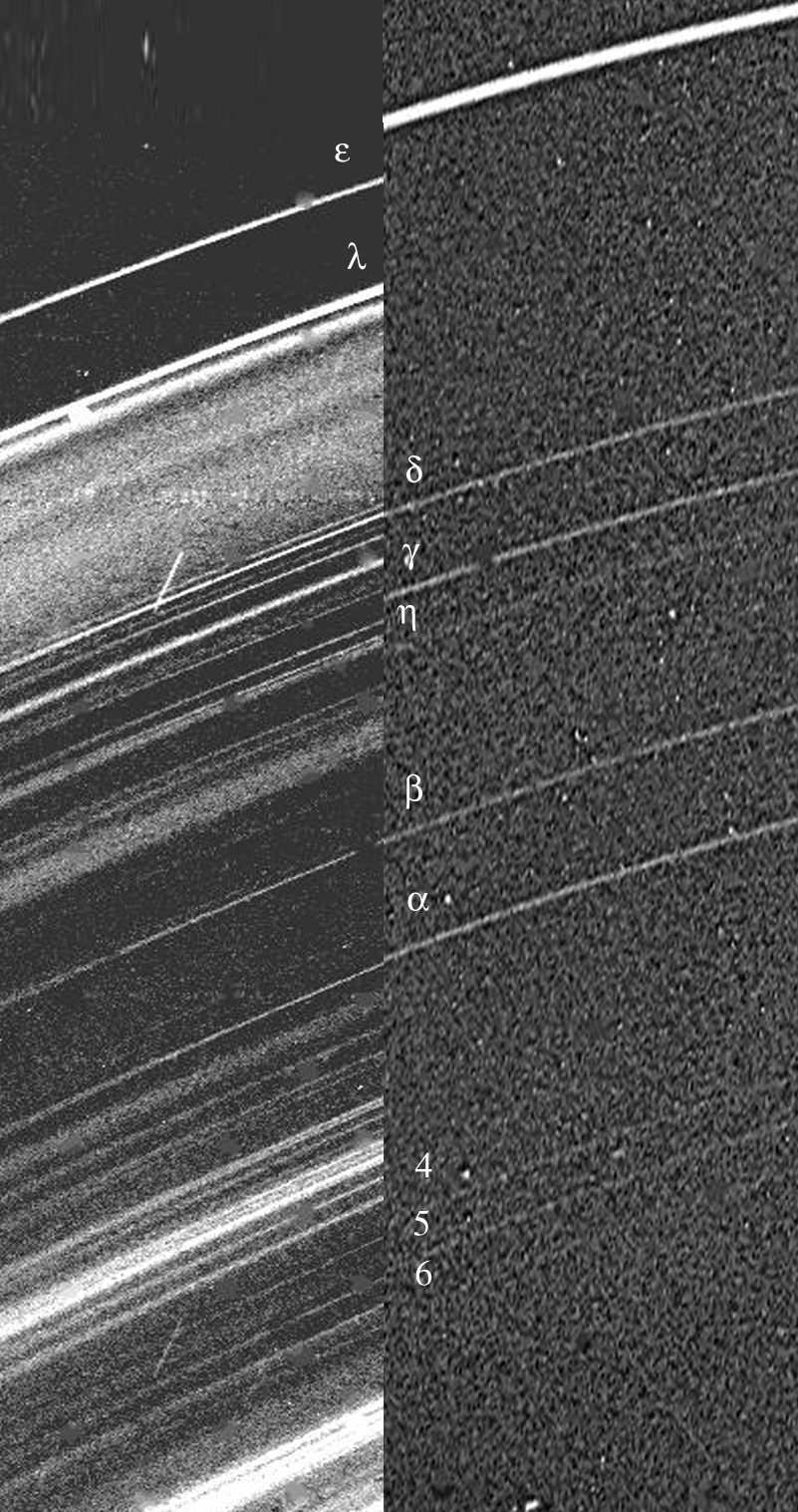}
\caption{This composite image of Uranus' main rings in forward-scattered (left) and back-scattered (right) light shows that a network of dust structures is interleaved with the planet's dense main rings.  The disjoint in the $\epsilon$ ring is due to its eccentricity.  As the left-hand image is the only high-phase image ever successfully taken of Uranus' rings (by the post-encounter \Voyit{~2}), the detailed workings of the dust structures remains largely unknown. 
Credit: NASA and Wikimedia Foundation. 
\label{UranusHiLoPhase}}
\end{center}
\end{figure*}

All of Uranus' main rings are narrow, and many are eccentric and/or inclined, unlike the broad disk of Saturn.  On the other hand, many of Uranus' rings are dense and sharp-edged, unlike the diffuse rings of Jupiter.  Thus, the Uranian system represents a third paradigm for ring systems, one that truly deserves the label of ``rings'' in the plural.  The main set of 10~narrow rings (including all the named rings except dusty $\zeta$, $\nu$, and $\mu$) occupies a fairly small radial range from 1.64~to 2.00~$R_\mathrm{U}$ from Uranus' center (\Fig{}~\ref{UranusRingsMoons}), 
inward of Uranus' 13~small inner moons except that the innermost moon Cordelia is inward of the $\epsilon$~ring.  A panoply of unnamed dusty rings was seen interspersed with the main rings in the single high-resolution high-phase image taken by \Voyit{~2} (\Fig{}~\ref{UranusHiLoPhase}), and gaps in these have been cited as evidence for additional moons \citep{MT90,MT90erratum}.  The so-called ``Portia group'' of eight moons packed into an annulus from 59,100~to 76,500~km from Uranus' center (i.e., 2.31~to 2.99~$R_\mathrm{U}$) appears from orbital simulations to be dynamically unstable on timescales of $10^6$ to $10^8$ years \citep{DLiss97,SL06}.  The dusty $\nu$~ring, which lies between two of the moons in this group, may well be the detritus of a recent significant collision, perhaps the disruption of a moon.  Outward beyond the Portia~group, the $\mu$~ring is centered on the orbit of Mab \citep[see Section~\ref{ThinDusty}]{SL06}.  

Given their dynamical instability, the Portia~group of Uranian moons are probably constantly evolving by means of occasional collisions followed by the re-accretion of material into new moons \citep{SL06,DawsonDDA10,FS12}, and may have looked rather different from its present configuration over most of solar system history, though what effect this might have on the main rings is unknown.  The contrast between the main Uranian rings and the Portia~group may simply be the difference between a particle population dominated by disruption and one dominated by accretion.  The mean surface density of the Portia~group region is $\sim 45$~g~cm$^{-2}$, calculated by spreading the moons' mass evenly over the annulus containing them. 
This is comparable to the surface density of dense rings such as Saturn's A~ring, though direct comparisons to the Uranian rings are difficult as the masses of the latter are poorly known.
The Roche critical density (see Section~\ref{Roche}) for the boundary between the two regions is 1.2~g~cm$^{-3}$, 
which possibly indicates a high rock fraction, especially considering the high internal porosity inherent in accretion of small bodies with low central pressures.  

The composition of the Uranian rings is almost entirely unknown, as \Voyit{} did not carry an infrared spectrometer with enough spatial resolution to detect the rings.  However, it is clear from their low albedo that at least the surfaces of the ring particles cannot be primarily water ice.  Color imaging indicates that the Uranian rings are dark at all visible wavelengths, indicating a spectrum similar to that of carbon. 

Most of Uranus' rings have been given Greek letters ($\alpha$, $\beta$, $\gamma$, $\delta$, $\epsilon$, $\eta$, $\lambda$, $\zeta$, $\mu$, and $\nu$) in order of their discovery, except for three which are labeled with numbers (4, 5, and~6).  This idiosyncratic system can be traced back to their simultaneous discovery by two different research groups, one of which \citep{ElliotNature77} labeled the rings with Greek letters while the other \citep{MillisNature77} numbered them.  The former system was given priority for future use, but three of the rings had not been observed by the former research group and thus retained as their labels the numbers given to them by their discoverers \citep{MWC07}.  

A comprehensive review of the Uranian rings, up to and including the \Voyit{~2} encounter, was published in two parts discussing the rings' structure \citep{FrenchChapter91} and particle properties \citep{EspoChapter91}. 

\subsection{The rings of Neptune \label{Neptune}}

\begin{figure*}[!t]
\begin{center}
\includegraphics[width=8cm]{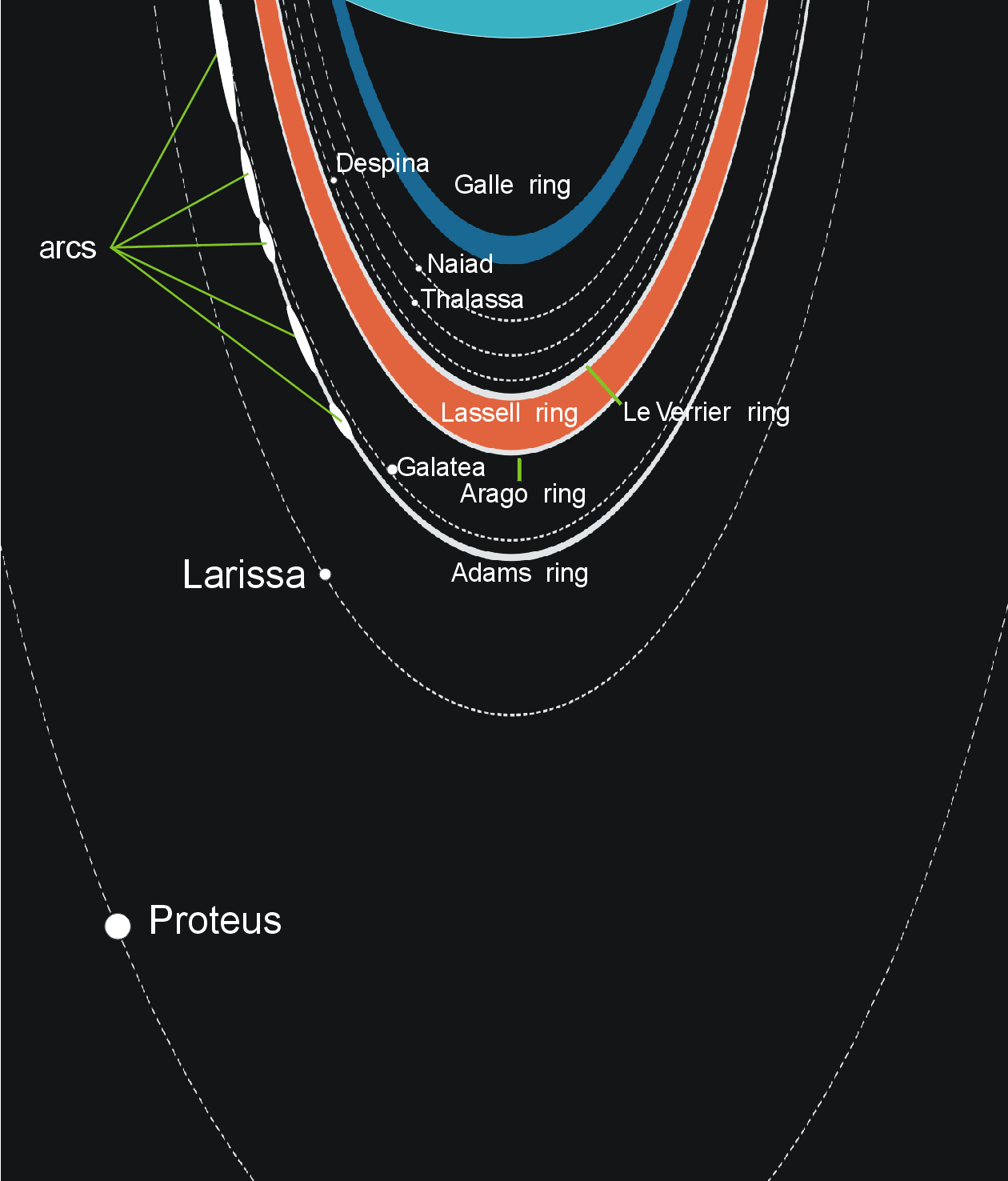}
\caption{In Neptune's ring system, uniquely, narrow rings and diffuse rings and moons are all interspersed together.  Credit: Wikimedia Foundation.  
\label{NeptuneRingsFig}}
\end{center}
\end{figure*}

Neptune's ring system, like that of Uranus, consists primarily of a few narrow rings, though Neptune's are generally less dense, higher in dust, less sharp in their edges, and farther from their planet than those of Uranus.  Neptune's rings are named for individuals associated with the 1846 discovery of Neptune.  The Le Verrier, Arago, and Adams rings are narrow, while the Galle and Lassell rings are tenuous sheets of dust (\Fig{}~\ref{NeptuneRingsFig}).  The Adams ring, Neptune's most substantial, is best known for its series of arcs, the first ever discovered, which are discussed in Section~\ref{Arcs}.  

Unlike those of Saturn and Uranus, Neptune's ring system is thoroughly interwoven with known moons.  This is at least partly enabled by the fact that Neptune's rings are the farthest from their planet, in terms of planetary radii, and have the lowest value of $\rho_{\mathrm{Roche}}$ at the \textit{inner} edge of the ring system (\Fig{}~\ref{rochedens}).  Still, the presence of rings, rather than accreted moons, must indicate that the natural density of accreted objects is lower than $\rho_{\mathrm{Roche}}$.  Thus, it may be that the ring-moons Naiad, Thalassa, Despina, and perhaps Galatea, either originated farther from the planet than their current position, or accreted in a formerly more dense ring. 

The composition of the Neptunian rings, like that of the Uranian rings, is unknown due to \Voyit{}'s inability to detect them in the infrared.  However, again like the Uranian rings, the low albedo of Neptunian ring particles makes it clear that at least their surfaces cannot be primarily water ice. 

A comprehensive review of the Neptunian rings, up to and including the \Voyit{~2} encounter, was given by \citet{PorcoChapter95}. 

\subsection{Unconfirmed ring systems}

The four giant planets are the only bodies known to have rings, as just described.  Here we discuss bodies for which rings have been seriously discussed but not observed.

\subsubsection{Mars \label{Mars}}

Mars has been predicted to have a tenuous ring system comprising dust grains ejected from its moons Phobos and Deimos by meteoroid impactors \citep[and references therein]{Ham96,KH97}.  Simulations by \citet{BurnsChapter01} indicate that Deimos' ring should be offset away from the Sun and tilted out of Mars' equatorial plane by the Sun's perturbations.  However, attempts to observe rings around Mars have been unsuccessful to date \citep{SHN06}, and the image quality has progressed to the point that some models can now be observationally excluded.  The lack of dust could be due to dust production rates being lower than expected, or the lifetimes of dust particles being shorter than expected.  Solar radiation pressure limits the lifetimes of dust particles (especially smaller ones) by driving their orbital eccentricity to values so high that they impact the planet.  \textit{In~situ} observations by Mars-orbiting spacecraft of anomalies in the solar wind magnetic field were interpreted in the 1980s as being due to Martian rings, but more extensive measurements by the magnetometer aboard \MGSit{} showed that observable fluctuations are likely due to well-known solar wind or bowshock phenomena \citep{Oieroset10}

Looking for rings around a solid planet like Mars at low phase is more difficult than similar observations at gas giant planets because of the former's lack of atmospheric methane.  Because methane has very strong absorption bands (e.g., at 2.2~$\mu$m), images of gas giants taken at selected wavelengths will see a greatly darkened planet, facilitating the detection of faint rings.  On the other hand, looking for Martian rings at high phase is less effective than for other dusty rings because particles smaller than $\sim 50$~$\mu$m are expected to be depleted due to radiation pressure \citep{Ham96,SHN06}

\subsubsection{Pluto \label{Pluto}}

Pluto, like Mars, could harbor a tenuous ring system of dust derived from its small moons Nix and Hydra and P4 \citep{SternPluto06}.  
Charon, as discussed in Section~\ref{ThinDusty}, is paradoxically less likely to be a major source of dusty rings because its gravitational field will more efficiently retain any dust ejected from its surface.  Observations to date \citep{StefflStern07} are not sensitive enough even to rule out the conservatively estimated normal optical depth $\tau_\perp < 10^{-6}$ suggested by \citet{SternPluto06}, though more sensitive observations were very recently obtained \citep{ShowCBET11}.  

Further Earth-based observations may improve on this sensitivity, and a clearer picture of Pluto's rings (or lack thereof) should come from imaging during the planned \NHit{} flyby in 2015, especially during the post-encounter period when the phase angle will be high. However, both of the handicaps discussed above for Mars, the lack of atmospheric methane and the loss of smaller dust grains due to radiation pressure, are also likely to hamper the detection of any Plutonian rings. 

\subsubsection{Rhea and other moons \label{Rhea}}

Rhea, the largest of Saturn's airless moons, is in the opposite situation from Mars and Pluto, with no clear theoretical prediction but a claim that rings have been observed.  \citet{Jones08} reported unusual electron-absorption signatures detected by the Magnetospheric Imaging Instrument (MIMI) during Rhea flybys of the \Cassit{} spacecraft, and attributed these signatures to rings.  What would be the first known ring system around a moon includes several narrow bands embedded within a broad diffuse cloud.  The MIMI instrument detects changes in the miasma of charged particles through which the spacecraft is constantly passing, enabling inferences regarding solid and magnetospheric structures that shape the plasma environment.  A good analogy is the way a person driving in a blinding rain can perceive having driven under a bridge by the sudden cessation of raindrops hitting the windshield.\fn{Credit: G.~H.~Jones in JPL podcast, 6 March 2008 (\texttt{http://www.jpl.nasa.gov/podcast/content.cfm?content=671}).}  

\begin{figure*}[!t]
\includegraphics[width=16cm]{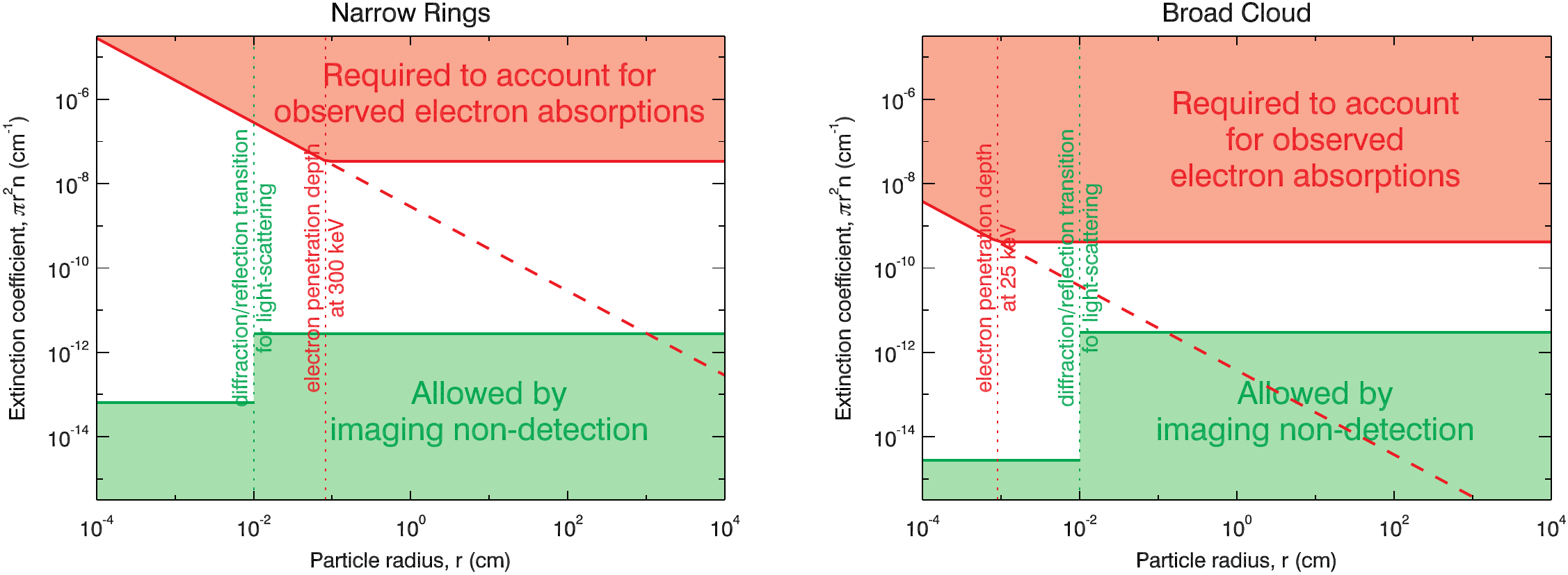}
\caption{Comparison of the radius $r$ and number density $n$ of particles, the latter expressed in terms of the extinction coefficient $\pi r^2 n$ of putative Rhea rings, inferred from charged-particle absorptions observed by \Cassit{}~MIMI \citep{Jones08} and imaging non-detection by \Cassit{}~ISS \citep{Rhearpx10}, shown in red and green, respectively.  Dashed lines indicate requirements previously claimed \citep{Jones08} for particle sizes larger than the electron penetration depth.  For narrow rings (\textit{left}), even allowing the latter claim, the combined observations require particles larger than 10~m in radius, indicating an unrealistic lack of smaller particles.  Furthermore, once allowance is made for the role of the electron absorption length (horizontal lower boundary to red area), the imaging non-detection rules out any form of absorption by solid material as the cause of the observed charged-particle absorptions for both narrow rings and broad cloud.  Figure from \citet{Rhearpx10}. 
\label{rhearpx_fig}}
\end{figure*}

However, an extensive search for Rhea rings using \Cassit{} ISS images \citep{Rhearpx10} places severe limits on any possible Rhea rings.  Using the calculations of \citet{Jones08}, which assume that a particle's ability to block electrons increases linearly with its mass (and thus its volume), the ISS observations limit the narrow rings to a characteristic particle size $r > 10$~m, and $\mu$m-sized dust in any significant abundance is unequivocally excluded.  This minimum particle size is unrealistically large given that such large particles must constantly be eroded to smaller sizes, which then would have been detected in the ISS~images.  Furthermore, \citet{Rhearpx10} pointed out that the electron penetration depth for the low-energy electrons used by \citet{Jones08} is much smaller than the suggested particle sizes, so that a particle's ability to block electrons should increase linearly with its surface area rather than its volume.  Thus, \citet{Rhearpx10} re-calculate the ring optical depth required to explain the \citet{Jones08} observations to be several orders of magnitude higher than the values excluded by the ISS observations, both for the narrow rings and the broad cloud (\Fig{}~\ref{rhearpx_fig}).  It seems most likely, therefore, that the signatures detected by the \Cassit{} MIMI instrument are due to some magnetospheric phenomenon, and not to rings of solid material around Rhea.  

Following the MIMI announcement that Rhea might have rings, \citet{SchenkDPS09} reported a clumpy and discontinuous chain of discrete bluish splotches, up to 10~km wide, aligned along more than half the circumference of a great circle inclined by $\sim$2$^\circ$ to Rhea's equator.  \citet{Schenk11}, formally presenting that finding after publication of the ISS non-detection, argued that the band may be a sign of a \textit{former} Rhean ring, even if none exists now.  However, it remains unclear whether a ring system is even plausible at Rhea (whose shape is triaxial rather than oblate), what would be the source of its material, whether a tenuous ring would assume the flatness (generally produced only in dense rings by collisional evolution, while tenuous rings generally are vertically thick) implied by the observed narrow band, and whether the observed color variations are the likely result of such impacting material. 

Among other Saturnian moons, Mimas and Tethys also have equatorial color features extending over $\pm 40^\circ$ and $\pm 20^\circ$ of latitude, respectively, much wider than Rhea's and possibly the result of magnetospheric bombardment \citep{Schenk11}.  Iapetus has a 13-km-high equatorial mountain range extending across $>$110$^\circ$ of longitude \citep{PorcoPhoebeIapetus05,GieseIapetus08}.  Several endogenic hypotheses have been suggested for this structure, but \citet{Ip06} argued that it was created by ring material falling out of Iapetan orbit, and \citet{Levison11} further developed the idea by proposing a giant impact to create the ring and sketching a scenario that would simultaneously account for Iapetus' surprisingly oblate shape.  While Iapetus may be a more hospitable site for rings than Rhea, given its oblate shape and large Hill sphere, it remains unclear just how or whether orbiting material would be incorporated into a mountain range, nor how such a ring would evolve and fall out.  

\subsubsection{Exoplanets}

While hundreds of planets have now been detected in orbit about other stars using methods including radial velocity, transits, astrometry, and direct imaging \citep{ExoplanetsBook10}, no exoplanet yet has a confirmed ring system.  The signature in a transit observation expected from an exoplanet ring system was discussed by \citet{BF04}, and a few exoplanet detections have been able to set meaningful upper limits on putative ring systems \citep[e.g.,][]{TBrown01}, but it is too early to do statistics on the frequency of exoplanet rings, especially since the available observations with sufficient sensitivity are generally for ``hot Jupiters'' that orbit very close to their stars.  There are many factors stacked against the detection of rings around hot Jupiters \citep{SC11}, including small Hill spheres, low obliquities\fn{Here, we refer to the inclination of the planet's equatorial plane with respect to the line of sight from Earth.} that would cause rings to be seen edge-on \citep[see also][]{Ohta09}, loss of ring particles to Poynting-Robertson drag and viscous drag from the planet's exosphere \citep[see also][]{Gaudi03}, and equilibrium blackbody temperatures too high for even refractory materials to remain solid.  However, each of these factors is mitigated for planets $\gtrsim$0.1~AU from their host stars \citep{SC11}.  Furthermore, warping of the ring planes (Laplace planes) of ``warm Saturns,'' diagnostic of their planetary oblateness (see Section~\ref{OrbElems}) may be discernible in transit lightcurves \citep{SC11}. 

The only known exoplanet for which a ring system has been specifically proposed is Fomalhaut~b, which was the first\fn{along with the three planets of the HR~8799 system, announced at the same time} exoplanet to be detected by direct imaging \citep{KalasFomB08}.  The orbit of Fomalhaut~b is $\sim 115$~AU from its star and maintains the inner edge of an eccentric debris belt that was known before the planet was (see Section~\ref{OtherDisks}).  The brightness of Fomalhaut~b in visible light, together with its dimness in the infrared, has led \citet{KalasFomB08} to suggest the planet has a large ring system, which would significantly increase its visible brightness via reflection without affecting its emitted infrared radiation.  However, a ring with a surface brightness similar to that of Saturn's A~ring while extending to the planet's Roche radius would reflect far too little flux to account for the observations, so it is necessary to invoke a disk extending to $\sim 35$~planetary radii, approximately the distance from Jupiter to its outermost large moon Callisto.  Such a large disk would not be stable against accretion (Section~\ref{Roche}) and thus would perhaps be more of a dynamically evolving proto-satellite disk than a stable ring system. 

A complex two-month-long eclipse observed for the star 1SWASP~J140747.93-394542.6 may have been due to a disk surrounding an otherwise-unknown planet, though it is also possible that the occulting disk instead adorns a low-mass stellar companion \citep{Mamajek12}. 

\section{Rings by type \label{RingsByType}}

\subsection{Dense broad disks \label{DenseBroad}}

\begin{center}
\begin{tabular}{|p{2cm}|p{3cm}|}
\hline
\textbf{Saturn:} & C ring \\
& B ring \\
& Cassini division \\
& A ring \\
\hline
\end{tabular}
\end{center}

Although the idea that Saturn's main ring system is composed of ``countless tiny ringlets'' continues to appear even in the professional literature, it is inaccurate.  It is much better to say that the main ring system is a nearly continuous disk with density that varies radially but only a few true gaps that would separate one ``ring'' from another.  Not only do waves travel radially through the disk (Section~\ref{SpiralWaves}), but material likely does as well through ballistic transport \citep{Durisen89,Durisen92} and direct migration \citep{Giantprops10}.  

\begin{figure*}[!t]
\begin{center}
\includegraphics[width=16cm]{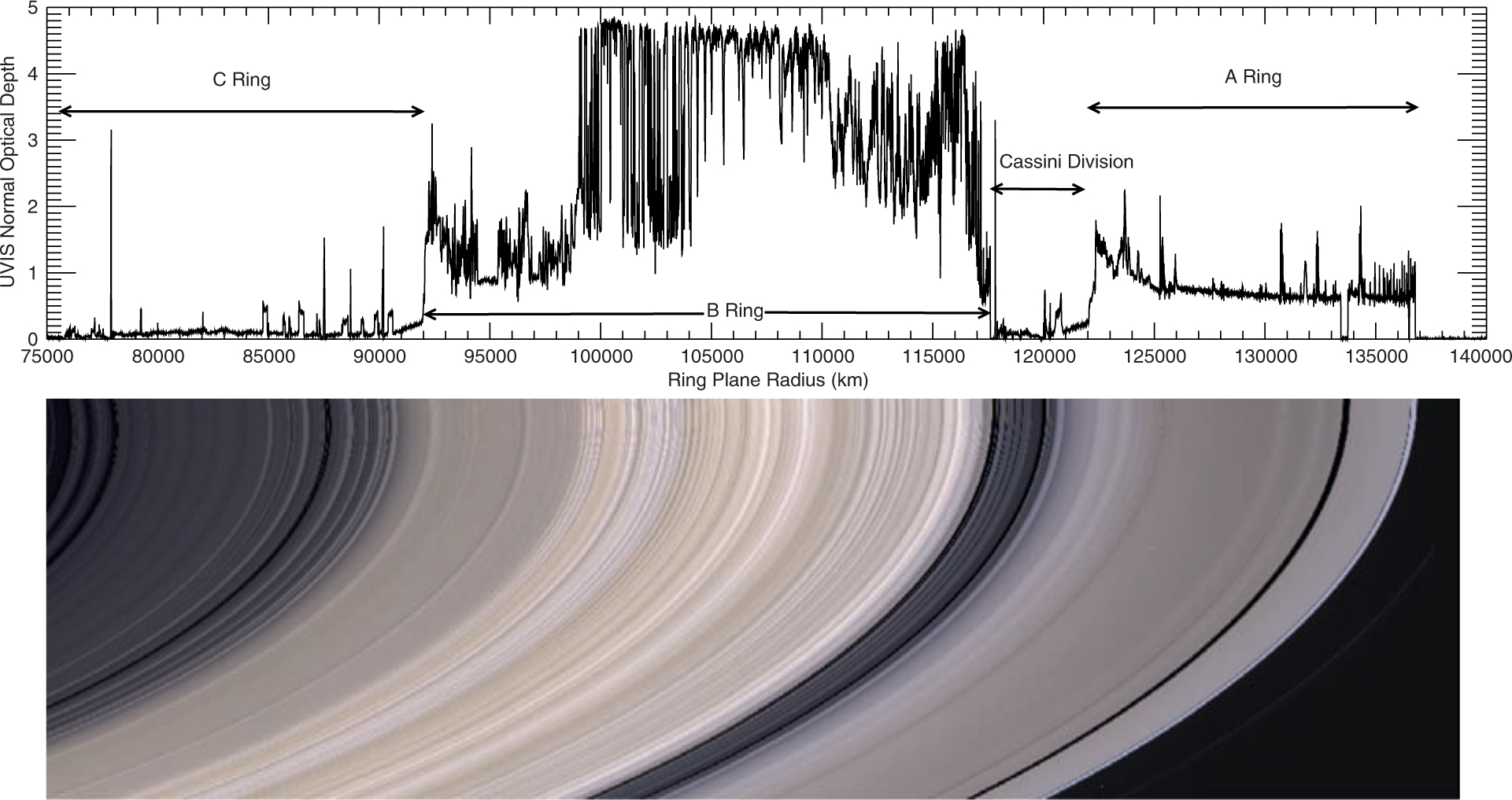}
\caption{An optical depth profile (top) and true-color image (bottom) of Saturn's main ring system.  Figure from \citet{ColwellChapter09}.  
\label{SaturnStrip}}
\end{center}
\end{figure*}

Although Saturn's ring system, taken as a whole, is the only dense broad disk known to us, its four main components are quite different from each other and thus still offer room for comparative study.  In addition to their wide differences in structure, the components vary in location and density.  The B~ring is by far the most dense, with measured $\tau$ of 5 or more, followed by the A~ring at $\tau \sim 0.5$, while the C~ring and Cassini~Division both have $\tau \sim 0.1$ (\Fig{}~\ref{SaturnStrip})  The B~ring's position astride the synchronous distance, at which the orbital period of particles equals the rotation period of the planet (and thus of its magnetic field) is thought to influence at least some of its properties (e.g., spokes), while the A~ring's outermost position causes it to be more susceptible to accretion-related processes.  The differences between the C~ring and the Cassini~Division, despite their apparent similarities in optical depth and composition, may also be primarily due to their different locations with respect to Saturn, its moons, and the B~ring. 

\subsubsection{Spiral waves \label{SpiralWaves}}

Spiral density waves and spiral bending waves are the most widespread well-understood phenomena in dense rings.  First described for the case of galaxies by \citet{LS64}, and applied to planetary rings by \citet{GT78a,GT78b,GT80}, they can arise in any disk subject to a periodic perturbation.  In Saturn's rings, the predominant mechanism is forcing from a perturbing moon.  At discrete locations in the disk, the orbits of individual particles are resonant with the forcing and become excited.  For a resonance with a moon, this happens when the resonance argument $\varphi$, the quantity that librates about a constant value at the resonant location, is of the form\fn{For brevity, this discussion is limited to inner Lindblad resonances and inner vertical resonances, where the disk is inward of the forcing moon.  Nearly all known spiral waves in rings are of this kind, though \citet{soirings} detected inwardly-propagating spiral density waves excited by outer Lindblad resonances with Pan.}
\begin{equation}
\varphi = (m+k) \lambda' - (m - 1) \lambda - X - k X' ,
\label{ResArgForm}
\end{equation}
\noindent where $m$ and $k$ are integers, primed quantities refer to the forcing moon, and $X$ is some integer combination of $\varpi$ or $\Omega$ (see Section~\ref{OrbElems} for orbital elements).  For any given value\fn{The azimuthal parameter $m$ gives the number of spiral arms in the resonant wave pattern, while $k+1$ is the ``order'' of the resonance, with first-order resonances generally being strongest, followed by second-order, etc.} of $m$ and $k$, and for any particular forcing moon, there is a unique location in the ring-plane at which the resonance occurs; this location can be found by differentiating \Eqn{}~\ref{ResArgForm} and setting the time-derivative $\dot{\varphi} = 0$, since $n$ ($= \dot{\lambda}$), $\dot{\varpi}$, and $\dot{\Omega}$ are all known functions of radial location in the ring-plane ($a$), and the orbital frequencies of the moon are known.  In fact, since $n$ and $n'$ are much larger than the others, the approximate location of the resonance can be found from them alone, and the resonance is generally labeled with the ratio ($m$+$k$):($m$-1).  At a Lindblad resonance (LR), where the identity of $X$ (though not necessarily $X'$) is $\varpi$, the eccentricity of the ring particle is excited, and a spiral density wave (a compression wave) propagates radially outward away from the resonant location.  At a vertical resonance (VR), where the identity of $X$ is $\Omega$, the inclination of the ring particle is excited, and a spiral bending wave (a transverse wave) propagates radially inward\fn{The exception is the nodal resonance, labeled -1:0, in which the mean motion of the ring particles is resonant with the forcing moon's nodal precession $\dot{\Omega}$.  This peculiar resonance has a negative pattern speed, and its bending wave propagates outward \citep{RL88}.} away from the resonant location.  For more details see, e.g., Section~10.3 of \citet{MD99}.  

When the perturbation is not too strong, the radial variation in surface density $\Delta \sigma(r)$ for a spiral density wave has the form
\begin{equation}
\Delta \sigma(r) \simeq A \xi \cos (\xi^2/2) e^{-(\xi/\xi_\mathrm{D})^3} , 
\label{DensWaveEqn}
\end{equation}
\noindent where $A$ is an amplitude, $\xi_\mathrm{D}$ is a damping constant, and the dimensionless radial distance from resonance $\xi$ is given by
\begin{equation}
\xi \equiv \left( \frac{ 3 (m-1) n_\mathrm{L}^2 r_\mathrm{L}}{2 \pi G \sigma_0} \right)^{1/2} \cdot \frac{r-r_\mathrm{L}}{r_\mathrm{L}} , 
\label{XiEqn}
\end{equation}
\noindent where $r_\mathrm{L}$ and $n_\mathrm{L}$ are the radius and mean motion at the location of exact resonance, $\sigma_0$ is the unperturbed surface density, and $G$ is Newton's constant.  This simplified equation is valid only for the downstream portion of the wave, $\xi \gtrsim 1$.  We have also removed the phase term, which causes the wave to have a spiral shape.  For the linear theory in its full form, see \citet{GT82}, \citet{Shu84}, or \citet{soirings}.  

\begin{figure*}[!t]
\begin{center}
\includegraphics[width=11cm]{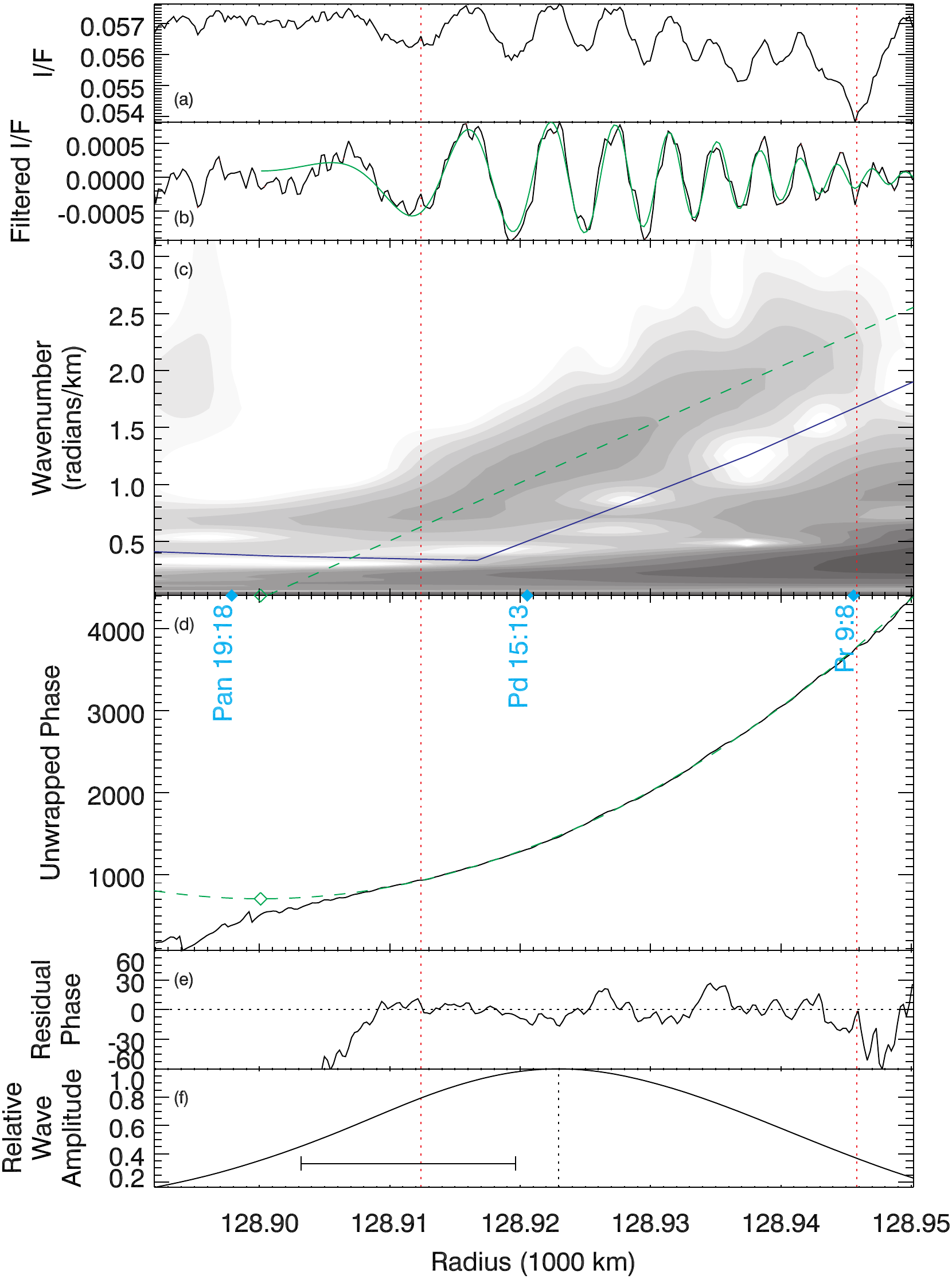}
\caption{A density wave fitting process carried out on a radial brightness profile from a \Cassit{}~ISS image.  a) Initial radial scan. b) High-pass-filtered radial scan, with the final fitted wave shown in green. c) Wavelet transform of radial scan, with blue line indicating the filter boundary, and the green dashed line indicating the fitted wave's wavenumber. d) Unwrapped wavelet phase, with green dashed line indicating the quadratic fit and green open diamond the zero-derivative point. e) Residual wavelet phase, showing that the interval used for the fit is the interval in which the phase behaves quadratically. f) Wave amplitude, the local maximum of which (vertical dotted line) gives $\xi_\mathrm{D}$; scale bar indicates the smoothing length of the boxcar filter.  Figure from \citet[q.v. for details of the wave-fitting process]{soirings}. 
\label{WaveletPlot}}
\end{center}
\end{figure*}

Spiral waves, especially weak ones, are useful structures that can be thought of as \textit{in~situ} scientific instruments placed in the rings.  \Eqn{}~\ref{DensWaveEqn} describes a sinusoid with frequency that increases linearly with distance from the resonance; the rate of increase is inversely proportional to the background surface density $\sigma_0$, and thus can be used to measure it.  The sinusoid oscillates within an envelope whose amplitude begins by increasing linearly, but then turns over and begins to decay as the exponential damping term takes over; the location of that turn-over is governed by the damping constant $\xi_\mathrm{D}$, which can thus be obtained and used to constrain the ring's dynamic viscosity \citep{GT78b,Shu84}.  Additionally, the mass of the perturbing moon can be obtained from the amplitude $A$, its orbital phase from the wave's phase term, and the absolute distance from Saturn's center from the resonance location $r_\mathrm{L}$, though these parameters are often already known too precisely for the density wave to make a significant contribution.  An algorithm for extracting these parameters from observed weak density waves (\Fig{}~\ref{WaveletPlot}) was described by \citet{soirings}, making use of the spatially localized frequency spectra provided by a wavelet transform \citep{Daub92,TC98}.  

\begin{figure*}[!t]
\begin{center}
\includegraphics[width=11cm]{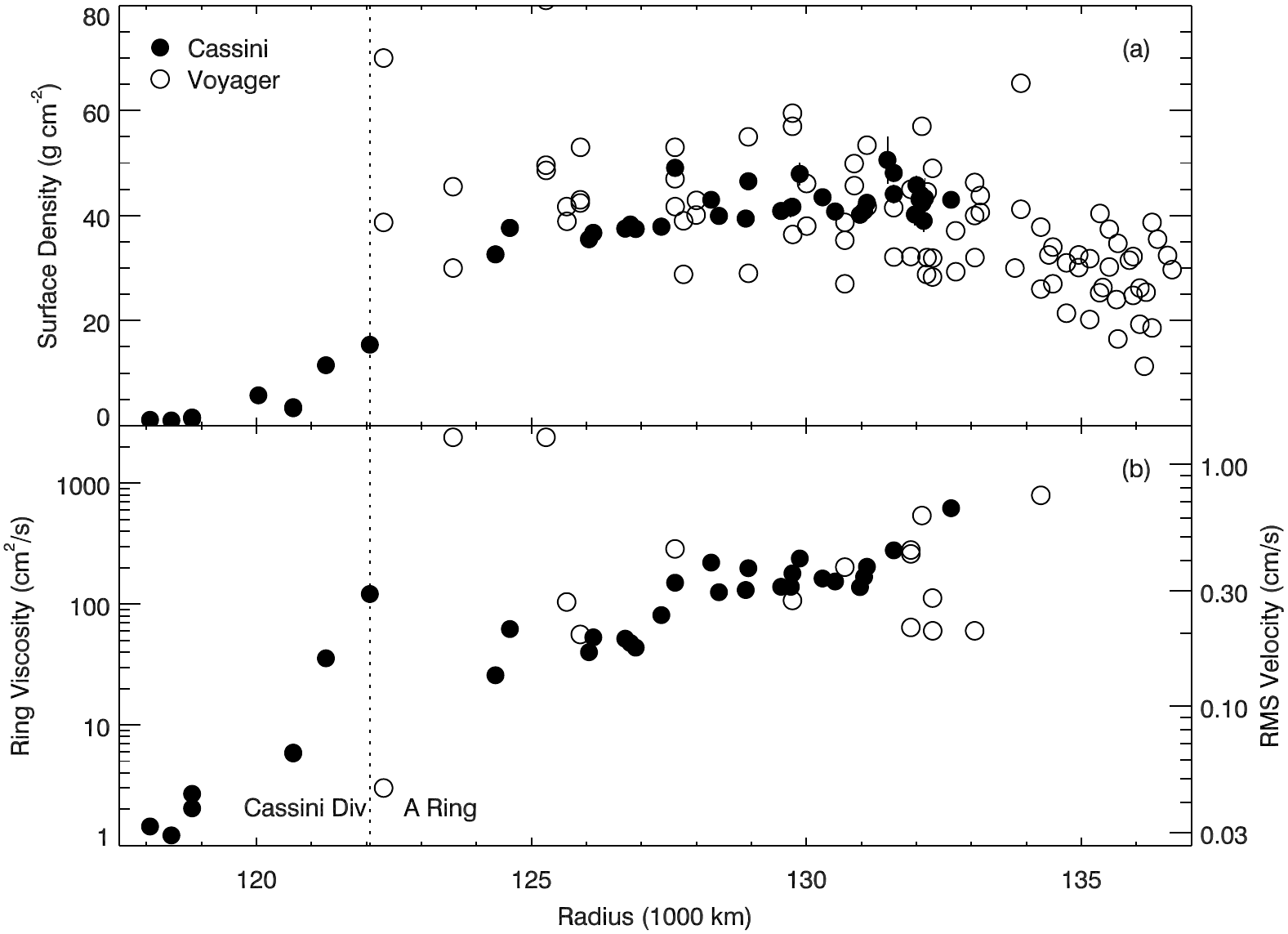}
\caption{a) Ring surface mass densities $\sigma_0$ from density wave analysis. b) Local ring viscosities; the associated rms velocity is within the ring plane, and likely enhanced by self-gravity wakes.  Open circles indicate \Voyit{} data, filled circles \Cassit{} data.  Based on figures from \citet{soirings} and \citet{ColwellChapter09}, q.v.~for details of individual measurements.  
\label{sigmanu}}
\end{center}
\end{figure*}

\begin{figure*}[!t]
\begin{center}
\includegraphics[width=12cm]{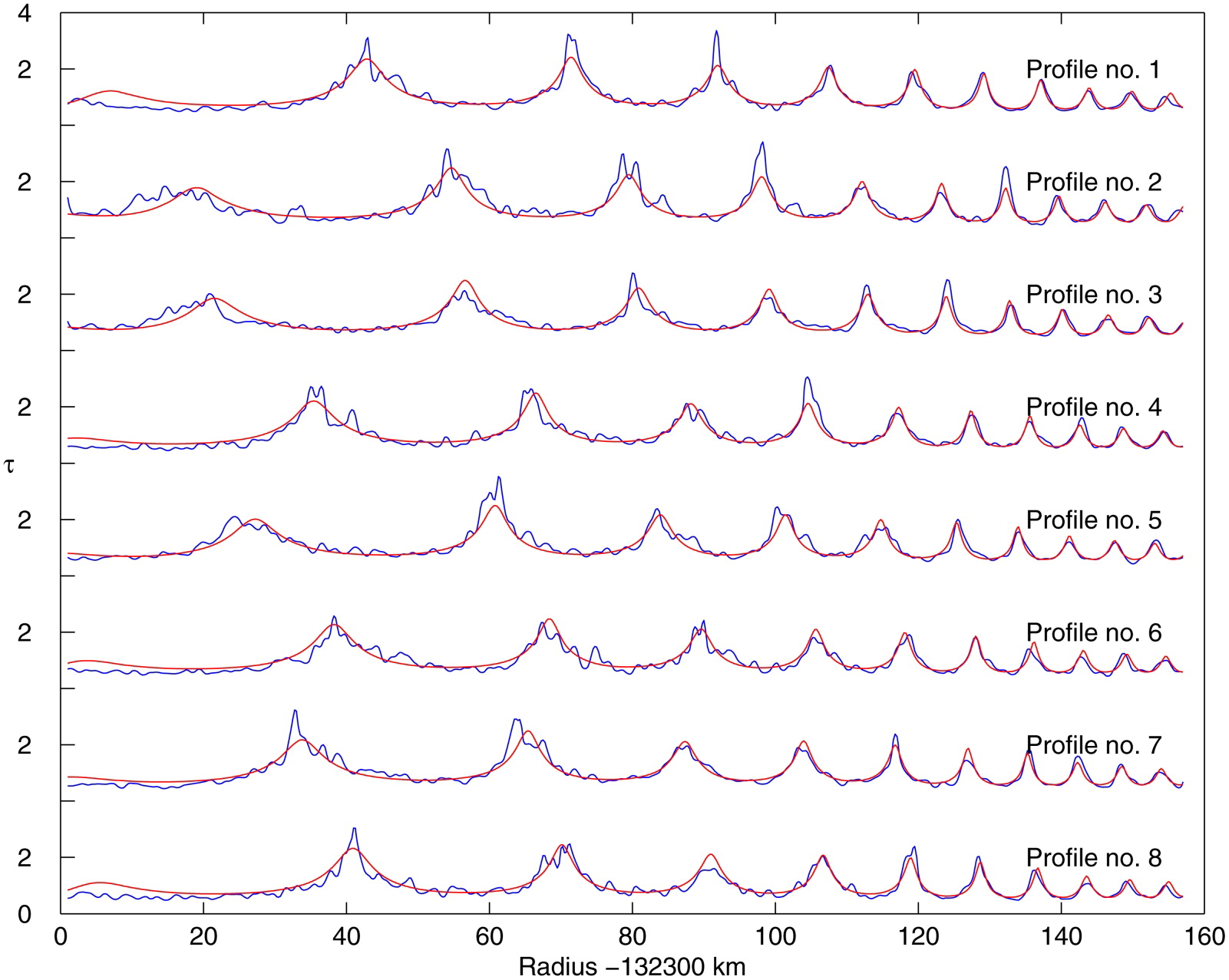}
\caption{Radial optical depth profiles from RSS occultations of the Mimas~5:3 spiral density wave (blue) with model fits (red).  Figure from \citet{Rapp09}. 
\label{NonlinearPlot}}
\end{center}
\end{figure*}

A profile of surface densities and viscosities in the Cassini Division and A~ring is shown in \Fig{}~\ref{sigmanu}.  The \Cassit{} measurements (filled circles) have much less scatter than the \Voyit{} measurements (open circles) not only because of greater measurement sensitivity but also because they rely on weaker waves that adhere more closely to the linear theory (\Eqn{}~\ref{DensWaveEqn}) but were not detectable in \Voyit{} data.  Non-linear spiral density waves occur when the amplitude of the density perturbation $\Delta \sigma$ becomes comparable to the background density $\sigma_0$.  In this case, instead of the quasi-sinusoidal profile of the linear case (\Eqn{}~\ref{DensWaveEqn}, \Fig{}~\ref{WaveletPlot}), the oscillation frequency no longer increases linearly with distance from the resonance, and additionally the wave morphology develops sharp narrow peaks and broad flat troughs.  Significant progress has been made on developing an analytical model to describe non-linear density waves \citep[for a description and list of references, see][]{SchmidtChapter09}.  Most recently, \citet{Rapp09} extracted model parameters from a series of RSS occultation profiles of the Mimas~5:3 density wave, the strongest (and thus most non-linear) density wave in Saturn's rings (\Fig{}~\ref{NonlinearPlot}).  

Spiral bending waves behave similarly to spiral density waves in their wavelength dispersion, but with several differences in other characteristics.  Because they are transverse waves, both damping and non-linearity arise less readily, which in principle makes them even more useful as ring diagnostics.  However, they are significantly less abundant than density waves, because they can only be excited by perturbing moons on inclined orbits.  Also, bending waves are harder to observe because they do not directly affect the optical depth.  Rather, because they appear as corrugations in the ring-plane, they are most readily seen when illuminated from a light source nearly in the ring-plane and oriented azimuthally so as to shine across waveforms rather than along them, so that the waveforms can maximize the effect of changing the path length and thus the observed $\tau$, even while $\tau_\perp$ remains nearly constant (see Section~\ref{OpticalDepth}).  In practice, these conditions are met by occultations with low elevation angle and in images taken near in time to Saturn's equinox (which occurs every $\sim 14.5$~yr).  In \Cassit{} images taken during the 2009 equinox, some of the strongest bending waves (whose peaks rise the highest out of the mean ring-plane) are seen to cast shadows.  

\subsubsection{Gap edges and moonlet wakes \label{GapEdges}}

There are 14 named gaps within Saturn's main rings: 4 in the C~ring, 8 in the Cassini Division, and 2 in the A~ring (\Fig{}~\ref{SaturnStrip}).  Both of the A~ring's gaps contain known moons that clear them by exerting a torque on nearby ring material, and some gaps in the C~ring coincide with known Lindblad resonances, but the other gaps are unexplained.  Furthermore, even the two gaps that do contain known moons exhibit a surprising amount of unexpected behavior at their edges.  

\begin{figure*}[!t]
\begin{center}
~{}~{}~{}~{}\vspace{0.3cm}
\includegraphics[width=15cm]{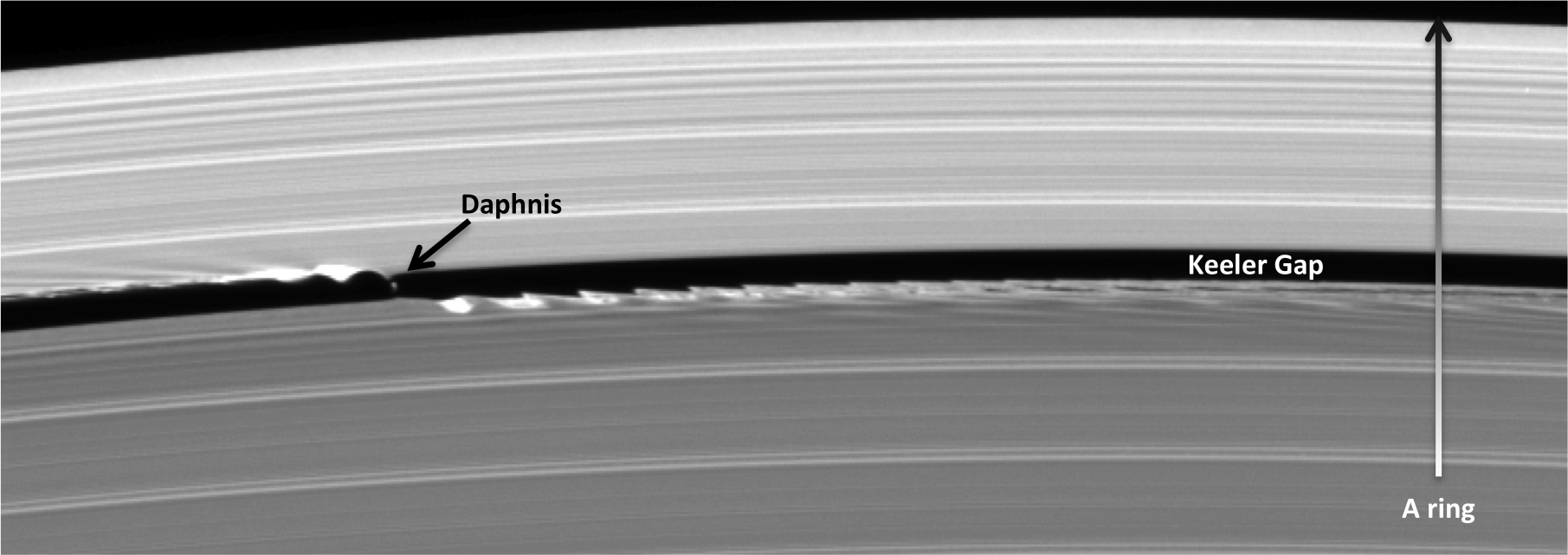}
\includegraphics[width=16cm]{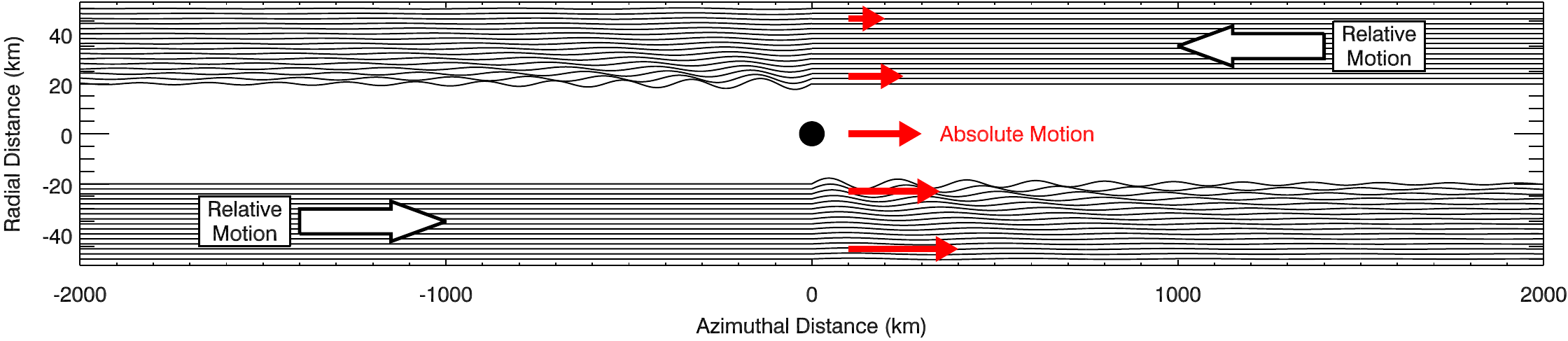}
\caption{(top) Wavy edges and moonlet wakes are seen at the edges of the Keeler Gap, surrounding the position of the gap-moon Daphnis.  (bottom) Streamline diagram showing $3 \pi \Delta a$ wavelength set up as ring material passes a gap-moon.  Different wavelengths at different radial distances $\Delta a$ set up the ``moonlet wakes'' pattern.  By Kepler's Third Law, inner material is moving faster than outer material (red arrows); by the same token, in the moon's frame of reference, inner material moves forward and outer material moves backward (white arrows).  
\label{StreamlinesFig}}
\end{center}
\end{figure*}

When a ring particle passes through conjunction with a nearby moon, the moon's gravity imparts an eccentricity as well as very slightly pushing the particle's semimajor axis away from its own.  On its now-eccentric orbit, the ring particle passes through periapse and apoapse once per orbit; during the same time period, an inward (outward) particle moves forward (backward) by a distance $3 \pi \Delta a$ in the moon's frame of reference (\Fig{}~\ref{StreamlinesFig}).  To derive this, consider Kepler's Third Law, which states that an object's orbital rate $n$ decreases with increasing semimajor axis $a$, specifically $n^2 a^3 = \mathrm{constant}$.  Differentiating this with respect to $a$, we find the equation for \textit{keplerian shear},
\begin{equation}
\frac{\ud n}{\ud a} = - \frac{3}{2} \frac{n}{a} .
\label{KeplerShearEqn}
\end{equation} 
\noindent At relative velocity $v = a \Delta n$, over one orbital period $P = 2 \pi / n$ the relative motion of a ring particle with respect to a perturbing moon is $vP = 2 \pi a \Delta n / n$.  Substituting \Eqn{}~\ref{KeplerShearEqn}, this becomes $3 \pi \Delta a$.  

The characteristic wavelength $3 \pi \Delta a$ is ubiquitous in the vicinity of a gap-moon.  Not only does the gap edge form waves with that wavelength, but streamlines with that wavelength penetrate into the disk (\Fig{}~\ref{StreamlinesFig}).  Because of the increasing wavelength, the distance between streamlines (which correlates with surface density) forms a pattern called ``moonlet wakes''\fn{Moonlet wakes have little in common with self-gravity wakes (Section~\ref{SGWs}), despite an unfortunate similarity in terminology.} that rotates with the gap-moon.  It should be emphasized that these structures are not properly called ``waves,'' as they do not propagate or otherwise dynamically evolve; rather, they are kinematic phenomena caused by the gap-moon organizing the orbital properties of the material around it into streamlines.  However, some of Pan's moonlet wakes do reach high enough densities that the mutual self-gravity of particles enhances the peaks \citep{Porco05} in a manner similar to non-linear density waves (Section~\ref{SpiralWaves}).  Before Pan or any other gap-moon had been discovered, the wavy edges of the Encke Gap were tracked by \citet{CS85}, and the nearby moonlet wakes by \citet{Show86}.  Interpretation of these observed features in terms of the simple theory just described allowed these authors to constrain the position of the moon causing them, which led \citet{Show91} to discover Pan in archival \Voyit{} images. 

Sharp ring edges can also be maintained at the locations of Lindblad resonances (see Section~\ref{SpiralWaves}).  The nature of the perturbation is similar to that in the impulse approximation described above, except that now the resulting streamline wavelength is exactly $1/m$ times the orbital circumference at that location, so that repeated conjunctions combine in a resonant effect.  In fact it can be shown that the resonant streamline wavelength $2 \pi a / m$ reduces to $3 \pi \Delta a$ in the case of large $m$ (which is to say, small $\Delta a/a$).  

For example, the outer edge of the B~ring is coincident with the 2:1 LR with Mimas, and the outer edge of the A~ring with the 7:6 LR with the co-orbital moons Janus and Epimetheus.  Both edges generally exhibit the expected 2-lobed and 7-lobed (respectively) structure, though there are significant deviations that have been attributed to the time-variable orbits of Janus and Epimetheus for the A~ring edge \citep{SP09} and the excitation of normal modes in the nearby disk for the B~ring edge \citep{SP10}.  Of the 8 gaps in the Cassini Division, those not containing ringlets all have circular outer edges, and those not associated with known moon resonances all have freely-precessing elliptical inner edges \citep{HedmanCassDiv10}.  The Keeler Gap in the A~ring also follows this pattern, with a nearly circular outer edge and an 32-lobed inner edge due to a resonance with Prometheus, though again the expected pattern is superposed with other structure that may be due to additional free or forced modes \citep{DPS05,Torrey08}.  

Some edges also exhibit vertical structure.  The moon Daphnis, at the center of the Keeler Gap, has an inclined orbit that ventures $\sim 9$~km 
above and below the ring plane \citep{Jake08}.  The resulting vertical corrugation of nearby portions of the gap edge were predicted and then seen by their shadows cast during the 2009 equinox \citep{Weiss09}.  A region of vertical structure, probably due to embedded moonlets on inclined orbits, has also been detected and tracked in the outer edge of the B~ring \citep{SP10}.  

\subsubsection{Radial structure \label{RadialStructure}}

Several varieties of azimuthally-symmetric radial structure are found in Saturn's rings.  In addition to the gaps discussed above, the two largest in size are a series of sharp-edged annuli in the outer C~ring with optical depths several times higher than the surrounding background, which have been dubbed ``plateaux'', and an undulating variation in optical depth in the inner B~ring.  Both of these have radial scales of $\sim 100$~km and have remained entirely unchanged during the 25~yr between \Voyit{} and \Cassit{} observations \citep{NichSOI08}, as well as remaining unexplained.  

\begin{figure*}[!t]
\begin{center}
\includegraphics[width=14cm]{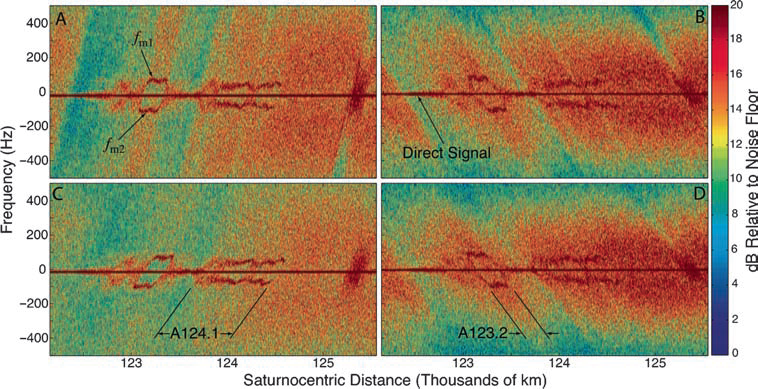}
\caption{Spectrograms of \Cassit{}~RSS data showing periodic microstructure in the inner A~ring.  The central horizontal line in each panel is the direct signal, while the side bands that occur at some locations are coherent diffracted signal from the periodic microstructure acting as a diffraction grating.  Figure from \citet{Thomson07}.
\label{ThomsonFig}}
\end{center}
\end{figure*}

Regions of short-wavelength axisymmetric waves have been found by occultations in the inner A~ring and in the B~ring \citep{Thomson07,Colwell07}.  The wavelengths range from 0.15~to 0.22~km, and are locally monochromatic -- the waveforms interacted with the spectrally coherent \Cassit{}~RSS radio occultation signal as if they were a diffraction grating \citep{Thomson07}.  This structure (\Fig{}~\ref{ThomsonFig}) has been explained in terms of viscous overstability, which occurs when a perturbation triggers an overly strong restoring force that resulting in continuing oscillations \citep[for details and references, see][]{SchmidtChapter09}.  Viscous overstability can occur when the ring's viscosity increases steeply with density, as naturally occurs in dense rings due to increasingly frequent collisions, and is sensitively affected by the strength of mutual self-gravity, the distribution of particle sizes, and the existence of self-gravity wakes.  Work is ongoing to characterize the appearance of overstability waves in \Cassit{} data, as well as their behavior in response to various environmental factors in simulations. 

Both the A~and B~rings have sharp drops in optical depth at inner edges, with a ``ramp'' region of gradually decreasing optical depth (with decreasing distance) inward of them.  These ramps have generally been considered to be the outermost portions of the Cassini Division and the C~ring, respectively (see \Fig{}~\ref{SaturnStrip}).  The similarities in morphology, for the inner boundaries of the only two truly dense ($\tau \gtrsim 0.5$) broad disk structures known, are striking.  The morphology of the ramp structure as well as the sharp edge at its outward boundary (which, in both known cases, is not correlated with any known resonance strong enough to explain it), as well as the apparent compositional similarity between the ramp material and the denser ring on the other side of the edge, has been explained in terms of ballistic transport, the radial movement of material due to micrometeroid bombardment \citep{Durisen89,Durisen92}.  However, doubt is cast on that hypothesis by recent measurements of the wavelength dispersion of a spiral wave (Section~\ref{SpiralWaves}) stretching across the sharp edge in optical depth between the Cassini Division ramp and the A~ring, which indicates that there may not be a corresponding sharp change in surface density at that location \citep{DPS09}.  

\subsubsection{Self-gravity wakes \label{SGWs}}

The boundary between disruption-dominated regions (in which disks are stable) and accretion-dominated regions (see Section~\ref{Roche}) is not a sharp one.  In the outer parts of the region of stability for disks, gravitational instabilities drive temporary accretion that is quickly disrupted again by tides, forming a disk micro-structure known as self-gravity wakes (SGWs).  This balance is characterized by Toomre's $Q$ parameter \citep[for details and references, see][]{SchmidtChapter09}, defined as 
\begin{equation}
Q = \frac{c_\mathrm{r} \kappa}{3.36 G \sigma} , 
\label{ToomreQ}
\end{equation}
\noindent where $c_\mathrm{r}$ is the radial velocity dispersion, $\sigma$ is the local surface density, and $\kappa \equiv n - \dot{\varpi} \approx n$ (see Section~\ref{OrbElems}).  Gravitational instability (i.e., accretion) is generally avoided as long as $Q$ is at least a few times unity, but can become prominent if random velocities are damped (lower $c_\mathrm{r}$) and/or surface density $\sigma$ is increased and/or at locations further from the planet (lower $\kappa$).  Balance between accretion and disruption, leading to SGWs, occurs when $Q \sim 2$.  Data from the damping of spiral density waves (Section~\ref{SpiralWaves}) indicate that $c_\mathrm{r}$ and $\sigma$ adjust themselves in order to maintain $Q \sim 2$ over a wide region of the A~ring \citep{soirings}.  

\begin{figure*}[!t]
\begin{center}
\includegraphics[width=8cm]{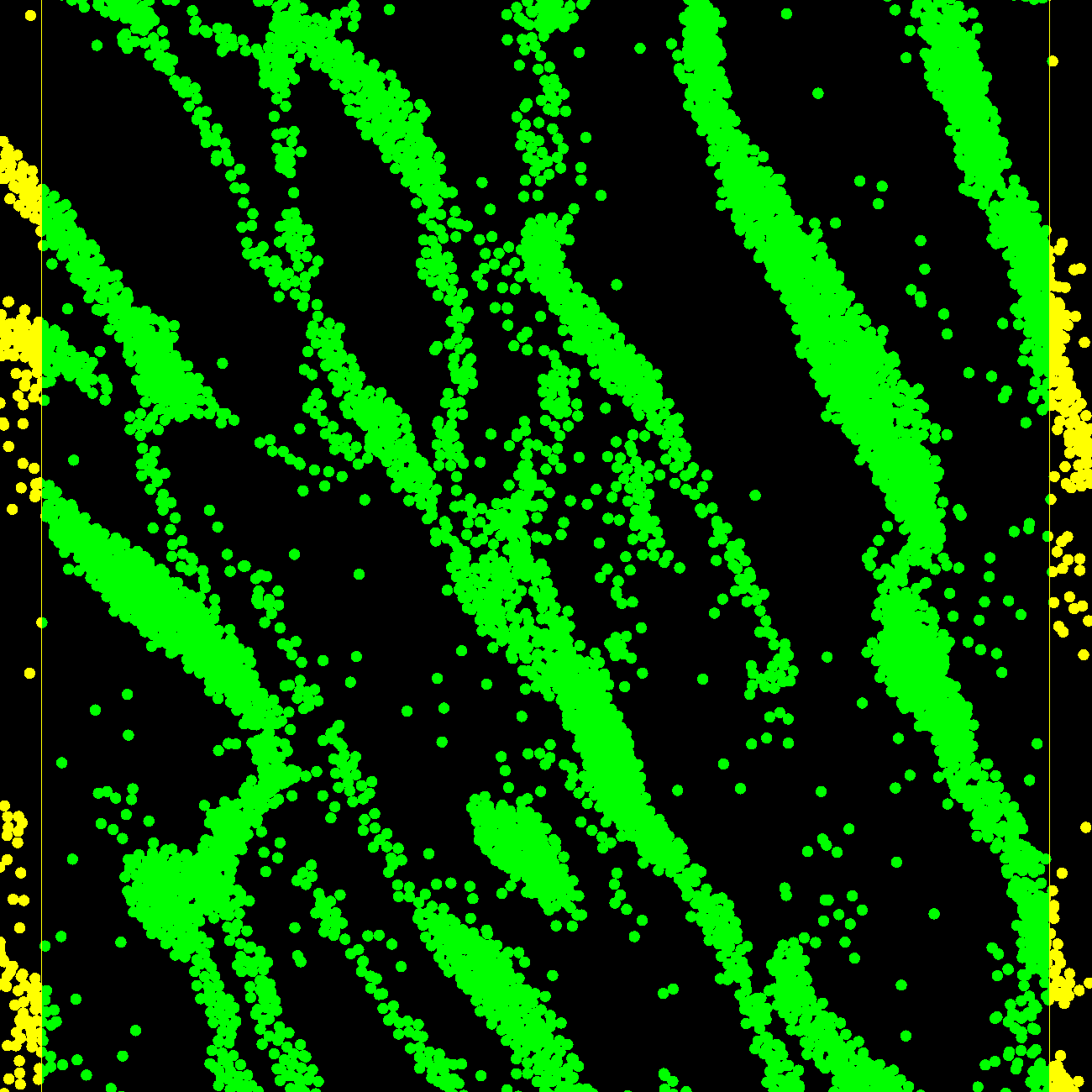}
\caption{Simulated self gravity wakes (SGWs).  This microstructure develops within dense rings as particles clump together under their mutual self-gravity but are ripped apart again by Saturn's tides.  Figure courtesy of R.~P.~Perrine and D.~C.~Richardson.
\label{SGWsFig}}
\end{center}
\end{figure*}

\begin{figure*}[!t]
\begin{center}
\includegraphics[width=16cm]{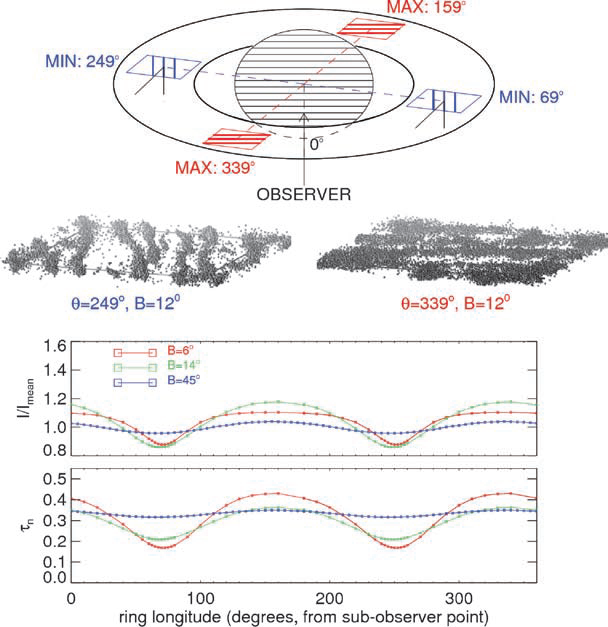}
\caption{The self-gravity wakes shown here \citep[from the simulations of][]{Salo04} are canted at an angle of $\sim 21^\circ$.  At low elevation angle $B$, the transparency is higher when viewed along the wake structures (blue) than when viewed across the wake structures (red), leading to an azimuthal brightness asymmetry that depends on longitude relative to the observer (lower panel).  Figure from \citet{SchmidtChapter09}.
\label{SGWsSchmidtFig}}
\end{center}
\end{figure*}

SGWs generally have a webbed structure that is elongated in a characteristic direction (\Fig{}~\ref{SGWsFig}), usually a few degrees to a few tens of degrees from azimuthal.  Perpendicular to the characteristic direction, there is a characteristic spacing given by the Toomre critical wavelength, 
\begin{equation}
\label{ToomreWavelength}
\lambda_\mathrm{cr} = \frac{4 \pi^2 G \sigma}{\kappa^2} .  
\end{equation}
\noindent Because of their non-axisymmetric structure and their finite vertical thickness, the brightness of a disk pervaded by SGWs depends on the observer's longitude.  An observer looking along the direction of the elongated wake structures will see more of the gaps between the wakes than an observer looking across the wakes (\Fig{}~\ref{SGWsSchmidtFig}), especially at low elevation angles.  \citet{Colombo76} were the first to suggest the presence of SGWs in Saturn's rings as an explanation for the observed azimuthal brightness asymmetry.  Today, observations of ring photometry combined with simulations of SGWs are further refining understanding of ring properties (see Sections~\ref{NumSims} and~\ref{CoeffRest}).  

\begin{figure*}[!t]
\begin{center}
\includegraphics[width=8cm]{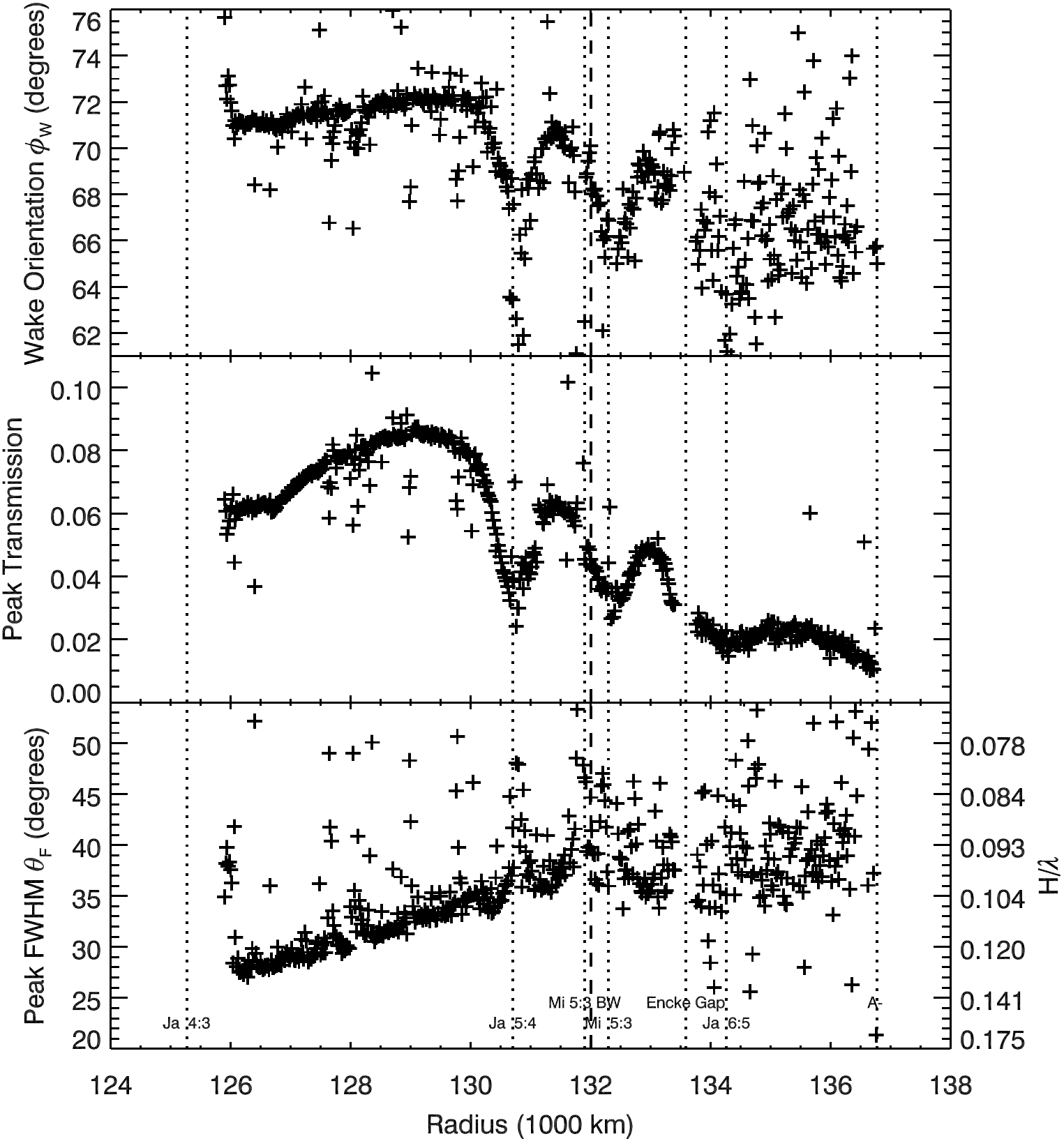}
\caption{Radial profiles of measured parameters for self-gravity wakes in the A~ring.  Work is ongoing \citep{HedmanDPS11} to understand the ``halos'' surrounding density waves (dotted lines) but not bending waves (dashed lines), within which the intensity and orientation of SGWs are altered, in addition to changes in spectral absorption by water ice and in the abundance of propellers (Section~\ref{Propellers}).  Figure from \citet{Hedman07}.
\label{SGWsHedmanFig}}
\end{center}
\end{figure*}

\begin{figure}[!t]
\begin{center}
\includegraphics[width=8cm,keepaspectratio=true]{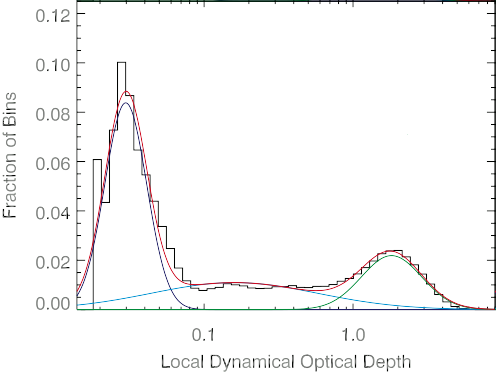}
\caption{A histogram of local dynamical optical depth $\tau_\mathrm{dyn}$ calculated using a local density estimation method for a ring patch containing self-gravity wakes \citep{Anparsgw10}.  A least-squares fit was made to a sum of three gaussians.  The input surface density was 50~g~cm$^{-2}$, and the coefficient of restitution law (see Section~\ref{CoeffRest}) is that of \citet{BGT84} using $v^* = 0.001$~cm~s$^{-1}$.  Figure modified from \citet{Anparsgw10}.
\label{dcr_histfit4}}
\end{center}
\end{figure}

Repeated stellar occultations with varying geometries are another way to gather information about SGW structure.  Such observations from \Cassit{} UVIS \citep{Colwell06,Colwell07} and \Cassit{} VIMS \citep{Hedman07,NH10} have been interpreted in terms of simple models with a bimodal distribution of optical depths, treating the SGWs as nearly opaque with optical depth $\tau_\mathrm{wake}$, and the intervening space as characterized by a constant lower optical depth $\tau_\mathrm{gap}$.  Both teams have produced resulting data sets of various wake parameters as a function of radial location in the disk (\Fig{}~\ref{SGWsHedmanFig}). However, \citet{Anparsgw10} investigated the density contrast and the distribution of densities in simulated SGWs and found instead a trimodal distribution (\Fig{}~\ref{dcr_histfit4}).  In their histograms (\Fig{}~\ref{dcr_histfit4}), the high-$\tau$ peak corresponds to the $\tau_\mathrm{wake}$ assumed in simpler models, while the low-$\tau$ peak is practically transparent by contrast.  In such a regime, the photometry of SGWs (especially for occultations and for images of the unlit side of the rings) is largely dominated by the mid-$\tau$ peak, which simulated movies identify as former wakes in the process of disruption. 
The areal average of the mid-$\tau$ and low-$\tau$ regions can be identified with the $\tau_\mathrm{gap}$ values inferred from simpler models, except in the case of very low elevation angle \citep{Anparsgw10}, thus preserving the usefulness of the previous UVIS and VIMS studies.  Recent preliminary results from a very-high-resolution UVIS occultation appear to give empirical confirmation that the distribution of surface densities in SGW-dominated disk regions is trimodal \citep{SremcevicAGU09}.  

SGWs may have a dramatic effect on the relationship between ring optical depth and surface density.  Simulations by \citet{Robbins10} found that increased surface density merely added more mass to the already-opaque wakes and only weakly increased the overall optical depth.  They estimated that the mass of the B~ring may be higher than \Voyit{}-era estimates by a factor of 10 or more, approaching twice the mass of Mimas.  See Section~\ref{AgeOrigin} for a discussion of the impact of this finding on the age and origin of Saturn's rings. 

\subsubsection{Propellers \label{Propellers}}

\begin{figure}[!t]
\begin{center}
\includegraphics[width=14cm,keepaspectratio=true]{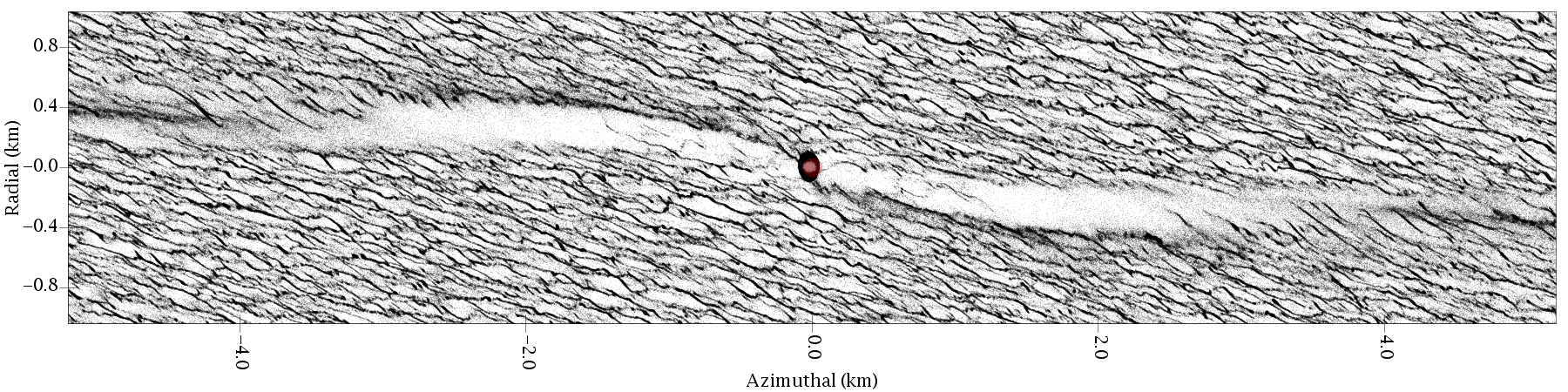}
\caption{Simulated ``propeller'' disturbance in the A~ring due to an embedded moonlet.  Note the texture of self-gravity wakes in the unperturbed regions, and the near-horizontal regions of depletion (white) and enhancement (black) in density due to the moonlet.  Figure courtesy of M.~C.~Lewis. 
\label{r4step2180}}
\end{center}
\end{figure}

\begin{figure}[!t]
\begin{center}
\includegraphics[width=14cm,keepaspectratio=true]{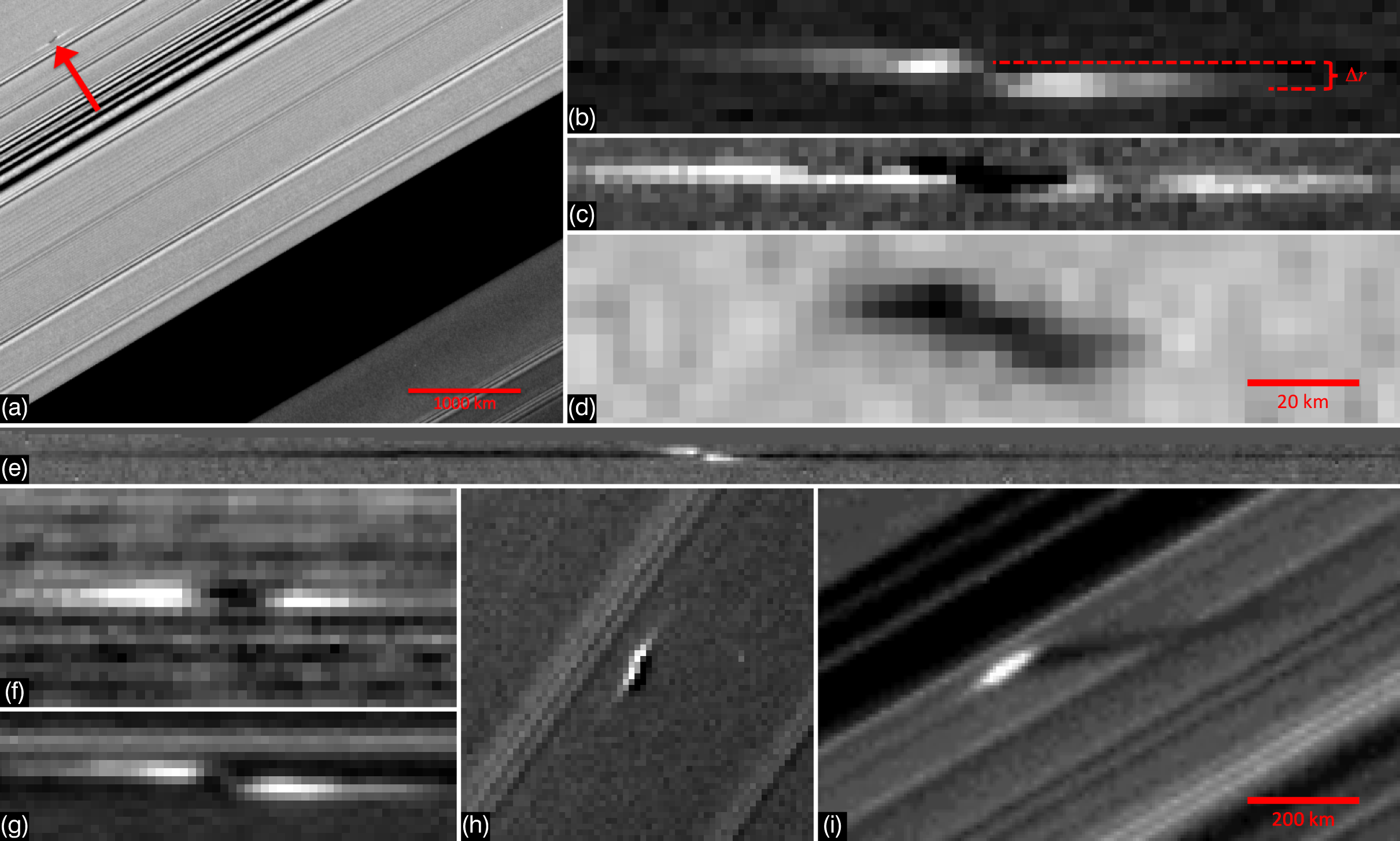}
\caption{Propellers as seen in selected \Cassit{}~ISS images.  Panel (a) shows a propeller in context of the Encke Gap and several density waves.  Panel (b) illustrates the radial offset $\Delta r$ between the two azimuthally-aligned lobes, proportional to the size of the unseen central moonlet.  Panels (b), (c), and (d) show three views of the same propeller at the same scale, demonstrating how its appearance changes with viewing geometry.  Non-equinox views are on the lit (b,e,g) or unlit (a,c,d,f) face of the rings while, panels (h) and (i) show propellers casting shadows near the saturnian equinox.  The scale bar in panel (d) also applies to panels (b), (c), (f), and (g).  The scale bar in panel (i) also applies to panels (e) and (h).  Figure from \citet{Giantprops10}.
\label{apjl2010_fig1}}
\end{center}
\end{figure}

A disk-embedded moon that is too small to open a full circumferential gap may still create a local disturbance in the disk.  Because of Kepler's Third Law, the radially inward portion of the disturbance is carried forward and leads the moon while the radially outward portion trails the moonlet (\Fig{s}~\ref{r4step2180} and~\ref{apjl2010_fig1}).  Due to this characteristic shape, such moonlet-caused disturbances have been named ``propellers''.  A propeller-shaped disturbance can be thought of as a moon's unsuccessful attempt to form a full circumferential gap, which is frustrated by local ring viscous processes that limit the gap's azimuthal extent.  Following predictions based on theory and modeling \citep{SS00,SSD02,Seiss05}, propellers in a range of sizes have been observed in \Cassit{} images \citep{Propellers06, Sremcevic07,Propellers08,Giantprops10}.  In all cases, only the propeller-shaped disturbance is directly seen, while the responsible moon at the center is inferred.  

\begin{figure*}[!t]
\begin{center}
\includegraphics[width=8cm]{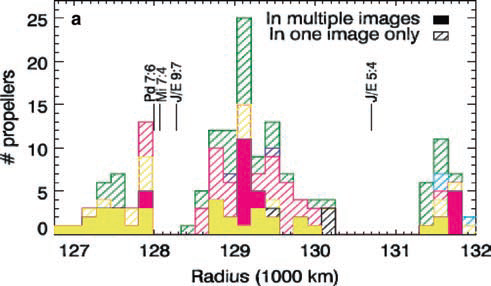}
\caption{A histogram of the abundance of propellers, as a function of radius, in the Propeller Belts of the mid-A~ring.  Figure from \citet{Propellers08}.
\label{PropellerHistogram}}
\end{center}
\end{figure*}

The occurrence of observed propellers is confined to only a few annular regions within the A~ring.  The three ``Propeller Belts'' in the mid-A~ring are located between 127,000~and 132,000~km from Saturn's center (\Fig{}~\ref{PropellerHistogram}) and are separated by propeller-poor ``halo'' regions, centered on large density waves, in which other ring properties are also altered (\Fig{}~\ref{SGWsHedmanFig}).  It is not known whether the radial variations in observed abundance of propellers are due to variations in their origins, survival, observability, or a combination of the three.  

The Propeller Belts contain numerous small propellers with the radial offset parameter $\Delta r$ (\Fig{}~\ref{apjl2010_fig1}) ranging from 0.3~to 1.4~km and azimuthal extent of up to several~km \citep{Propellers08}.  A separate class of ``giant propellers'' has been found in the outermost A~ring, outward of the Encke Gap (133,700~km from Saturn's center) and thus called the ``\textit{trans}-Encke'' population, with measured $\Delta r$ as high as 6~km and azimuthal extent of up to several thousand~km \citep{Giantprops10}.  

Simulated propeller structures have both density-depleted and density-enhanced regions (\Fig{}~\ref{r4step2180}), and the interpretation of observed propellers in terms of relative density has been a point of controversy.  In simulated propellers, the radial offset $\Delta r$ is consistently close to 4 times the central moon's Hill radius (\Eqn{}~\ref{HillRadiusEqn}), and thus is diagnostic of its mass.  The first observed propellers, which were seen during \Cassit{}'s Saturn Orbit Insertion (SOI) maneuver and are thus the smallest known but also the best-resolved in the Propeller Belts, are relative-bright with respect to nearby unperturbed regions of the A~ring \citep{Propellers06}.  Although this photometry is consistent with a region of enhanced optical depth, the morphology of the SOI~propellers is very similar to that of density-depleted regions in propeller simulations.  The photometry of larger propellers in the Propeller Belts also turned out to be generally consistent with enhanced optical depth \citep{Sremcevic07,Propellers08}.  Hypotheses to explain the apparent correlation between high optical depth and depleted density include the temporary liberation of ring-particle regolith within the propeller structure \citep{Sremcevic07,HalmeDPS10} and the disruption of self-gravity wakes within the propeller structure \citep{Anparsgw10}.  In giant propellers, for the first time, relative-dark and relative-bright regions are sometimes seen in the same propeller structure (\Fig{}~\ref{apjl2010_fig1}), enabling density-enhanced and density-depleted regions to be disentangled in some cases \citep{Giantprops10}.  However, the structure of giant propellers may be qualitatively different than that of the smaller propellers in the Propeller Belts, so parallels should be drawn cautiously. 

The giant \textit{trans}-Encke propellers are larger and more prominent than those in the Propeller Belts, and also much less numerous.  Taken together, these factors allow giant propellers to be studied individually and tracked over a period of years.  In the Propeller Belts, the swarm of particles is such that, even if the same object were seen on multiple occasions, it would be very difficult to have much certainty that it was the same object due to the many nearby similar objects.  This criterion may be useful for drawing a distinction, should one wish to do so, between a ``moon'' and a ``moonlet''.  Basing such a distinction upon size is arbitrary and lacks any wide agreement, as there is no major physical threshold crossed by objects in the km-size range.  We suggest that any object be called a ``moon'' if it can be singled out for long-term study, while a ``moonlet'' is a member of a population or swarm that prevents individual tracking.  This may still not be a bright line, but at least it's based on a physical property.  The question of whether propeller-causing objects deserve to be considered as full-fledged moons is further complicated by the fact that, though their positions have been tracked over long periods, they are not directly seen but hidden within the surrounding propeller-shaped disturbance. 

The propeller population follows a very steep particle-size distribution, with a power-law index $q \sim 6$ \citep{Propellers08,Giantprops10} for propeller moonlets of size between 30~m and 300~m.  By contrast, the vast majority of ring mass is concentrated in the continuum particles of size between 1~cm and 10~m, with a much shallower power-law index $q \sim 2.75$ \citep{Zebker85,CuzziChapter09}

\begin{figure*}[!t]
\begin{center}
\includegraphics[width=8cm]{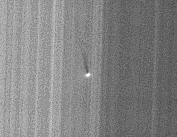}
\caption{The feature known as S/2009~S~1, identified by the shadow it cast during equinox, was seen only in this image.  Figure from \citet{SP10}.
\label{Bprop}}
\end{center}
\end{figure*}

\citet{SP10} reported a single observation of what appears to be an embedded moon of radius 0.3~km in the outermost parts of the B~ring, based on the shadow it cast onto the (vertically much thinner) ring while the 2009 equinox brought nearly edge-on ring illumination (\Fig{}~\ref{Bprop}).  However, though this object's size is comparable to that inferred for the largest propeller moons, no propeller structure is apparent around this object.  \citet{MK11} investigated the possibility that the B~ring's high surface density, with accompanyingly strong self-gravity wakes, might prevent the formation of a propeller structure.  They found that propellers form when the moonlet's Hill~radius $r_\mathrm{Hill}$ (\Eqn{}~\ref{HillRadiusEqn}) is larger than the Toomre critical wavelength $\lambda_\mathrm{cr}$ (\Eqn{}~\ref{ToomreWavelength}).  While this condition might hold if a moon of the observed size were in the densest parts of the central B~ring, the surface densities inferred by \citet{SP10} for the outer B~ring are not high enough.  Possible explanations include that the surface densities in the outer B~ring are higher than thought, that the observed object is not a moonlet but perhaps an impact cloud or a large SGW clump, that the observed bright feature does indeed include an unresolved propeller structure, or that propellers require a (not well understood) photometric balance in order to be observed and that that balance is not met in the outer B~ring. 

\begin{figure}[!t]
\begin{center}
\includegraphics[width=16cm,keepaspectratio=true]{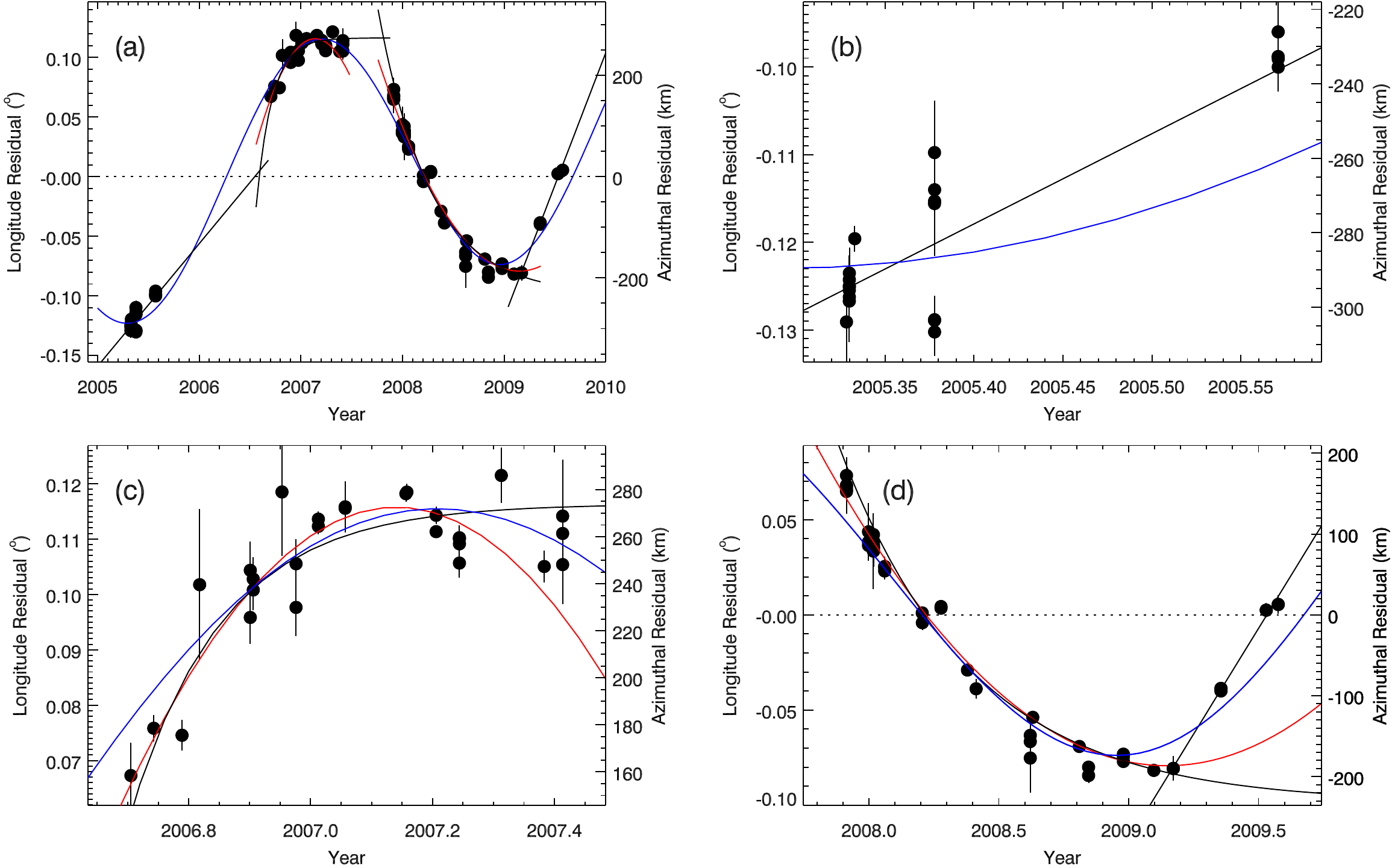}
\caption{Non-keplerian motion of the propeller ``Bl\'eriot'' over 4~years.  Panel (a) contains all the data, while panels (b), (c), and (d) contain subsets of the data shown in greater detail.  The blue line indicates a linear-plus-sinusoidal fit to all the data, while the red lines indicate piecewise quadratic fits corresponding to a constant drift in semimajor axis and the black lines indicate exponential fits.  Figure from \citet{Typeiprops12}.  
\label{bleriot_orbit}}
\end{center}
\end{figure}

Orbital tracking of \textit{trans}-Encke propellers has revealed a surprising non-keplerian component to their motion \citep{Giantprops10}.  The best-observed example had its semimajor axis increase (as detected by tracking its longitude as a function of time, not its actual radial position) by a rate as high as $+0.11$~km~yr$^{-1}$ between 2006 and 2007, then decrease by $-0.04$~km~yr$^{-1}$ from 2007 to 2009 (\Fig{}~\ref{bleriot_orbit}).  Although this profile is similar to a sinusoid, no resonance with a known moon has been found to be capable of producing such behavior.  Another mechanism that would produce a sinusoidal residual is the so-called ``frog resonance'' \citep{PC10}, in which the propeller moonlet interacts primarily with the mass at either end of the propeller gap.  The propeller moonlet plausibly librates with the observed amplitude and period in a simple version of the model, but it remains unclear whether the moon-formed gap responds sluggishly enough to allow the moon to librate within it.  

Other hypotheses for non-keplerian motion of propellers are based on the concept of ``Type~I migration.''  As classically formulated for protoplanetary disks \citep{Ward86,Ward97,PapaloizouPPV07}, the angular momentum exchange at inner Lindblad resonances between a disk and an embedded mass fails to exactly cancel with that at outer Lindblad resonances, resulting in a differential torque that leads to inward migration of the embedded mass.  However, classical Type~I migration depends crucially on the gas component of the disk, which causes Lindblad resonance locations to shift asymmetrically.  For the case of planetary rings, which are strictly particulate, \citet{Crida10} re-derived the equations for Type~I migration from first principles, using analytical arguments and numerical simulations to trace the angular momentum exchange between streamlines of continuum ring particles and the embedded moon.  For the case of a homogeneous disk, \citet{Crida10} found an asymmetric torque that is always inward and is one to two orders of magnitude too weak to explain the observed non-keplerian motion.  Building upon this work, \citet{RP10} considered temporal variations in the disk, proposing that the propeller moon is constantly perturbed by stochastic variations in the disk's surface density due to self-gravity wakes, finding that a random walk in the propeller moon's semimajor axis can result.  On the other hand, \citet{Typeiprops12} considered spatial variations in the disk, proposing that an externally-produced radial surface density profile results in an equilibrium semimajor axis to which the propeller moon will return after episodic kicks. 

Continued observations should distinguish among these models.  The leading models interpret the existing data as a sinusoid \citep{PC10}, episodically-initiated exponentials \citep{Typeiprops12}, and a pure random walk \citep{RP10}, and thus they differently predict the qualitative nature of future data.  However, all of the leading models imply that propeller-moons, which are the first objects ever discovered to orbit while embedded in a disk rather than in free space, are directly interacting with the disk, a phenomenon that has long been an integral part of disk models but has never before been directly observed.  

\subsubsection{Spokes and impacts \label{Spokes}}

``Spokes'' are near-radial markings on Saturn's B~ring, likely composed of dust levitating above the ring-plane due to electromagnetic forces \citep[for details and references, see][]{HoranyiChapter09}.  Spokes usually form on the dawn side of the rings where ring particles are just coming into full sunlight, and have a correlation with periodicities believed to originate with Saturn's magnetic field.  Susceptibility to electromagnetic forces surely is connected to the fact that spokes appear at radial locations near to (and often astride) that of synchronous orbit, where the orbital period of a ring particle matches the rotation period of the planet, and thus also of the magnetic field.  Spokes appear to be a seasonal phenomenon, correlated with a low elevation angle of the Sun above the ring-plane, probably due to variations in the generation of plasma by photocharging near the rings.  The Hubble Space Telescope tracked the decline and disappearance of spokes during ``Saturnian October/November'' in the late 1990s \citep{McGhee05}, and they remained absent during the first 1.5~yr of \Cassit{} operations at Saturn before reappearing in ``Saturnian January/February'' in late 2005 \citep{Mitchell06}.  Since their reappearance, \Cassit{} images have tracked the morphology, photometry, evolution, and temporal variability of spokes \citep{Mitchell12}. 

Numerous theories for the formation of spokes have been put forward, but none has gained definitive acceptance.  The most popular mechanism is that of \citet{GM83}, who interpret spokes as dust levitated above the ring plane by an electromagnetic disturbance initiated by a micrometeroid impact.  Following the inference of \citet{Smith82} from \Voyit{~2} images that spokes with a radial dimension of thousands of km form on a timescale of several minutes, the \citet{GM83} model includes a propagation speed for the the plasma cloud $\gtrsim 20$~km~s$^{-1}$, though this result was criticized by \citet[see reply by \citealt{MT05}]{FG05}.  Other ideas for the rapid appearance of spokes include near-simultaneous impact of a broad assemblage of particles generated elsewhere; the most developed model of this type suggests an electron beam generated by saturnian lightning \citep{HM81,Jones06}, though a dispersed cloud originating from an impact has also been suggested \citep{HamiltonDPS06}.  On the other hand, despite executing a number of high-frequency imaging sequences designed to do so, \Cassit{} has not observed the rapid formation of spokes, but did observe spokes growing from negligible to strong over $\sim$hr timescales \citep{Mitchell12}. 

During the 2009 saturnian equinox, dust clouds evolving under keplerian shear were observed in the A~and C~rings and attributed to micrometeoroid impacts \citep[see also Section~\ref{Detectors}]{AGU09Impacts}.  Ongoing analysis of these impact clouds may lead to new constraints on the interplanetary micrometeoroid population, as well as better understanding of the relationship between impacts and spokes. 

\subsection{Dense narrow rings \label{DenseNarrow}}

\begin{center}
\begin{tabular}{|p{2cm}|p{4.8cm}|}
\hline
\textbf{Saturn:} & Titan ringlet \\
& Maxwell ringlet \\
& Bond ringlet (``1.470~R$_\mathrm{S}$'') \\
& Huygens ringlet \\
& ``Strange'' ringlet \\
& Herschel ringlet (``1.960~R$_\mathrm{S}$'') \\
& Jeffreys ringlet \\
& Laplace ringlet (``1.990~R$_\mathrm{S}$'') \\
\hline
\textbf{Uranus:} & 6 ring \\
& 5 ring \\
& 4 ring \\
& $\alpha$ ring \\
& $\beta$ ring \\
& $\eta$ ring \\
& $\gamma$ ring \\
& $\delta$ ring \\
& $\epsilon$ ring \\
\hline
\end{tabular}
\end{center}

Dense narrow rings are a unique assemblage of matter, behaving as a coherent self-contained object on planetary lengthscales yet ephemerally thin (1~to 100~km in radial width, compared to $\sim 100,000$~km in diameter) and not in a gravitational ground state (unlike a planet, one would collapse if it stopped moving).  The formation and proximate causes of these highly organized dynamical systems are almost entirely unknown, and even their present dynamics are only partly understood.  For uranian narrow ringlets, which constitute the majority of the known examples, their highly time-variable properties were only dimly revealed by the single snapshot provided by the \Voyit{~2} flyby, with temporal resolution provided by Earth-based occultations, while analysis of the more extensive \Cassit{} data set of saturnian narrow ringlets is still in progress.  

Nearly all known dense narrow rings are either non-circular or inclined to the main ring plane, or both.  
Their radial widths range from $\sim 1$~km (many examples) up to $\sim 100$~km for the Maxwell~ringlet and the $\epsilon$~ring.  Furthermore, nearly all dense narrow rings have edges that are quite sharp; as with dense disks, the existence and stability of such sharp edges requires some confinement mechanism to counteract the natural process of radial viscous spreading.  Confinement may be due to an external moon or to the ring's own self-gravity, and in some cases to processes yet to be understood.  A detailed table of ringlets and their properties was given by \citet{ColwellChapter09} for Saturn's rings and by \citet{FrenchChapter91} for Uranus'.  For general details and references on the dynamics of narrow rings, see \citet{FrenchChapter91} and \citet{SchmidtChapter09}.  

The ``shepherding'' mechanism by which a moon opens a gap or maintains an edge in a disk (Section~\ref{GapEdges}) can also occur with two shepherd moons on either side of a narrow ringlet \citep{GT79u}.  As previously discussed, this can occur through repeated impulses on nearby material or more distantly through a resonance.  The only example of a dense narrow ring with a known shepherd on either side is Uranus' $\epsilon$ ring, though Saturn's F~ring (see Section~\ref{ThickDusty}) is a variation on that idea.  The Titan ringlet in Saturn's C~ring is at the location of an apsidal resonance with Titan, where the ring particle's precession frequency $\dot{\varpi}$ is commensurate with Titan's orbital motion, so that the eccentric ringlet always keeps its apoapse pointed towards Titan.  The Bond ringlet in Saturn's outer C~ring is associated with a 3:1 Lindblad resonance with Mimas, and a few edges of uranian rings coincide with resonances, but the details of the interaction are yet to be understood in all cases. 

Any ringlet of finite width ought to have a different precession rate $\dot{\varpi}$ at its inner and outer edges, which should smear out the orientation of ring particle orbits and prevent the ringlet from appearing eccentric.  However, this effect can be counteracted by the ring's own gravity \citep{GT79epsilon,GT81,BGT83elliptical}, possibly combined with viscous and collisional effects \citep{DM80,CG00,ME02}.  A purely gravitational model requires a positive ``eccentricity gradient,'' which is to say that the eccentricity monotonically increases from the ring's inner edge to its outer edge.  

A number of ringlets indeed appear to precess about their planet as a rigid body.  Observations of some of these, such as the Maxwell~ringlet and the $\alpha$~and $\beta$~rings, are consistent with a pure freely-precessing ellipse (i.e., an $m=1$ mode) as described by theory, and the Maxwell~ringlet even has a clearly positive eccentricity gradient \citep{SpitaleDDA06}, while other ringlets appear to have additional components to their motion.  The Huygens ringlet appears to have an additional $m=2$ mode, possibly influenced by the Mimas~2:1 resonance that governs the nearby outer edge of the B~ring, as well as an $m=6$ mode of unknown origin \citep{SpitaleDDA06}.  The $\delta$~ring also has an $m=2$ mode, while the $\gamma$~ring has an $m=0$ mode, which is to say a radial oscillation \citep{FrenchChapter91}.  Higher-$m$ modes, some corresponding to known moons, have also been found in several Uranian rings by \citet{ShowDPS11}, who also pointed out that non-detections of shepherd moons at Uranus have become significant enough that shepherding is unlikely to be the dominant mechanism of ring confinement. 

Spiral density waves (Section~\ref{SpiralWaves}) can also occur in dense narrow rings, at least those broad enough (generally a few~km) for a wave to develop, but not many examples have been identified.  A density wave due to the Pandora 9:7 LR appears to propagate through the Laplace ringlet \citep{ColwellChapter09}.  A possible density wave was detected in the \Voyit{} stellar occultation of Uranus' $\delta$~ring, but it cannot be confirmed in the absence of data at multiple longitudes and/or times, and furthermore there is no known moon at the proper place to raise such a wave. 

In Saturn's rings, gaps are named by the IAU but ringlets are not.  In most cases, the main ringlet within a particular gap is given the same name as its gap (though many favor the name ``Titan ringlet'' for the ringlet in the Colombo Gap, for its strong association with a resonance with Titan), but the existence of multiple ringlets in one gap requires some naming creativity.  For example, the Huygens Gap contains a second dense narrow ringlet outward of the main Huygens ringlet, which has been informally nicknamed the ``Strange'' ringlet in part because of its unusually high inclination \citep{SpitaleDPS08} and in part as a complement to the ``Charming'' ringlet (Section~\ref{ThickDusty}).  Five ringlets with non-circular features were identified in \Voyit{}-era publications \citep[e.g.,][]{French93} only by their distance from Saturn's center in units of Saturn radii; of these, three can now be easily named for the gap containing them as shown in the accompanying table.  However, the former ``1.495~$R_\mathrm{S}$~ringlet'' is classified by \citet{ColwellChapter09} as a plateau or embedded ringlet because it is adjacent to the continuum C~ring rather than fully contained in the Dawes Gap, while the former ``1.994~$R_\mathrm{S}$~ringlet'' is now considered part of the continuum Cassini Division between two narrow gaps, rather than a ringlet that nearly fills its gap, judging from the characteristic pattern of empty gaps having circular outer edges and resonant inner edges \citep[see also Section~\ref{GapEdges}]{HedmanCassDiv10}. 

\subsection{Narrow dusty rings \label{ThickDusty}}

\begin{center}
\begin{tabular}{|p{2cm}|p{3.5cm}|}
\hline
\textbf{Saturn:} & ``Charming'' ringlet \\
& Encke ringlets \\
& F ring \\
\hline
\textbf{Uranus:} & $\lambda$ ring \\
\hline
\textbf{Neptune:} & Le Verrier ring \\
& Arago ring \\
& Adams ring \\
\hline
\end{tabular}
\end{center}

The smallest particles in dense rings are usually swept up by larger ones and incorporated into their regolith \citep{CuzziChapter09}, and so such rings are largely dust-free, which is to say that they have few particles smaller than mm-~to cm-size and thus are not strongly forward-scattering in their interactions with light.  Therefore, the dustiness of a few structures that do occur within Saturn's and Uranus' main rings is a likely indicator of relative dynamism and/or youthfulness that either prevents dust from being swept up or has not allowed time for it to be swept up yet.  In Neptune's rings, on the other hand, the lack of any ring component with optical depth exceeding 0.1 may account for the general dustiness of the system. 

In Saturn's main rings, high dust fractions are found exclusively in a small number of narrow dusty ringlets that occur in the larger empty gaps \citep{HoranyiChapter09}.  The only gap with multiple dusty ringlets is the Encke Gap, which three ringlets share with the moon Pan.  These are riddled with ``clumps'' (azimuthal brightness variations) and ``kinks'' (radial offsets) that drift slowly with respect to each other \citep{Burns05cheat,HedmanDDA11,HoranyiChapter09}.  The Encke ringlets and the ``Charming'' ringlet within the Laplace Gap are known to be ``heliotropic,'' which is to say that they have an eccentricity forced by solar radiation pressure that causes their apoapses to always point towards the Sun \citep{HedmanDPS07,HedmanCharming10}.  The ``Charming'' ringlet is smooth, lacking clumps or kinks; the best-studied of the heliotropic rings, it also has free eccentricity and free inclination components in addition to its solar-forced eccentricity \citep{HedmanCharming10}. 

The F~ring is the granddaddy of narrow dusty ringlets, being by far the best studied as well as the largest.  For a review and references, see \citet{ColwellChapter09}.  Located a few thousand~km off the outer edge of Saturn's main rings, and with both significant vertical thickness and inclination ($\gtrsim 10$~km in both cases), the F~ring effectively frustrates any attempt to see the rest of the main rings in edge-on viewing geometries.  Its high fraction of forward-scattering dust makes it by far the brightest component of Saturn's ring system when viewed at high-phase geometries.  

The core of the F~ring contains a large amount of dust enveloping an unseen belt of km-size moonlets, inferred from their absorption of charged particles \citep{CB88}, from the characteristic ``fan'' structures they create in surrounding dust \citep{Murray08,Beurle10}, from direct detection by occultations \citep{Espo08,Hedman11f}, and from shadows cast during the 2009 saturnian equinox \citep{Beurle10}.  Several dusty lanes or ``strands'' accompany the core on either side, nearly parallel to it though \citet{Charnoz05} pointed out that the most prominent strands can be laid end-to-end to form a one-armed kinematic spiral.  Nascent strands, or ``jets,'' have been associated with collisions between moonlets and the core \citep{Murray08}.  Despite being entirely composed of ``clumps'' and ``kinks'' so numerous that they cannot be individually tracked as they are in the Encke~ringlets, the F~ring core nevertheless maintains over decadal timescales the shape of a freely precessing eccentric inclined ellipse; the orbital solution formulated to account for \Voyit{} and other pre-\Cassit{} data \citep{Bosh02} has, somewhat surprisingly, remained a good predictor of the core's position through the \Cassit{} mission \citep{Murray08,Albers12}.  However, decade-scale time variations in the core's internal structure, as well as in the surrounding dust, have clearly taken place between the \Voyit{} and \Cassit{} visits \citep{ColwellChapter09,ShowDPS09}.

\begin{figure*}[!t]
\begin{center}
\includegraphics[width=16cm]{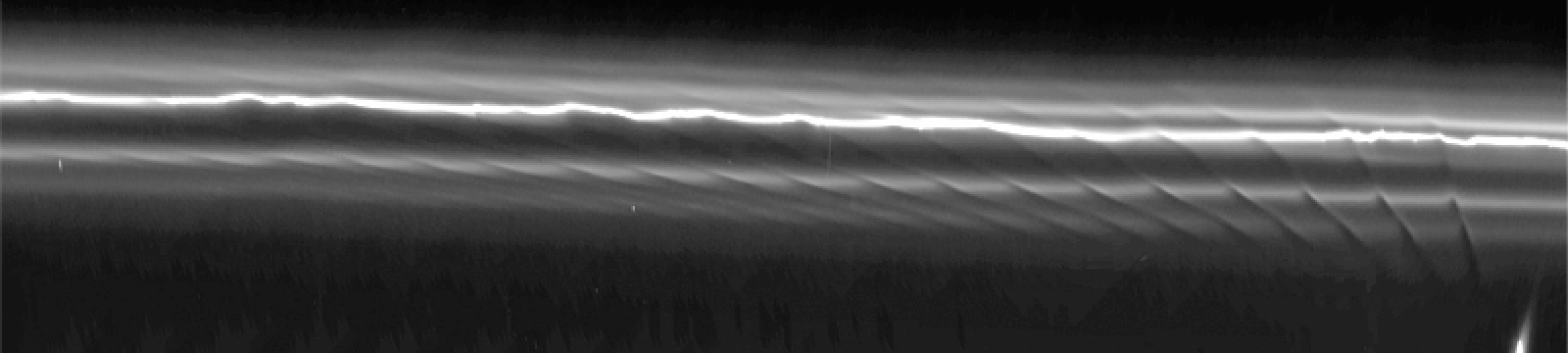}
\caption{Sheared channels in the F~ring created as Prometheus (lower right) dips into the ring.  The horizontal dimension of this \Cassit{}~ISS mosaic covers 60$^\circ$ of longitude ($\sim$150,000~km) and the vertical (radial) dimension is 1,500~km.  Figure from \citet{Murray05}.
\label{FRingFig}}
\end{center}
\end{figure*}

The F~ring is one of only two narrow rings (along with the $\epsilon$~ring) to have known ``shepherd'' moons orbiting on either side of it.  However, it has not been conclusively shown how (or even that) the moons Prometheus and Pandora actually constrain the F~ring in its place, and in fact they appear to stir the ring up at least as much as to maintain it.  Simulations by \citet{Winter07} found the moons to be responsible for both strong chaos and significant radial confinement.  Both moons create at each closest approach a ``streamer-channel'' in nearby dust strands that subsequently moves downstream and shears under keplerian motion \citep[\Fig{}~\ref{FRingFig},][]{Murray05}.  During the 2009 saturnian equinox, which coincided with the once-in-17-yr alignment\fn{The periodicity of this alignment was recalcluated by \citet{Chavez09}, using updated orbital data.} of Prometheus' apoapse with the F~ring's periapse, minimizing their mutual closest approach distance \citep{Chavez09}, moonlets inferred from their shadows had a clear correlation of abundance with longitude relative to Prometheus \citep{Beurle10}, indicating that Prometheus is directly triggering accretion within the F~ring, the products of which may then be what collides with the core to form new jets and strands \citep{Beurle10}.  All of these interwoven phenomena make the F~ring the solar system's foremost natural laboratory for direct observation of accretion and disruption processes. 

Like the F~ring, the $\lambda$~ring is by far the brightest component of its planetary ring system when viewed at high phase angles, due to its high fraction of forward-scattering dust.  But the $\lambda$~ring appears to be significantly simpler and more sedate than its saturnian cousin, and its low detectability at lower phase and in occultations indicates that it is poor in macroscopic particles. 

Among dense ring systems, Neptune's has a much higher dust fraction than those of Saturn and Uranus, and is unique in having no significant dust-free regions.  The three main rings of Neptune lack sharp edges and are generally more tenuous, with only the Adams~ring reaching optical depths even as high as~0.1, and are consequently less well observed.  The only post-\Voyit{} observations, other than of the Adams ring and its arcs (see Section~\ref{Arcs}), are a marginal detection of the Le~Verrier~ring reported by \citet{Sicardy99}, which if confirmed would require it to be brighter than expected from \Voyit{} data, and a clear detection by \citet{dePaterNeptune05} that found the Le~Verrier~ring's brightness to be consistent with \Voyit{} measurements.  Possible explanations include temporal brightening in the Le~Verrier~ring followed by a return to its previous state, or unexpected spectral properties, or (a possibility they admit) that the \citet{Sicardy99} detection was affected by image artifacts. 

\subsection{Diffuse dusty rings \label{ThinDusty}}

\begin{center}
\begin{tabular}{|p{2cm}|p{4.5cm}|}
\hline
\textbf{Jupiter:} & Halo ring \\
& Main ring \\
& Amalthea Gossamer ring \\
& Thebe Gossamer ring \\
\hline
\textbf{Saturn:} & D ring \\
& Roche division \\
& Janus/Epimetheus ring \\
& G ring \\
& Methone ring \\
& Pallene ring \\
& Anthe ring \\
& E ring \\
& Phoebe ring \\
\hline
\textbf{Uranus:} & $\zeta$ ring \\
& $\nu$ ring \\
& $\mu$ ring \\
\hline
\textbf{Neptune:} & Galle ring \\
& Lassell ring \\
\hline
\end{tabular}
\end{center}

Every known ring system has a diffuse component that is optically thin and composed of $\mu$m-size dust particles.  The dynamics and evolution of diffuse dusty rings are made more complex by the importance of forces other than gravity \citep{Burns79}.  Dust particles are commonly low enough in mass that static electrical charging makes them susceptible to electromagnetic forces comparable to gravity.  They also have high ratios of surface area to volume, which makes them susceptible to pressure from solar radiation, including Poynting-Robertson drag.  The evolution induced by these additional forces shortens the lifetimes of dust particles so that the stability of diffuse dusty rings is only of a dynamical variety.  The ring consists of particles that originated from a source and are on their way to a sink; it may indeed look the same at a different time, if the sources and sinks have not significantly changed, but the particles comprising it will be different. 

A comprehensive review of dusty ring systems was published by \citet{BurnsChapter01}. 

\begin{figure}[!t]
\begin{center}
\includegraphics[width=8cm]{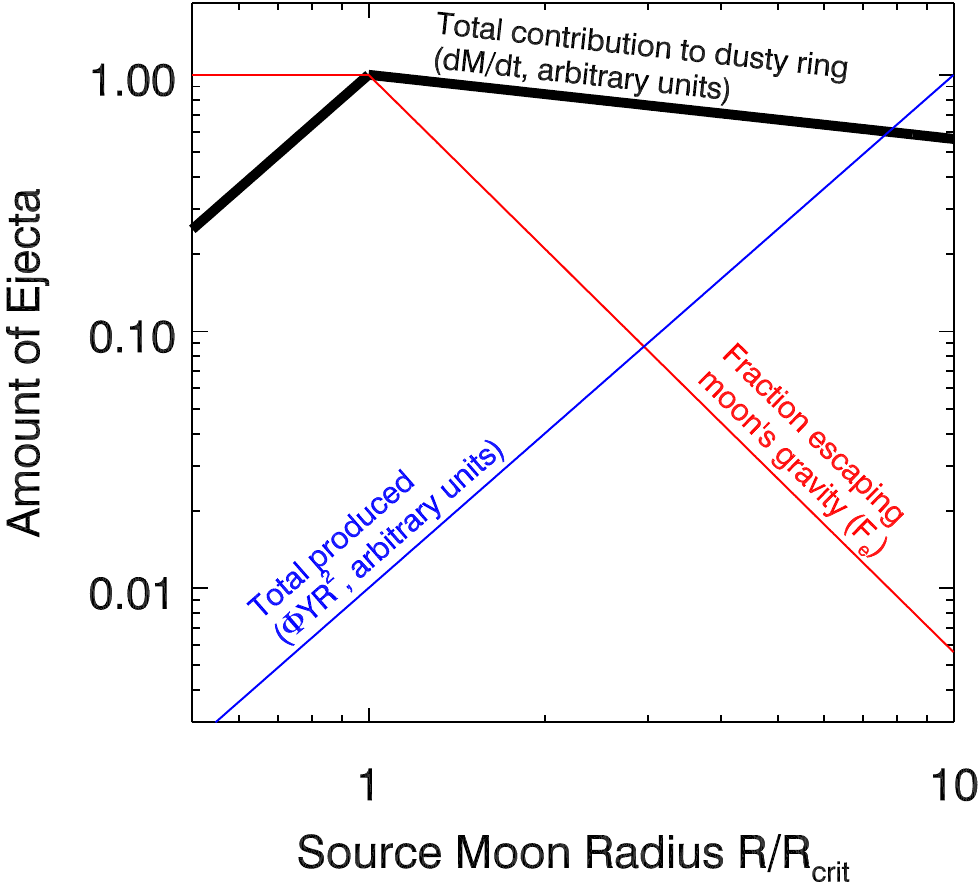}
\caption{Supply rate to a dusty ring via impact ejecta ($\ud M / \ud t$) is plotted in black, as a function of source moon radius $R$.  This quantity is the product (\Eqn{}~\ref{DustRate}) of the total ejecta produced ($\Phi Y R^2$, plotted in blue in arbitrary units) and the fraction of ejecta that escapes the moon's gravity and achieves planetary orbit ($F_\mathrm{e}$, plotted in red).  Moons smaller than $R_\mathrm{crit}$ lose all of their ejecta, while larger moons produce more ejecta but allow a smaller fraction of it to escape.  The optimal moon radius for supplying dust to a ringlet varies with surface properties, but is $R_\mathrm{crit} \sim 10$~km for icy moons with soft regolith \citep{Burns99}.
\label{burns99plot}}
\end{center}
\end{figure}

The primary mechanism for producing orbiting dust particles is micrometeoroid bombardment of larger orbiting bodies.  The rate at which a moon of radius $R$ supplies mass to a dusty ring is given by \citep{Burns99}
\begin{equation}
\frac{\ud M}{\ud t} \sim \Phi Y F_\mathrm{e} R^2 , 
\label{DustRate}
\end{equation}
\noindent where $\Phi$ is the impactor flux, $Y$ is the yield of ejected mass as a fraction of projectile mass, and $F_\mathrm{e}$ is the fraction of ejected mass that achieves planetary orbit.  The yield $Y$ depends on the material properties of the moon's regolith, while the fraction $F_\mathrm{e}$ of ejecta that can escape the moon's gravity depends on the moon's escape velocity and the distribution of ejecta velocities.  Empirically, the fraction of ejecta with velocity higher than $v$ goes as $(v_\mathrm{crit}/v)^{9/4}$, where $v_\mathrm{crit}$ is the minimum speed at which material is ejected, between 10~and 100~m~s$^{-1}$ where lower values are for softer moon's regoliths \citep{Burns84,Burns99}.  The escape velocity of a spherical moon with bulk density $\rho$ goes as $v_\mathrm{esc} \propto \rho^{1/2} R$ and also ranges between 10~and 100~m~s$^{-1}$.  So for $v_\mathrm{esc} < v_\mathrm{crit}$ we have $F_\mathrm{e} = 1$ and $\ud M / \ud t \propto R^2$, but when $v_\mathrm{esc} > v_\mathrm{crit}$ we have $F_\mathrm{e} \propto R^{-9/4}$ and $\ud M / \ud t \propto R^{-1/4}$.  That is, there turns out to be an optimal moon radius $R_\mathrm{crit} \sim 10$~km for supplying dust to a ringlet, with smaller moons intercepting fewer impactors and thus producing less dust, while larger moons allow a smaller fraction of the produced dust to escape (\Fig{}~\ref{burns99plot}).  However, these trends are only for general estimation.  A more detailed treatment must consider the mechanics of ejecta production, including not only surface mechanics but also the changes in the ejecta velocity profile with increasing $R$ as impact formation moves from the strength regime to the gravity regime \citep[e.g.,][]{JRichardson07}.  Also, larger moons are more efficient at sweeping up escaped particles during subsequent encounters \citep{Agarwal08,Kempf10}. 

All four known planetary ring systems appear to have an optically thin dusty ring as their innermost component --- Jupiter's Halo~ring, Saturn's D~ring, Uranus' $\zeta$~ring, and Neptune's Galle~ring.  Dust rings often extend inward of their sources because Poynting-Robertson drag in particular enforces an inward evolution of dust particles, as does resonant charge variation inward of synchronous orbit \citep{BurnsChapter04}.  Both Jupiter's Halo/Main~ring and Saturn's D~ring have likely source material at their outer edges (namely, Metis and Adrastea, and the C~ring, respectively), although at least the D~ring has internal radial structure \citep{Hedman07d} that may also require embedded sources (i.e., undiscovered moons).  The sources of the poorly observed $\zeta$~ring and Galle~ring are not known, and may also include embedded moons very close to the planet.  Some of these dust sheets may extend all the way down to the planet's cloud tops, although Saturn's D~ring appears to have a clear gap of 5,000~km between its inner boundary and the cloud tops \citep{Hedman07d}, through which the \Cassit{} spacecraft is slated to fly in 2017 during its end-of-mission maneuvers \citep{SealBuffington09}.  Models of Jupiter's rings also find an empty region above the cloud tops, through which the \Junoit{} mission plans to fly in 2016, though this has not been confirmed with definitive observations.  

Outward movement of dust is possible, and has been invoked to explain an extension of the Gossamer ring beyond the orbit of Thebe \citep{HK08}.  The mechanism proposed for outward movement is a ``shadow resonance'' in which dust particles moving through their planet's shadow temporarily lose their electrical charge and are carried outward by the momentum of their electromagnetically-influenced orbits.  In other locations, such as Saturn's G~ring, dust evolution is primarily outward because of ``plasma drag'' from abundant charged particles co-rotating with the planet's magnetic field, which orbit at speeds much faster than the dust's keplerian velocity since the G~ring is far outward of synchronous orbit \citep{Hedman07g}. 

Not all dusty rings consist of ejecta from a single dominant source moon.  Several are tenuous extensions of nearby dense rings, such as the D~ring, the dusty sheet in the Roche~division, and the Lassell~ring, which all lie inward of their likely sources, respectively, the C~ring, the F~ring, and the Arago~ring.  The Phoebe~ring may also be derived from multiple sources, as it is difficult for models to account for the ring's mass from impacts onto Phoebe alone because Phoebe is so large that it should retain much of its ejecta.  However, although Phoebe itself has no known collisional family, disruptive impacts are known to have broken other large irregular satellites into pieces that are separately observed today, and it is quite plausible that km-size pieces have been ejected from Phoebe and continue to share similar orbits.  Since, as shown above, km-size moons are actually the most efficient sources of orbiting dust, the observed dust densities can be explained by such a distributed population of source bodies \citep{VSH09}. 

\begin{figure*}[!t]
\begin{center}
\includegraphics[width=12cm]{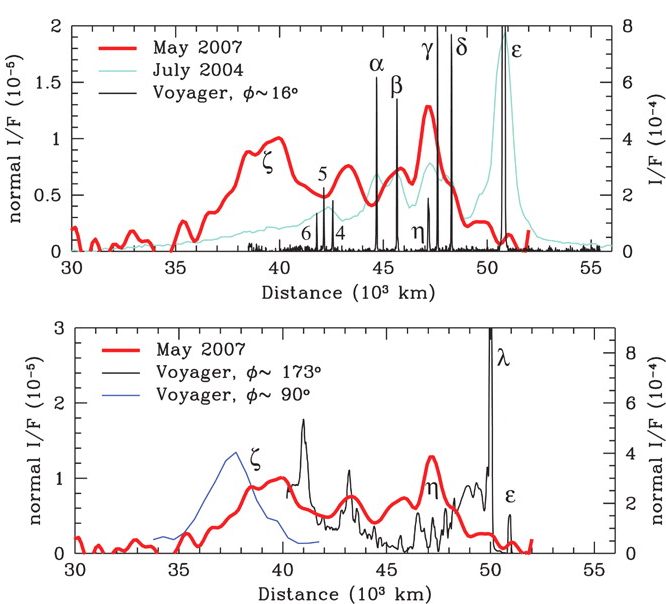}
\caption{Radial brightness profiles for Uranus rings from \Voyit{} in 1986 and from the Hubble Space Telescope near (2004) and at (2007) the Uranian equinox.  Figure from \citet{dePater07}.
\label{UranusEqx}}
\end{center}
\end{figure*}

The Uranian rings had a very different look (\Fig{}~\ref{UranusEqx}) when the Uranian equinox (an event occurring every 42~years) was observed in 2007 by the Hubble Space Telescope \citep{dePater07}.  With the Sun and the Earth on opposite sides of the Uranian ring plane, the brightest features in this rare view of the unlit side of the rings had relatively low optical depth but macroscopic particles.  The dense $\epsilon$~ring practically disappeared because its high optical depth made it opaque, while the dusty $\lambda$~ring was also dim due to the low phase angle.  The brightest feature was at the location of the $\eta$~ring, and can probably be identified with a low-$\tau$ ``extension'' previously seen adjacent to that ring.  Some of the dusty rings seen at high-phase had bright counterparts in HST's unlit-side view, as did a similar ``extension'' to the $\delta$~ring, while others did not.  Thus, the latter are likely pure dust rings while the former contain a component of larger particles.  A broad innermost ring, which had been given the provisional designation 1986~U2R based on \Voyit{} images, was also seen in HST's unlit-side view and named the $\zeta$~ring.  However, the core of the $\zeta$ ring as seen by HST is several thousand~km outward of its position in \Voyit{} images.  While acknowledging the possibility that overlapping particle populations with different optical properties would explain both the mid-phase \Voyit{} observations and the HST~data, \citet{dePater07} argue it is more likely that the $\zeta$~ring has changed dramatically in the intervening 20 years. 

\begin{figure*}[!t]
\begin{center}
\includegraphics[width=6cm]{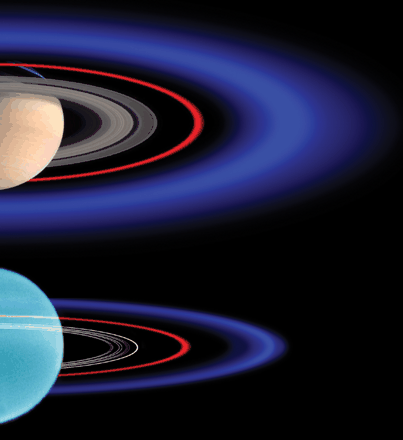}
\caption{A schematic view of the outer rings of Saturn and Uranus, in which each system has been scaled to a common planetary radius.  Highlighted are the red coloration of Saturn's G~ring and Uranus' $\nu$~ring, both of which appear to be typical dusty rings, and the unusual blue coloration of Saturn's E~ring and Uranus' $\mu$~ring.  Figure from \citet{dePaterOneRing06}.  
\label{OneRingFig}}
\end{center}
\end{figure*}

While most dusty rings are either gray or red in their spectral properties, two known rings have a prominent blue color:  Saturn's E~ring and Uranus' $\mu$~ring (\Fig{}~\ref{OneRingFig}).  Each of these rings is centered on the orbit of a moon -- Enceladus and Mab, respectively \citep{dePaterOneRing06}.  For rings made of $\mu$m-sized particles, which are close in size to the wavelengths of visible and infrared light, a blue spectral trend is best explained by a particle-size distribution (regardless of particle composition) that is either narrowly centered on a particular size or characterized by a very steep power law \citep{SCL91,dePater04eg}.  For Enceladus and the E~ring, the mechanism by which this happens seems well understood: ice particles are spewed from Enceladus south-polar geysers at a variety of speeds, with smaller particles being more easily accelerated by gas in the plume, leading to higher velocities that enable them to join the E~ring \citep{Schmidt08,HedmanPlume09,Kempf10}.  However, it's harder to imagine this mechanism applying to the $\mu$~ring.  Enceladus, with a diameter of $\sim 500$~km, is already so small that the source of sufficient internal heat to account for its plume is a matter of much debate \citep[e.g.,][]{MW07,MW08a,MW08b}, and Mab has approximately 1/10 the diameter (and thus 1/1,000 the mass) of Enceladus.  A satellite the size of Mab ought to be a good source of dust liberated by collisions (see above) but that should lead to shallower size distributions and a redder spectral trend like those of most dusty rings.  

\subsection{Ring arcs and azimuthal clumps \label{Arcs}}

\begin{center}
\begin{tabular}{|p{2cm}|p{3.25cm}|}
\multicolumn{2}{l}{\textit{\textbf{Arcs}}} \\
\hline
\textbf{Saturn:} & G ring \\
& Methone ring arc \\
& Anthe ring arc \\
\hline
\textbf{Neptune:} & Galatea ring arc \\
 & Adams ring \\
\hline
\end{tabular}
\end{center}

\begin{center}
\begin{tabular}{|p{2cm}|p{3.25cm}|}
\multicolumn{2}{l}{\textit{\textbf{Azimuthal clumps}}} \\
\hline
\textbf{Jupiter:} &  Main ring \\
\hline
\textbf{Saturn:} & Encke Gap ringlets \\
& F ring \\
\hline
\end{tabular}
\end{center}

Neptune's iconic ring arcs, known since \Voyit{} and strongly suspected even earlier from occultation data, have now been joined by several Saturnian structures among the ranks of ring arcs.  All share the common characteristic of an azimuthally confined region of enhanced brightness embedded within a fainter circumferential ring.\fn{For Saturn's Anthe and Methone arcs, as for Neptune's Galatea arc, the circumferential ring is too faint to have been detected as yet but likely consists of material recently escaped from the arc-confining mechanism.}  All known arcs orbit at the appropriate keplerian rate for their distance from planet center.  But there are also significant differences among known arcs, as they cover a wide range of densities and are caused by at least two different mechanisms.  

Left to itself, any clump of material orbiting a planet should spread out into a ring on a fairly short timescale.  This is a direct result of keplerian shear (\Eqn{}~\ref{KeplerShearEqn}), by which two objects with semimajor axes $a$ and $a+\delta a$ will have mean motions $n$ and $n+\delta n$, where $\delta n = -(3n/2a) \delta a$, and the time for one to ``lap'' the other by an entire orbit is 
\begin{equation}
T_{\mathrm{spread}} = \frac{2 \pi}{\delta n} = - \frac{4 \pi}{3 (GM)^{1/2}} \frac{a^{5/2}}{ \delta a} , 
\end{equation}
\noindent where we have used the precise form of Kepler's Third Law, $n^2 a^3 = GM$.  For the giant planets, $(GM)^{1/2}$ is of order $10^4$~km$^{3/2}$~s$^{-1}$, and a typical distance from the planet is $a \sim 10^5$~km.  Therefore, even for a compact initial clump $\delta a \sim 1$~km, the spreading time is only a few decades and is inversely proportional to $\delta a$.  

This exercise demonstrates that azimuthal variations in a ring's mass are not intrinsically stable.  Therefore, the several examples that exist of observed ring arcs must be dynamically generated or maintained.  The two most prominent mechanisms for this are resonant confinement and asymmetric injection of mass into the ring.  

Corotation resonances have resonant arguments (see Section~\ref{SpiralWaves}) of the form 
\begin{equation}
\varphi = (m+k) \lambda' - m \lambda - k X' ,
\label{CorotResArgForm}
\end{equation}
\noindent where, as for \Eqn{}~\ref{ResArgForm}, $m$ and $k$ are integers that label the resonance as ($m$+$k$):$m$, primed quantities refer to the forcing moon, and $X$ is either $\varpi$ or $\Omega$.  For a corotation eccentricity resonance (CER), $X$ is $\varpi$ and the resonance strength is proportional to the perturbing moon's eccentricity, while for a corotation inclination resonance (CIR), $X$ is $\Omega$ and the resonance strength is proportional to the perturbing moon's inclination.  

Corotation resonances differ from their Lindblad and vertical cousins (see Section~\ref{SpiralWaves}) in that the lowest-order resonances involve the eccentricity or inclination of the perturbing moon, rather than that of the particle being perturbed.  Thus, rather than pumping up the eccentricities and inclinations of ring particles and driving waves, corotation resonances tend to azimuthally confine particles into an orbit commensurate with that of the perturbing moon.  This mode of confinement was first proposed for Neptune's ring arcs by \citet{GTB86}.  

\subsubsection{Neptune's Adams ring}

\begin{figure*}[!t]
\begin{center}
\includegraphics[width=16cm]{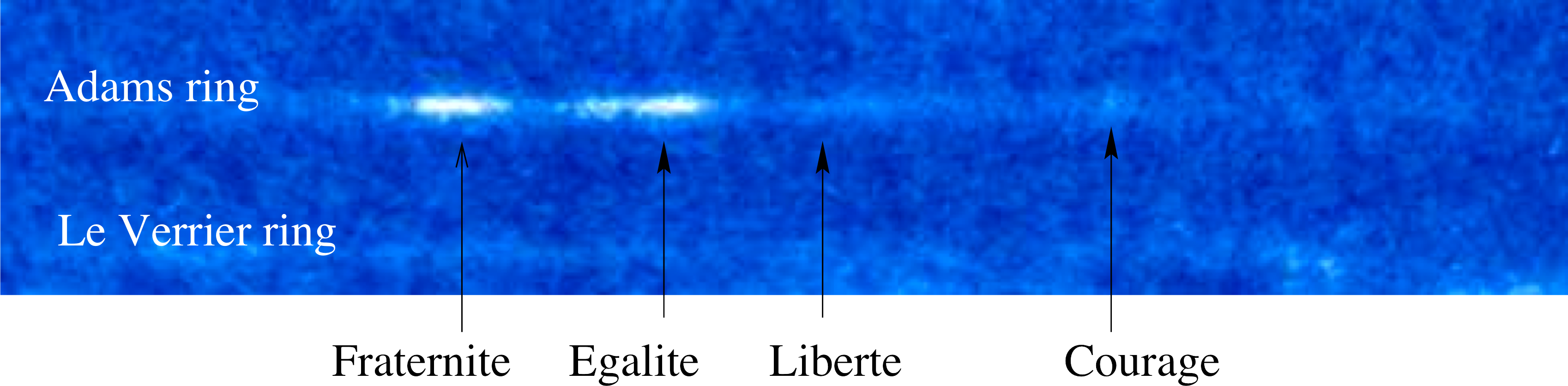}
\includegraphics[width=10cm]{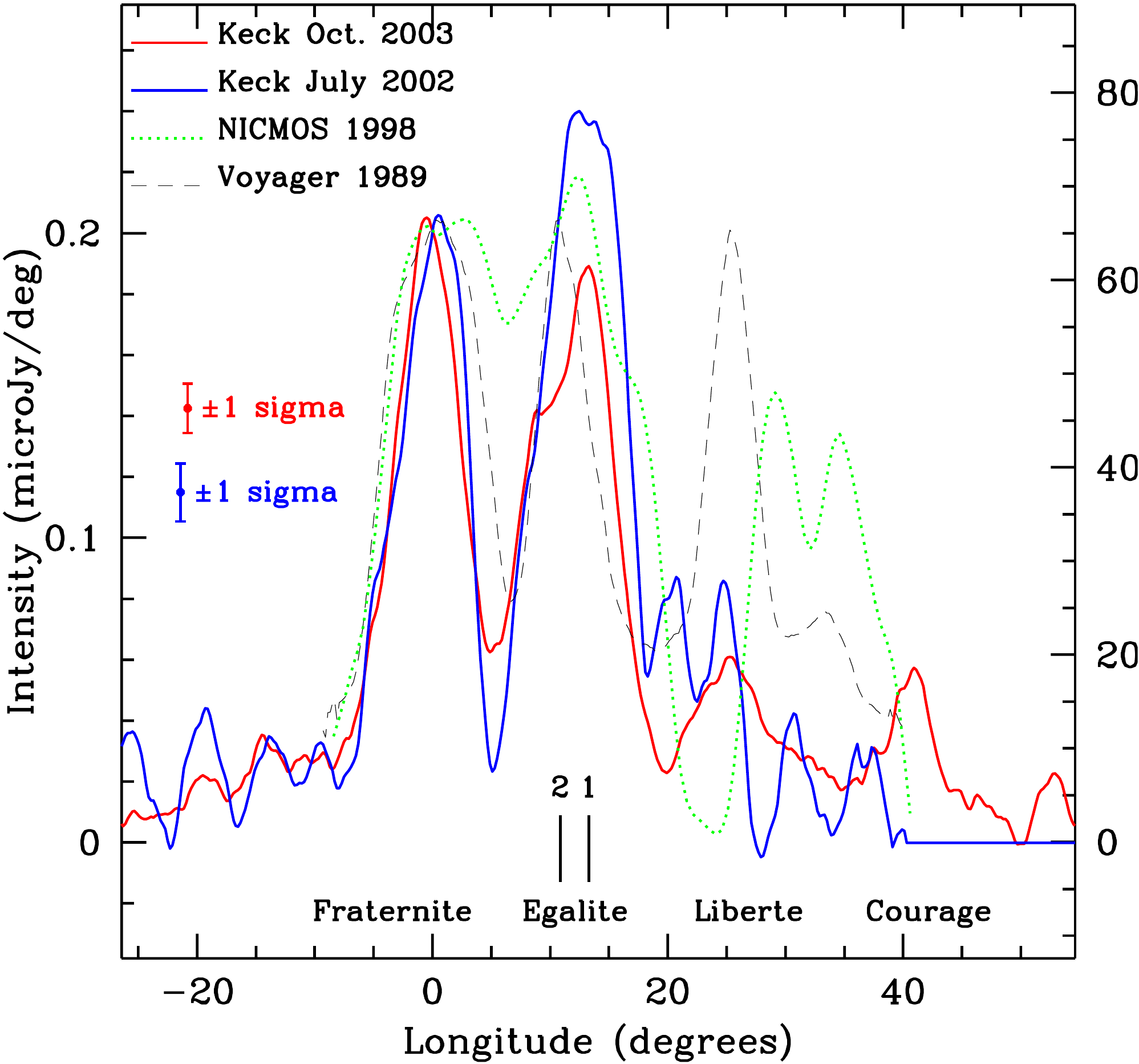}
\caption{(upper) Reprojected ground-based image of the Adams ring (with arcs) and the Le~Verrier ring acquired in October 2003.  (lower) Azimuthal profiles of the Adams arcs from four separate observations.  The leading arc Libert\'e was seen by \Voyit{~2} (black dashed line) to be as bright as the other two main arcs, but subsequently dimmed.  Figure from \citet{dePaterNeptune05}.  See \Fig{}~\ref{NeptuneRingsFig} for schematic view. 
\label{AdamsArcs}}
\end{center}
\end{figure*}

The first-discovered and best-known set of ring arcs are found in Neptune's Adams ring.  These are the densest components of Neptune's ring system (with $\tau_\perp \sim 0.1$) and were oftentimes the only component detectable in pre-\Voyit{} Earth-based occultations, leading to much confusion as to whether the detected signatures were rings at all until \Voyit{~2} settled the question.  There are five arcs occupying $\sim 20^\circ$ out of a region extending over $\sim 40^\circ$ of longitude (\Fig{}~\ref{NeptuneRingsFig} and~\ref{AdamsArcs}).  The three main arcs were named for the French revolutionary slogans Libert\'e, Egalit\'e, and Fraternit\'e, then closer inspection showed a bifurcation in Egalit\'e and a dimmer fourth arc that was named Courage.  

The first integrated dynamical explanation of the arcs was given by \citet{Porco91}, who combined the theoretical framework of \citet{GTB86} with observations showing that the Adams ring lies very close to a 43:42 resonance with the moon Galatea.  \citet{Porco91} suggested that the azimuthal confinement mechanism was due to Galatea's 43:42 corotation inclination resonance (CIR) while radial spreading due to particle collisions was prevented by Galatea's nearby 43:42 outer Lindblad resonance (OLR).  This explanation requires both Libert\'e and Egalit\'e to stretch over multiple consecutive corotation sites (each of which is only\fn{Because inclination resonances cannot be first-order \citep[see, e.g., Section~10.3.3 of][]{MD99}, the CIR actually functions as an 86:84 resonance, which is why the number 86 appears here.  For the CER discussed by \citet{NP02}, the corotation sites would be twice as long.} $360^\circ/86 = 4.2^\circ$ long) and predicted that the arcs would orbit at the pattern speed of the 43:42 CIR.  A synthesis of all \Voyit{} and Earth-based data \citep{Nich95} found two possible solutions for the arcs' pattern speed, one of which was consistent with the Galatea 43:42~CIR, and detailed dynamical work \citep{FS96,HP97} supported the plausibility of resonant confinement by this mechanism.  However, the re-acquisition of the arcs with Earth-based imaging \citep{Dumas99,Sicardy99} called the resonant pattern speed into question, favoring instead \citet{Nich95}'s other solution.  An attempt to salvage the Galatea model was made by \citet{NP02}, who proposed an alternative model employing the 43:42 corotation eccentricity resonance (CER) rather than the CIR.  They furthermore invoked the mass of the ring arcs themselves to adjust the resonant pattern speed to match the observations, thus deriving a value for the arcs' mass that is required for their model to work.  Further Earth-based observations \citep{dePaterNeptune05} not only confirmed that the arcs are moving at the ``wrong'' pattern speed for the original \citet{Porco91} model, but showed significant changes in the arcs' structure over the 20~years in which they have been observed in detail (\Fig{}~\ref{AdamsArcs}).  The brightness of Libert\'e, already diminished in the 1998 observations, had further declined until it was dimmer than Courage, while Courage had moved forward in longitude relative to the other arcs.  Egalit\'e has also undergone less dramatic shifts in its morphology and longitude, while Fraternit\'e appears largely stable. 

The Adams arcs remain an enigma.  While the model invoking resonant confinement by Galatea has appeared promising, the details have not come together as hoped.  The original model by \citet{Porco91} was appealing because its predicted pattern speed appeared to match very closely with the observations; however, the re-working of the model by \citet{NP02} is less convincing because it requires the invocation of an additional free parameter in order to fit the data.  Furthermore, now that other examples of resonantly confined ring arcs have come to light (see Section~\ref{GRingArcs}), all of which have the morphology predicted by \citet{GTB86}, with maxima at the center of a corotation site and brightness falling off long before the site's edges are reached, the objection can be raised anew that Egalit\'e and Fraternit\'e span multiple corotation sites while showing no clear evidence of internal minima at the expected spatial frequency.  On the other hand, the azimuthal structure in Jupiter's Main ring (Section~\ref{JupiterArcs}), though far less well-observed, appears similar to the Adams arcs in that the spatial frequencies do not match those of nearby corotation resonances.  Perhaps the corotation-resonance model will be vindicated in the end, or perhaps the answer will be more like the shepherding model of \citet{LissauerArcs85}, which invokes yet-undiscovered moons embedded within the Adams ring, or perhaps the Adams arcs will be found not to be long-term stable structures. 

\subsubsection{Jupiter's Main ring and other azimuthal clumps \label{JupiterArcs}}

In the core of Jupiter's Main ring, \NHit{} images found one close pair of azimuthal clumps and another family of three to five clumps \citep{Show07}.  The semimajor axes of the $\alpha$ and $\beta$ clump families, measured from their orbital rates, fall less than 1~km from, respectively, the Metis 115:116 and 114:115 corotation inclination resonances (CIR), the same kind of resonance originally invoked by \citet{Porco91} for Neptune's Adams arcs.  Given the distance between the two semimajor axes and the distance between the resonances, \citet{Show07} estimate the probability of this happening by coincidence to be 4\%.  Finally, to complete the analogy with Neptune's Adams arcs, the 1.8$^\circ$ azimuthal spacing of the clumps does not correspond to the expected $360^\circ / 230 = 1.56^\circ$ for these resonances \citep{Show07}.  

The Adams ring and the Main ring may turn out to be more like Saturn's F~ring and the ringlets in the Encke Gap (Section~\ref{ThickDusty}), which have rich azimuthal structure that is clearly not associated with a resonant spatial frequency and is most likely due to embedded source moons. 

\subsubsection{Saturn's G ring and other moon-embedded arcs \label{GRingArcs}}

Saturn's G~ring was, for a long time, the ring that should not be there.  Saturn's main disk was clearly a long-term stable structure, the D~ring clearly derived from it, the E~ring associated with Enceladus, and the F~ring confined by Prometheus and Pandora.  Yet the G~ring, composed of dust grains that cannot persist over long time periods, had no apparent mechanism for its needed continuous generation.  

\begin{figure*}[!t]
\begin{center}
\includegraphics[width=7.6cm]{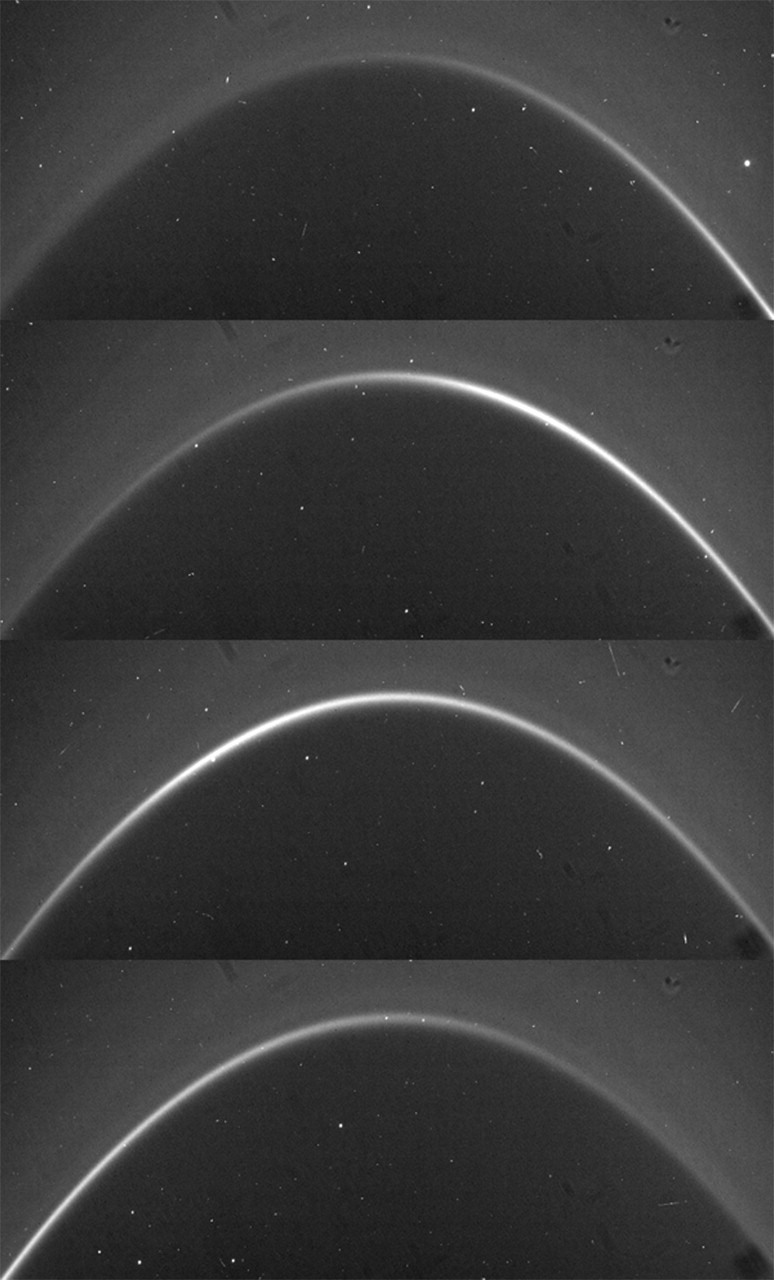}
\includegraphics[width=14cm]{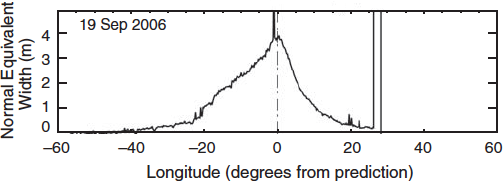}
\caption{(top) Images and (bottom) azimuthal profile of the G~ring arc, centered on the Mimas~7:6 corotation site.  The peaky shape in the azimuthal profile indicates a significant fraction of arc material is tightly bound to the resonance (i.e., has low libration amplitude).  Figure from \citet{Hedman07g}. 
\label{GRingProfile}}
\end{center}
\end{figure*}

The first step towards addressing this question came with the discovery of a relatively bright arc within the G~ring \citep{Hedman07g}.  The arc is brighter than the rest of the G~ring by a factor of several (but still at $\tau_\perp \sim 10^{-6}$), and is radially much narrower ($\sim 250$~km).  It is very plausible that the rest of the G~ring is composed of grains evolving away from the arc under the influence of non-gravitational forces.  Furthermore, the arc's orbital rate matches the 7:6 corotation eccentricity resonance (CER) with Mimas, again invoking the \citet{GTB86} mechanism.  The azimuthal profile of the arc is roughly triangle-shaped (\Fig{}~\ref{GRingProfile}), consistent with a source body with only a small-amplitude libration about the exact resonance, but has broad wings that plausibly fill the corotation site's length of $360^\circ/7 = 51^\circ$ in longitude. 

A further piece of the puzzle came with the discovery of Aegaeon \citep{HedmanAegaeon10}, a 1-km moon orbiting within the G~ring arc.  Aegaeon is probably only the largest of a population of source objects within the arc, as electron-absorption measurements indicate massive objects spread over a distance much larger than Aegaeon itself \citep{Hedman07g}.  

Two other arcs, even more tenuous than the G~ring's, surround the small moons Anthe and Methone, both situated between Mimas and Enceladus.  A third moon in this vicinity, Pallene, appears to have a very tenuous ring but not an arc.  The Methone arc was first detected by charged-particle absorptions \citep{Roussos08}, and all three structures were then seen in imaging data \citep{HedmanArcs09}.  Both Anthe and Methone are in resonance with Mimas, in the 11:10 and 15:14 CERs, respectively, and these resonances continue to confine the arc material once it has left the source moons \citep{Agarwal09}.  The azimuthal length of the Anthe and Methone arcs are both approximately half that of the available corotation site \citep{HedmanArcs09}, but that may simply reflect the point at which the faint signal falls below detectability. 

Because these ring arcs at Saturn are so much more tenuous than Neptune's, they are essentially collisionless.  This allows them to be confined by the corotation resonance alone, which is necessary because the associated Lindblad resonances are radially farther away than they are for Neptune's case, both because of Saturn's higher $J_2$ compared to Neptune and because the resonances have lower azimuthal parameter $m$, and thus are not available to perform the radial confinement that was an essential part of the \citet{GTB86} model for the collisional Adams ring. 

Finally, we note that \Voyit{~2} imaged a faint unnamed ring sharing the orbit of the moon Galatea, which may be an arc given its intermittent detectability \citep{SC92,PorcoChapter95}.  Although little is known about this structure, it can now be placed in context with the Anthe and Methone arcs and may very well be driven by similar mechanisms. 

\subsection{Rings as detectors  \label{Detectors}}

Planetary rings have shown themselves to be useful, in many cases, as detectors of planetary processes around them.  

Spiral structures in the D~ring, and in the similar tenuous dusty sheet in the Roche Division, are driven by Lindblad resonances with the rotation period of the planet's magnetic field.  These structures are only seen in these tenuous sheets populated by tiny grains that are easily charged and thus subject to electromagnetic forces.  The multiple pattern speeds required to explain the resonant structures are a major source of information for the complex rotation of Saturn's enigmatic magnetic field \citep{HedmanOrganizing09}. 

Evidence for vertical corrugations in Jupiter's Main ring were first seen in \Galit{} images \citep{Ockert-Bell99,Burns99}, but were not well-understood.  Next, \Cassit{} images of Saturn's D~ring showed evidence of a vertical corrugation whose radial wavelength is decreasing with time, a trend easily accounted for with a model of differential precession that begins with the ring as an inclined flat sheet $\sim 20$~yr before \Cassit{}'s arrival at Saturn \citep{Hedman07d}.  Images obtained during Saturn's 2009 equinox showed the vertical corrugation pattern extending far into the C~ring, and the variation of the wavelength with radius confirms the previous result that the pattern originated in 1983 when an event of some kind caused the ring to be tilted by $\sim 10^{-7}$ radians (a few meters) with respect to the Laplace plane \citep{HedmanCorrugation11}.  Finally, similar analysis of the corrugations in Jupiter's Main ring yields a superposition of wavelengths implying two tilting events, one in July~1994 and one in 1990 \citep{ShowCorrugation11}.  July~1994 is, of course, the date of the impact of comet Shoemaker-Levy~9 (SL9) into Jupiter, leading to the hypothesis that both corrugation patterns carry records of the impact of a spatially-dispersed cloud, which is one mechanism for spreading the impulse over a large portion of the ring.  The cloud of dust surrounding SL9 would likely fill the bill at Jupiter, and either a similar cometary system or a meteor stream could be the cause of the Saturn event. 

Saturn's rings showed themselves capable of more direct detections of impactors during the 2009 equinox event.  Bright markings, canted with respect to the azimuthal direction, were seen on both the A~ring and the C~ring during the few days before and after the Sun's passage through the ring mid-plane \citep{AGU09Impacts}.  Assuming that these are dust clouds evolving under keplerian shear (\Eqn{}~\ref{KeplerShearEqn}), one can calculate the time elapsed since the cloud was radially aligned.  In one case, the same cloud was seen 24 hours apart, and the ages derived from keplerian shear differed by the same interval, confirming this interpretation of the observed structures.  Also, in very high-phase high-resolution images of the C~ring (not during equinox), small streaks appear that are probably transitory dust clouds produced by impacts \citep{AGU09Impacts}.  Both of these discoveries have the potential to use rings observations to constrain the influx of interplanetary impactors, though the derivation of impactor size from observed dust-cloud parameters has yet to be worked out. 

Spiral waves and wavy gap edges are both the result of gravitational forcing due to moons.  In the case of Janus and Epimetheus, the nature of the forcing changes with time, and the ring maintains a record of that change.  Initial steps towards understanding the nature of the rings' record have been made for both density waves \citep{jemodelshort} and wavy edges \citep{SP09}.  In the case of spiral density waves, the group velocity at which information propagates through the wave is slow enough that information can be gained about the state of the co-orbital moons as much as ten years prior to the time of observation.  

\section{Experimental rings science \label{Experiments}}

\subsection{Numerical simulations \label{NumSims}}

\begin{figure*}[!t]
\begin{center}
\includegraphics[width=8cm]{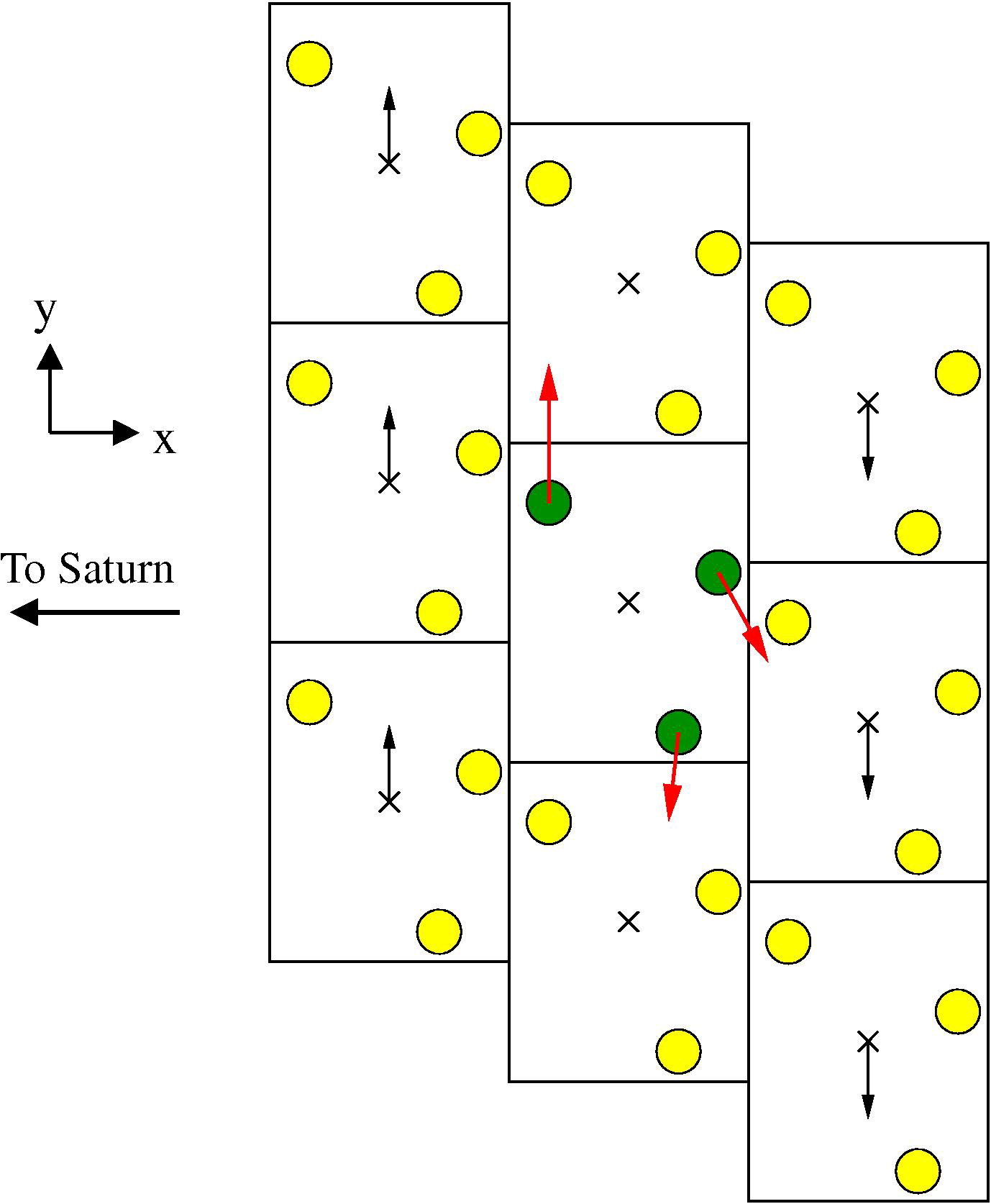}
\caption{Schematic representation of a ring-patch simulation with sliding boundary conditions.  The simulation cell (center, with green particles) is replicated on all sides, with the replicant cells (with yellow particles) positioned according to relative keplerian velocities (\Eqn{}~\ref{KeplerShearEqn}).  In this representation, increasing radius ($+\hat{x}$) is to the right and keplerian orbital motion ($+\hat{y}$) is up.  Figure from \citet{Perrine11}. 
\label{SlidingPatchFig}}
\end{center}
\end{figure*}

Advances in simulation techniques, alongside advances in computing hardware and software, have allowed numerical simulations to become an increasingly important part of the study of planetary rings.  Modern $n$-body dynamics exploded after \citet{WH91} published a symplectic mapping that uses Hamiltonian methods to calculate perturbations as a deviation from keplerian orbit, rather than as deviations from motion in a straight line, and thus requires fewer correction events per unit of simulated time.  This, along with ever-increasing computer power, allowed large $n$-body simulations to become routine; among the many solar-system applications is the case of dusty rings evolving under gravity and other forces.  For dense rings, though, the number of mutually interacting particles is too large for simulations that follow particles along their full orbits about the planet.  Instead, for cases that do not focus on large-scale azimuthal structure, a very productive line of simulations follows a relatively small ``patch'' of the ring with sliding boundary conditions as devised by \citet{WT88}.  The patch is surrounded by ``mirror'' patches (\Fig{}~\ref{SlidingPatchFig}); those on either side in the azimuthal direction are stationary, so that a particle that leaves the patch on one side simply reappears on the opposite side, but the neighboring patches in the radial direction slide past according to keplerian shear (\Eqn{}~\ref{KeplerShearEqn}).  Even with their greatly reduced spatial dimensions, ring-patch simulations routinely include so many mutually-interacting particles that they push the limits of the available computing power, and several approaches to efficiently accounting for their mutual interactions have been devised. 

The first great success of ring-patch simulations, and a continuing area of their usefulness, is the characterization of self-gravity wakes (SGWs) (Section~\ref{SGWs}).  First described by \citet{JT66} for the case of galaxies, it was understood by the 1980s that this non-axisymmetric structure due to a balance between accretion and disruption was likely present in Saturn's A~and B~rings and was likely responsible for the observer-centered quadrant azimuthal asymmetry observed in the A~ring's brightness \citep{Colombo76,Franklin87,DP89}.  SGWs were successfully produced in ring-patch simulations by \citet{Salo92,Salo95} and by \citet{Richardson94}, and simulations continue to be a vital tool for understanding the mechanics and structure of SGWs and comparing their simulated photometric properties to observational data, as well as other ring properties including the rotational states and thermal properties of ring particles, and the mechanics of sharp edges and propellers \citep[for details and references, see][]{SchmidtChapter09}. 

An entirely different class of simulations are the semi-analytical streamline models of \citet{Hahn07,Hahn08} and \citet{Hahn09}, which use the theoretical framework of \citet[and references therein]{BRL94} to probe the distribution of surface density, pressure, and viscosity of a ring in the vicinity of a sharp edge. 

\subsection{Physical experiments and the coefficient of restitution \label{CoeffRest}}

What happens when two ring particles collide?  Gentle collisions, with velocities of order mm~s$^{-1}$, occur constantly (frequencies of order the orbital frequency) within dense rings.  A law describing the coefficient of restitution $\varepsilon$ (the ratio between outgoing and incoming kinetic energies for two colliding particles in their center-of-mass reference frame) is an essential input for numerical simulations.  A number of physical experiments have been conducted to measure the coefficient of restitution directly, but these must begin with assumptions as to the shape, porosity, and surface friction of ring particles.  Consequently, the results of these experiments have been inconsistent.  More recently, comparisons between simulations and data have attempted to arrive empirically at a favored coefficient of restitution law, from which ring-particle properties can then be inferred. 

\citet{Bridges84} conducted ground-breaking experiments for determining the collisional properties of icy objects, using a double-pendulum apparatus to achieve the exceedingly low collision velocities seen in ring systems.  They found relations for $\varepsilon_\mathrm{n}$, as a function of mutual incoming velocity $v_\mathrm{n}$ (the subscript denotes normal collisions), for frosty ice spheres at temperatures of $\sim 200$~K.  Their result, 
\begin{equation}
\varepsilon_\mathrm{n} = \left( \frac{v_\mathrm{n}}{v_\mathrm{c}} \right)^b ,
\label{BridgesEqn}
\end{equation}
\noindent with exponent $b=-0.234$ and critical velocity $v_\mathrm{c} = 0.0077$~cm~s$^{-1}$, has been widely used in numerical simulations ever since.  Further experiments broadened the range of particle properties tested.  \citet{Hatzes88,Hatzes91} found that a frost layer on the surface can greatly increase the lossiness of a collision (i.e., depress $\varepsilon$), even leading to sticking at low collision velocities.  \citet{Dilley96} varied the mass of the incoming particles and found that smaller ice balls led to lossier collisions.  \citet{Supulver95}, found less lossy collisions overall (\Eqn{}~\ref{BridgesEqn} with $b=-0.14$ and $v_\mathrm{c} = 0.01$~cm~s$^{-1}$) for frost-free spheres at $\sim 100$~K and also found that glancing collisions are less lossy than normal collisions.  Initial results from experiments in a micro-gravity environment were given by \citet{ColwellMicrograv08} and \citet{Heisselmann10}, while \citet{Durda10} measured the coefficient of restitution for two 1-m granite spheres colliding at low speeds. 

All collision experiments thus far have represented ring particles as hard spheres of H$_2$O~ice.  The actual shape of ring particles is not known; they are far too small to be sphericalized by hydrostatic equilibrium, and might be expected to be roughly ellipsoidal if they take the shape of their Roche lobes.  Their surfaces may be smoothed by the numerous gentle collisions they experience, after the manner of pebbles in a stream bed, but this also is not known.  Finally, ring particles are almost certainly not as hard as solid ice, as their outer layers at least are probably quite porous (Section~\ref{Roche}). 

\begin{figure*}[!t]
\begin{center}
\includegraphics[height=7cm]{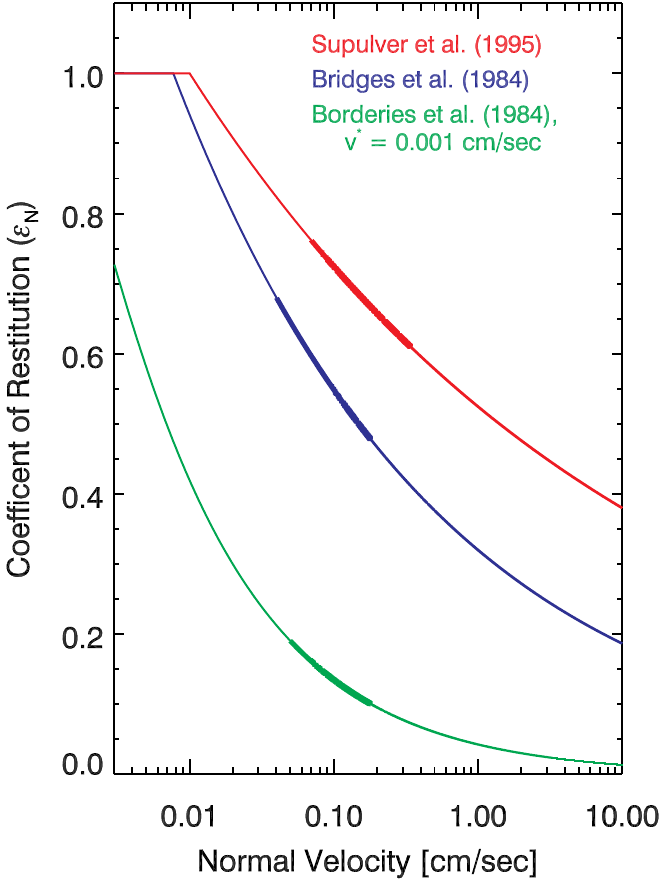}
\includegraphics[height=7cm]{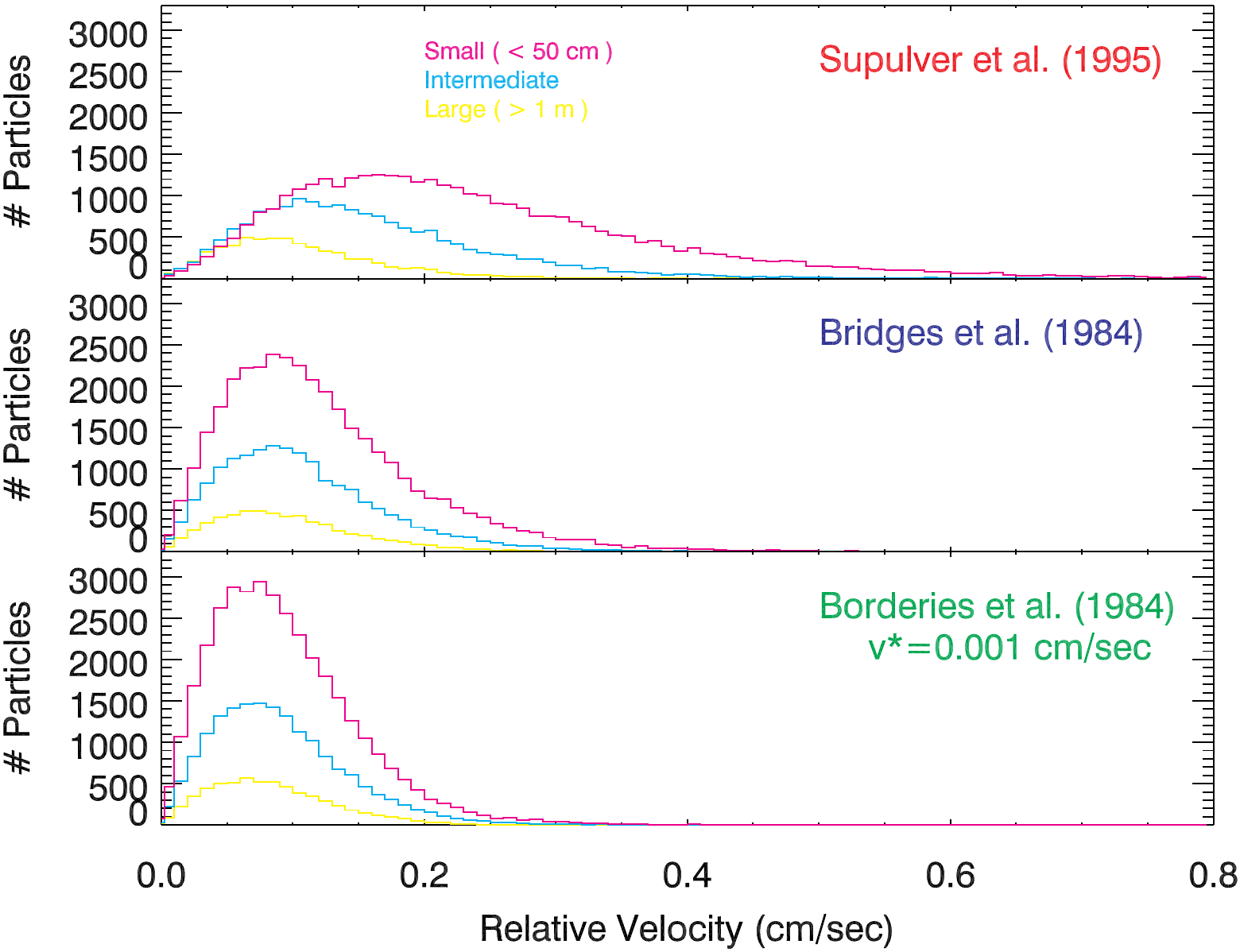}
\caption{(left) Coefficients of restitution, as a function of incoming velocity, according to different input laws.  The bold portion of each line indicates the velocity range for most simulated small particles.  (right) Velocity dispersion profiles for simulated ring-patches using the same input laws.  Figures modified from \citet{Porco08}. 
\label{CoeffRestPlot}}
\end{center}
\end{figure*}

Eventually, the role of physical experiments will ideally shift from determining a preferred restitution law based on an assumption of ring-particle properties, to using an empirically indicated restitution law to infer ring-particle properties.  A step towards this goal was taken by \citet{Porco08}, who compared numerical simulations with \Voyit{} imaging data of the azimuthal asymmetry to arrive at a preliminary conclusion favoring a lower coefficient of restitution (i.e., lossier collisions) than given by the prevailing \citet{Bridges84} law.  They favored instead a law with the form \citep{BGT84}
\begin{equation}
\varepsilon = \left[ -\frac{2}{3} \left( \frac{v^*}{v} \right)^2 + \left[ \frac{10}{3} \left( \frac{v^*}{v} \right)^2 - \frac{5}{9} \left( \frac{v^*}{v} \right)^4 \right]^{1/2} \right]^{1/2}
\end{equation}
\noindent This law is based on the \citet{Andrews1930} theory of colliding spheres, but the more flexible formulation of \citet{BGT84} allows $v^*$ to be a free parameter adjusting the law's lossiness.  \citet{Porco08} favored a value of $v^* = 0.001$~cm~s$^{-1}$; however, a full treatment of the question, incorporating the substantial \Cassit{} data set, has yet to be completed.  A plot of coefficient of restitution distributions derived from a variety of laws is shown in \Fig{}~\ref{CoeffRestPlot}.

In addition to the azimuthal photometric asymmetry, both the photometry of propellers (Section~\ref{Propellers}) and the sharpness of gap edges (Section~\ref{GapEdges}) have the potential of being used to constrain the physical properties of ring particles by combining data with simulations, though work on these lines is just beginning.  Propellers are large enough to appear in images with a reasonable amount of detail, but small enough that particle-particle interactions play a significant role in determining their structure \citep[e.g.][]{LS09}, although their poorly understood photometry \citep{Giantprops10}, possibly due to vertical structure, may make it difficult to use them as a standard.  Occultation data infer edges in Saturn's rings as sharp as 10~to 20~m \citep{ColwellDPS10}, significantly sharper than the edges obtained in simulations using the standard \citet{Bridges84} restitution law \citep{WeissThesis05}, while preliminary results from simulations with lossier restitution laws yield sharper edges more in line with observations (J.~E.~Colwell, personal communication, 2010). 

\subsection{Spectroscopic ground truth}

The project of characterizing the optical properties of materials occurring in the outer solar system is ongoing.  A large amount of spectroscopic data now exists for the rings of Jupiter and Saturn, in spectral ranges from the infrared to the ultraviolet, that contain numerous bands that are potentially diagnostic of ring-particle composition and/or particle size and state.  Laboratory measurements of candidate ring-forming materials are compared to these observations in an effort to constrain the chemical and physical composition of the rings.  For details and references, see \citet{CuzziChapter09}. 

\section{Age and origin of ring systems \label{AgeOrigin}}

The question of the age of a ring system is actually two separate questions, one being the age of the material and the other the age of the structure currently in place.  This distinction is particularly important for diffuse dusty rings (Section~\ref{ThinDusty}).  For example, if Amalthea, Thebe, Adrastea, and Metis were to suddenly disappear, or to suddenly cease emitting dust, Jupiter's rings would dissipate on a timescale on the order of $10^5$ years at the most \citep{BurnsChapter01}.  Thus, the age of individual particles in a diffuse ring is on the order of the mean residence time, while the age of the overall structure is related to how long the sources have been in place. 

The distinction between material age and overall structural age also turns out to be useful for Saturn's main rings.  Several aspects of the rings are difficult to reconcile with a ring age comparable to that of Saturn \citep[see][and references therein]{CharnozChapter09}, including 1) the $\gtrsim 95$\% water-ice composition of the A~and B~rings is difficult to reconcile with the constant pollution of the rings by infall of interplanetary micrometeoroids; 2) the same infall should significantly erode ring particles, especially in regions of lower optical depth; and 3) exchange of angular momentum with moons currently confining ring edges can only be ``rewound'' for $\sim 10^7$~yr.  On the other hand, the disruption of a sun-orbiting object containing sufficient mass in such a way as to form a ring system is an unlikely event given the recent state of the solar system, and Saturn is actually the least likely of the planets to capture an interloper because the balance of mass and solar distance gives it the smallest Hill sphere among the giant planets, leading to some doubts as to the viability of the young-rings scenario \citep{CharnozLHB09}.  

One potentially viable theory suggests that the B~ring core is ancient, dating from the first Gyr of the solar system in which collisions were much more frequent \citep{CharnozLHB09}, but that much of the specific organization of material in Saturn's rings is only $\sim 10^7$~yr old.  Recent indications that the B~ring's mass has previously been underestimated \citep[see Section~\ref{SGWs} and][]{Robbins10} provide an increased buffer against interplanetary pollution \citep{EE11}, as well as making the ring precursor body even larger and thus a recent ring origin even more unlikely.  Erosion of ring particles can be counteracted by an accretion-driven recycling process \citep{Durisen89,EspoAGU06} and is also slowed if ring particles have spent most of their history in a denser structure than those where they are now found.  The current ring-moons may also have formed within the rings and emerged only recently \citep{Charnoz10}.  Many questions still remain, however \citep{CharnozChapter09}.  The mass of the B~ring needs to be better known, as may well happen as a result of the planned close passes during \Cassit{}'s end-of-mission maneuvers \citep{SealBuffington09} both through direct sensing of the ring's gravitational pull and through its interaction with charged particles and gamma rays \citep[the latter improving on the \Pionit{~11} measurement reported by][]{Cooper85}.  Also needed are improvements in knowledge of the interplanetary impactor flux throughout the history of the rings.  The proposed recycling mechanism has yet to be carefully described and modeled.  And any theory for the formation of Saturn's rings, whether the age be young or old, requires careful separation of water ice from any silicate and/or metal that must have been part of any body accreting directly from the protoplanetary nebula. 

\begin{figure*}[!t]
\begin{center}
\includegraphics[width=12cm]{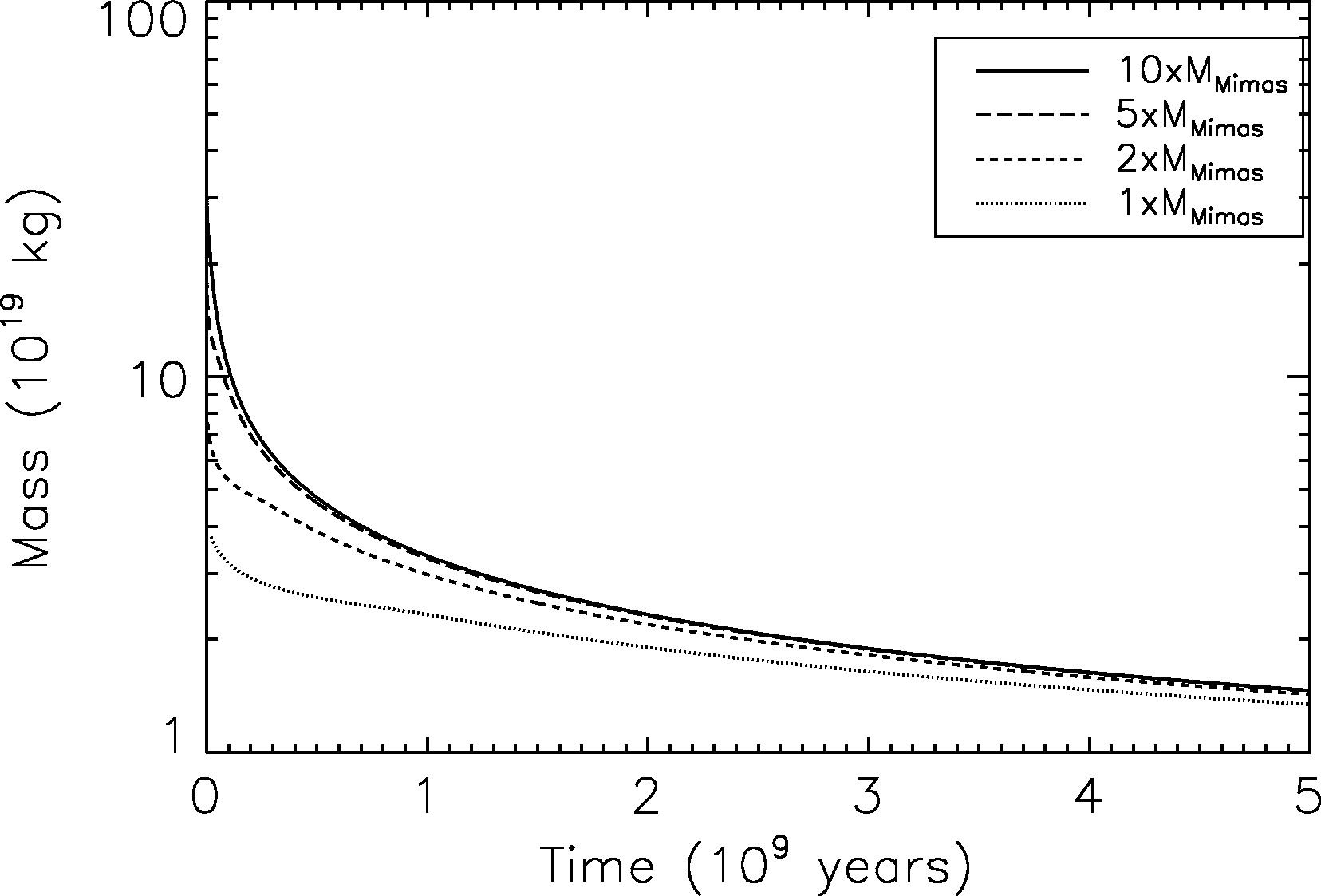}
\caption{Viscous spreading models predict that disk mass after 5~Gyr of evolution is relatively insensitive to the initial disk mass.  Figure from \citet{Salmon10}. 
\label{ViscousSpreadPlot}}
\end{center}
\end{figure*}

\citet{Canup10} recently suggested that Saturn's rings might have been formed by the disruption of a Titan-sized ($\sim$2500-km radius) proto-moon near the end of the planetary formation period during which the circum-planetary gas disk is present.  Due to its size, the proto-moon is differentiated, and as it spirals inward due to interaction with the disk, it first passes the Roche radius for ice (see Section~\ref{Roche}) and its icy mantle is stripped away.  Furthermore, because the incompletely consolidated Saturn would have been $\sim$50\% larger than at present, the proto-moon is engulfed by the planet before reaching the Roche radius for its rocky core, thus explaining the rings' icy composition.  The initially resulting ring is $\sim$1000 times more massive than that of today, but \citet{Salmon10} showed that viscous spreading of a ring over the age of the solar system can lead to a ring like that of today with relatively little sensitivity to the ring's initial mass (\Fig{}~\ref{ViscousSpreadPlot}).  Super-sizing the argument of \citet{Charnoz10}, \citet{Canup10} suggest that some of Saturn's mid-size moons (with radii of hundreds of km) may have been spawned by the viscous spreading of this massive ring, accounting for their relatively low densities (cf. Section~\ref{Roche}), and \citet{Charnoz11} in turn have explored this scenario in more detail.  The details of the \citet{Canup10} simulations are conducted in the context of that author's theory for circum-planetary disks \citep[e.g.,][]{CW02,CW06}, which has been criticized by some \citep[e.g.,][]{ME03a}.  However, the basic outline of the \citet{Canup10} story appears to have plausible merit even beyond its specific theoretical context. 

The rings of Uranus and Neptune, with their lower masses and much darker surfaces, have more plausibly remained little changed over the age of the solar system.  The uranian rings, as well as the nearby group of moons, may be part of a system that has oscillated between accretion and disruption for many~Gyr (Section~\ref{Uranus}).  But why are the rings where they are?  Is the dynamical environment such that a uniform supply of material at all distances from the planet would result in the rings as we see them today?  Or was the supply of source material somehow confined to the locations at which we now see rings?  On the planetary level, why does Saturn have its glorious broad dense disk while Uranus and Neptune have much more modest systems and Jupiter has only moon-generated dust?  None of these questions has a clear answer as yet. 

\section{Rings and other disks \label{OtherDisks}}

Planetary rings are just one variety among many disk-shaped systems known to astronomers, but they are the only variety that is not exceedingly far away in time or space or both and thus that is available for close inspection \citep{BC06}.  Other astrophysical disks include proto-planetary, proto-lunar, and proto-satellite disks as theorized for the origins of our solar system and as presently observed as gas disks, dust disks, and debris disks at other stars.  They also include accretion disks for binary stars, black holes, active galactic nuclei, etc. 

Some examples of fruitful cross-pollination between planetary rings studies and other astrophysical disks follow.  The interpretation of the inner edge of the Fomalhaut disk in terms of a resonant confinement as seen in planetary rings \citep{KalasFomB05} was validated by the discovery of Fomalhaut~b \citep{KalasFomB08}.  The dynamics of eccentric rings like Uranus' $\epsilon$ ring have been extended and applied to astrophysical eccentric disks such as ``superhump'' binary star systems \citep{Lubow10}.  Both spiral waves (Section~\ref{SpiralWaves}) and self-gravity wakes (Section~\ref{SGWs}) were first proposed as physical processes likely to occur in galactic disks, with specific applications to planetary rings coming a decade or more later.  But the shoe is now on the other foot, with direct observations of spiral waves and of SGWs in planetary rings having become so detailed that they can potentially inform understanding of similar processes occurring in galaxies.  A similar process may soon occur with free unstable normal modes, which have long been seen in numerical simulations of proto-planetary disks \citep{Laughlin97} and have now likely been observed directly in Saturn's B~ring \citep{SP10}.  Finally, the orbital evolution of disk-embedded ``propeller'' moons and the nature of their interaction with the disk is only just beginning to be directly observed in Saturn's rings (Section~\ref{Propellers}), potentially shedding light on the evolution of planetesimals and other disk-embedded masses. 

\acknowledgements{I thank Mark Showalter, Joe Burns, Josh Colwell, Jeff Cuzzi, Jonathan Fortney, Doug Hamilton, Matt Hedman, Doug Lin, Phil Nicholson, and John Weiss for helpful conversations.  I additionally thank Robin Canup, John Cooper, Estelle Deau, Larry Esposito, and Rob French for valuable comments on the manuscript.  I acknowledge funding from NASA Outer Planets Research (NNX10AP94G), NASA Cassini Data Analysis (NNX08AQ72G and NNX10AG67G), and the Cassini Project.}

\end{document}